\documentclass[pra,aps,amsmath,amssymb,twocolumn,floatfix,superscriptaddress]{revtex4-2}
\usepackage{amsmath}
\usepackage{graphicx}
\usepackage{color}
\usepackage{amssymb}
\usepackage[utf8]{inputenc}
\usepackage[T1]{fontenc}

\usepackage{hyperref}
\hypersetup{colorlinks,
 linkcolor=blue,%
 citecolor=blue,%
 urlcolor=blue
}
\usepackage{array}
\usepackage{comment}
\begin{document}
\title{Emergence of spin-mixed superstripe phases in spin-orbit coupled spin-1 condensate}

\author{Sanu Kumar Gangwar}
\affiliation{Department of Physics, Indian Institute of Technology, Guwahati 781039, Assam, India} 

\author{Rajamanickam Ravisankar}
\affiliation{Department of Physics, Saveetha School of Engineering, Saveetha Institute of Medical and Technical Sciences, SIMATS, Saveetha University, Chennai, 602 105, Tamil Nadu, India} 

\author{Paulsamy Muruganandam}
\affiliation{Department of Physics, Bharathidasan University, Tiruchirappalli 620024, Tamilnadu, India}
\affiliation{Department of Medical Physics, Bharathidasan University, Tiruchirappalli 620024, Tamilnadu, India}

\author{Pankaj Kumar Mishra}
\affiliation{Department of Physics, Indian Institute of Technology, Guwahati 781039, Assam, India}

\date{\today}
\begin{abstract} 
We numerically investigate the ground state phases and quench dynamics of spin-orbit coupled spin-1 Bose-Einstein condensates with ferromagnetic and antiferromagnetic interactions. For finite Rabi coupling, the system exhibits zero-momentum, elongated zero-momentum, and stripe phases, while in the limit $\Omega\rightarrow 0$, the superstripe wave phases emerge. Varying the attractive density-density ($c_0$) and spin-exchange ($c_2$) interactions induces the transition from stripe and superstripe phases to the elongated zero-momentum phase, characterized by miscibility and polarization. The condensate remains miscible in the zero-momentum phase and partially miscible for the elongated, stripe, and superstripe phases. Quenching in Rabi coupling stabilizes the condensate, while spin-orbit coupling quenching leads to fluctuations or the formation of complex patterns, with the stripe phase exhibiting the vanishing of the zeroth spin component over time. Our results may offer valuable insights into engineering exotic quantum phases in ultracold atomic gases. %
  
\end{abstract}


\maketitle
\section{Introduction}
Spin-orbit (SO) a coupling between a quantum particle's spin and its orbital angular momentum has emerged as a fundamental element in the condensed matter Physics, playing a crucial role in the realization of many of the complex quantum phases such as topological insulators~\cite{Hasan2010, Liang2011}, the quantum Hall effect~\cite{Kato2004, Konig2007}, and spintronics~\cite{Igor2004}. 
For atomic gases the SO coupling was artificially engineered in the laboratory experiment by Lin {\it et al.}~\cite{Lin2011} for spin-$\frac{1}{2}$ Bose-Einstein condensates (BECs) by using two of the three $F=1$ hyperfine states of $^{87}$Rb, effectively simulating the equal strength Rashba~\cite{Rashba1984} and Dresselhaus~\cite{Dresselhaus1955} couplings. The realization of SO coupling BECs in laboratory experiments has spurred an abundance of research in the field of ultracold atoms. Since its realization, a variety of novel phenomena, such as Dicke phase transition~\cite{Hamner2014}, observation of Zitterbewegung~\cite{Qu2013}, Faraday patterns~\cite{Zhang2022}, supersolid phase~\cite{Martone2021}, and elementary excitation~\cite{Chen2022}, have been reported in SO coupled spinor BECs. More recently, this approach has been extended to the spin-1 BECs employing all three $F = 1$ hyperfine states, generating SO coupled  spin-1 BECs~\cite{Campbell2016, Luo2016}. 

Over the years, numerical mean-field simulations based on the Gross-Pitaevskii equation (GPE) have proven to be a powerful tool for investigating complex phases and their stability in ultracold Bose gases. This framework has been extensively employed to study vortex formation~\cite{Bulgakov2003, Kato2011, Adhikari2021}, soliton dynamics~\cite{Katsimiga2021, Adhikari2021sym, Zhang2023}, and collective excitations~\cite{Sun2016}. Recent studies have explored various SO coupling regimes in quasi-1D BECs, revealing the existence of mobile vector solitons whose structure and dynamics depend on the spin-dependent interactions. In particular, ferromagnetic (FM) interactions support time-reversal symmetry-breaking solitons with single-peak density profiles~\cite{Gautam2015Mob}, whereas antiferromagnetic (AFM) interactions lead to multi-peak structures such as dark–bright–dark or bright–dark–bright solitons. These solitons remain dynamically stable and mobile, exhibiting quasi-elastic collisions at high velocities with conserved total density~\cite{Adhikari2020stable}.

Another key feature of SO coupled spinor condensates is the interplay between spin-dependent interactions, SO, and Rabi coupling, which governs miscibility and phase separation. Gautam \textit{et al.} \cite{Gautam2014} showed that in spin-1 BECs, any finite SO coupling induces phase separation in FM interactions in the absence of Rabi coupling. In contrast, SO coupling promotes miscibility in the AFM case. In trapped systems, the harmonic confinement generally favors miscibility, though AFM interactions may still exhibit phase separation above a critical SO coupling strength. Rabi coupling, in both FM and AFM regimes, enhances miscibility. Similar phase-separated structures have been reported in spin-2 SO coupled BECs under broken time-reversal symmetry \cite{Gautam2015S2}.

Quench dynamics in ultracold atomic systems have emerged as a powerful tool for probing nonequilibrium phenomena, including quantum phase transitions~\cite{Bookjans2011, Yang2019} and topological defect formation~\cite{Saito2007}, and collective excitations~\cite{Mishra2021, Gangwar2024}. In SO coupled BECs, sudden parameter changes have revealed rich dynamical behavior. For instance, Liu \textit{et al.}~\cite{Liu2019} observed a second-order transition to an immiscible phase with domain wall formation, consistent with the Kibble-Zurek mechanism. Zitterbewegung arising from inter-band oscillations following a Hamiltonian quench was reported in Ref.~\cite{Qu2013}. Deng \textit{et al.}~\cite{Deng2016} employed a time-dependent Bogoliubov approach to study the evolution of condensate fraction and momentum distribution after a quench, showing the emergence of a quasi-steady state. On the other hand, Ravisankar \textit{et al.}~\cite{Ravisankar_2020} demonstrated control of spin transport and identified diverse quench-induced features such as broken oscillations, breather states, miscibility transitions, soliton dynamics, and spin trapping.

There are relatively few works that explore the intricate relationship between the spin structure and the diverse phases of complex SO coupled BECs. For a spin-1/2 condensate, Ho and Zhang~\cite{Ho2011} identified a density-modulated stripe phase with zero magnetization under a special symmetric and spin-mixed limit. Later, Yu~\cite{Yu2016} has extended this analysis to spin-1 condensates in quasi-1D geometries. Their study categorically distinguished various phases, including the plane wave (PW), zero momentum (ZM), and several types of stripe (STR) phases. Notably, STR1 corresponds to a spin-mixed stripe phase with density modulation and vanishing longitudinal magnetization, while STR2 and STR3 exhibit finite magnetization accompanied by spin-tensor modulations.

Along similar lines, Sun \textit{et al.}~\cite{Sun2016} reported the existence of ZM and PW phases, as well as a stripe phase with equal spin population, characterized by uniform spin polarization and density modulation, and a magnetized stripe phase featuring oscillating spin polarization. More recently, Gangwar \textit{et al.}~\cite{Gangwar2024, Gangwar2025} examined the dynamical properties of quasi-1D SO coupled spin-1 BECs, focusing on excitation spectra, modulation instabilities, and avoided crossings under ferromagnetic (FM) and antiferromagnetic (AFM) interactions. While these works provided rigorous instability analyses, a systematic investigation of the ground state phase structure remains lacking.

In the present work, we address this gap by presenting a comprehensive study of both the ground state phases and dynamical properties of quasi-1D SO-coupled spin-1 BECs, with particular attention to the interplay between SO coupling and Rabi coupling. We identify a rich variety of ground state phases, including the zero momentum (ZM), elongated zero momentum (EZM~I and EZM~II), stripe wave (SW), and superstripe wave (SSW) phases. Notably, the ZM, EZM~I, and SW phases arise exclusively in AFM interactions and only in the presence of finite Rabi coupling. In contrast, the SSW phase appears in both FM and AFM interactions when the Rabi coupling is absent ($\Omega=0$), but vanishes as $\Omega$ becomes finite. Interestingly, the SSW phase exhibits a vanishing spin-density vector and, although it emerges as a robust ground state across a broad parameter regime, it is dynamically unstable. Moreover, the analysis of spin-component separation and miscibility shows that the ZM phase remains fully miscible. In contrast, the other phases display partial demixing among spin components, indicating distinct spin textures and correlations characteristic of each phase.

The structure of our paper is as follows. In Sec.~\ref{secmodel}, we present the mean field model to explore ground state phases and quench dynamics. In the following Sec.~\ref{sec:resdis}, we provide details about the numerical simulation. In Sec.~\ref{trapped_ferro}, we present the emergence of different ground state phases and characterization in FM interactions, followed by AFM interactions in Sec.~\ref{trapped_antiferro}. In Sec.~\ref{grndintr}, we exhibit the effect of interactions on the ground state phases. Further, in Sec.~\ref{quench}, we present the quench dynamics across these phases. Finally, in Sec.~\ref{sec:sumcon}, we summarize our findings.%
\section{Mean-Field Model}
\label{secmodel}
We consider a quasi-1D SO coupled spinor spin-$1$ BEC realized by tight confinement in the transverse direction. The GPEs govern the dynamics of the system, incorporating the effects of SO coupling and interatomic interactions. The generalized coupled GPEs are given as follows,
\begin{subequations}
\begin{align}
\mathrm{i} \frac{\partial \psi_{\pm 1}}{\partial t} &=  \left[- \frac{1}{2} \frac{\partial^{2}}{\partial x^{2}} + V(x) + c_{0} \rho\right] \psi_{\pm 1} \mp \frac{ k_{L}}{\sqrt{2}} \frac{\partial \psi_{0}}{\partial x} \notag \\ &+ c_{2}(\rho_{\pm 1} + \rho_{0} - \rho_{\mp 1}) \psi_{\pm 1} + c_{2} \psi_{0}^{2} \psi_{\mp 1}^{*}+\frac{\Omega}{\sqrt{2}}\psi_{0}, \label{gpe03} \\
\mathrm{i} \frac{\partial \psi_{0}}{\partial t} &=  \left[- \frac{1}{2} \frac{\partial^{2}}{\partial x^{2}}+ V(x) + c_{0}(\rho)\right] \psi_{0} + 2c_{2} \psi_{0}^{*} \psi_{+1}\psi_{-1} \notag \\ & + c_{2}( \rho_{+1} + \rho_{-1})\psi_{0} + \frac{ k_{L}}{\sqrt{2}} \bigg[ \frac{\partial \psi_{+1}}{\partial x}  - \frac{\partial \psi_{-1}}{\partial x}\bigg] \notag \\ & +\frac{\Omega}{\sqrt{2}}(\psi_{1}+\psi_{-1}). \label{gpe04}
\end{align}
\end{subequations}
Here, $\psi_{+1}$, $\psi_{0}$, and $\psi_{-1}$ are the spinor wave functions corresponding to sublevels $m_{F} = +1, 0, -1$ of the hyperfine state $F = 1$. These wave functions satisfy the normalization condition
\begin{align*}
\int_{-\infty}^{\infty} d x\; \sum_{j=-1}^{1} \vert\psi_{j}(x) \vert^2  = 1,
\end{align*}
where $\rho_{j}$ denotes the density of the $j$th component of the condensate ($j=\pm1,0$), and $\rho = \vert \psi_{+1} \vert^{2} + \vert \psi_{0} \vert^{2} + \vert \psi_{-1} \vert^{2}$ is the total atomic density. The coupled system of equations in (\ref{gpe03}) and (\ref{gpe04}) is expressed in dimensionless form, using the characteristic time, length, and energy scales as $t = \omega_{x} \tilde{t}$, $x = \tilde{x}/l_{0}$, and $\hbar \omega_{x}$, respectively. The condensate wave function is defined as $\psi_{\pm 1,0} = \tilde{\psi}_{\pm 1,0} \sqrt{ l_{0}/N}$, where $l_{0} = \sqrt{\hbar / (m \omega_{x})}$ is the harmonic oscillator length for a given trap frequency $\omega_x$ along the $x$ axis.

The trap potential is $V(x)=x^{2} /2$. The density-density interaction strength is $c_{0} = 2 N l_{0}(a_{0} + 2a_{2})/ (3 l_{\perp}^{2})$, and the spin-exchange interaction strength is $c_{2} = 2 N l_{0}(a_{2} - a_{0})/ (3 l_{\perp}^{2})$, where $a_{0}$ and $a_{2}$ are the $s$-wave scattering lengths in the total spin channels $0$ and $2$, respectively. By tuning $c_{2} < 0$ ($ c_2 > 0$), one obtains FM (AFM) interactions of the condensate~\cite{Ueda2012, Stamper2013}. The transverse oscillator length is $l_{\perp} = \sqrt{ \hbar / (m \omega_{\perp})}$, with $\omega_{\perp} = \sqrt{\omega_{y}\omega_{z}}$. The SO and Rabi coupling strengths are $k_{L} = \tilde{k}_{L} / (\omega_{x} l_{0})$ and $\Omega = \tilde{\Omega} / (\hbar \omega_{x})$, respectively. In the above description, quantities with a tilde represent dimensionful parameters. In all results presented in this work, we use dimensionless parameters.


The energy Functional of the system is given by,
\begin{align}
E  = & \frac{1}{2} \int dx \bigg\{ \sum_{j} \left\vert \partial_{x} \psi_{j} \right\vert^{2} + 2 V(x) \rho + c_{0} \rho^{2}  \nonumber  \\  & + c_{2}[ \rho_{+1}^{2} + \rho_{-1}^{2} + 2( \rho_{+1}\rho_{0}+\rho_{-1}\rho_{0} -\rho_{+1}\rho_{-1} \nonumber  \\  & +\psi_{-1}^{*}\psi_{0}^{2}\psi_{+1}^{*}+ \psi_{-1}\psi_{0}^{*2}\psi_{+1})]  + \sqrt{2} \Omega[(\psi_{+1}^{*}+\psi_{-1}^{*})\psi_{0} \nonumber  \\  & +\psi_{0}^{*}(\psi_{+1}+\psi_{-1})]+ \sqrt{2} k_{L}[  (\psi_{-1}^{*}-\psi_{+1}^{*})  \partial_{x} \psi_{0}  \nonumber  \\  & + \psi_{0}^{*}(\partial_{x} \psi_{+1}- \partial_{x}\psi_{-1})]\bigg\} \label{ene02}
\end{align}
As the primary focus of our work is to differentiate phases based on spin-mixing and spin-demixing behavior, we start by precisely defining the key quantities: the miscibility order parameter, root mean square size, and spin-density vector (SDV) polarization of the condensate for SOC spin-1 BECs.

\paragraph{Miscibility order parameter:}
The miscibility order parameter ($\eta$) of the spin-1 BECs characterizes the overlap of the densities $\psi_{+1}$, $\psi_{0}$, and $\psi_{-1}$. Another overlap parameter $\eta_{np}$, which specifically represents the overlap between the densities of $\rho_{+1}$, and $\rho_{-1}$ of the condensate. The miscibility order parameter and overlap parameter are given as follows~\cite{Ravisankar_2021}, 
\begin{subequations}\label{eq:overlap}
\begin{align}\label{eqa:overlap}
\eta_{np} & =  4 \int (\sqrt{\rho_{+1} \rho_{-1}}) \, dx, \\
\label{eqb:overlap}
\eta & = \int (\sqrt{\rho_{+1} \rho_{0}} + \sqrt{\rho_{-1} \rho_{0}} + \sqrt{\rho_{+1} \rho_{-1}}) \, dx.
\end{align}
\end{subequations}

\paragraph{Root Mean Square size:}
The other important entity is the root mean square size ($x_{rms}$) of the condensate which specifically being used to characterize the spatial extent and associated phase transformation of its ground state. The $x_{rms}$ of the condensate is given as follows~\cite{SKSarkar2024},
\begin{align}
x_{rms} = \bigg( \frac{\int_{-\infty}^{\infty} (x - \langle x\rangle)^{2} \rho dx}{\int_{-\infty}^{\infty} \rho dx} \bigg)^{1/2}
\end{align}
with,
\begin{align}
\langle x \rangle = \frac{\int_{-\infty}^{\infty} x \rho dx}{\int_{-\infty}^{\infty} \rho dx}
\end{align}
where $\rho$ is the total density of the condensate.

\paragraph{Spin-density Vector:}
The spin-density vector (SDV) in  spin-1 spinor BECs plays a key role in analyzing the spin textures. It is expressed as $\mathrm{S} = \psi^{\dagger} \Sigma \psi$, where $\psi = [\psi_{+1}, \psi_{0}, \psi_{-1}]^{T}$ is normalized wavefunction and $\Sigma$ = ($\Sigma_{x}$, $\Sigma_{y}$, $\Sigma_{z}$) are the spin-1 Pauli matrices. The SDV components are given as~\cite{Ueda2012},
\begin{subequations}\label{eqn:sdv}
\begin{align}
\mathrm{S}_{x} & = \frac{1}{\sqrt{2}} \left[(\psi_{+1}^{*} + \psi_{-1}^{*}) \psi_{0} + (\psi_{+1} + \psi_{-1}) \psi_{0}^{*} \right] \\
\mathrm{S}_{y} & = \frac{\mathrm{i}}{\sqrt{2}}\left[(\psi_{-1}^{*} - \psi_{+1}^{*}) \psi_{0} + (\psi_{+1} - \psi_{-1}) \psi_{0}^{*}\right] \\
\mathrm{S}_{z} & = \vert \psi_{+1}\vert^{2} - \vert\psi_{-1}\vert^{2}
\end{align}
\end{subequations}

Here, we outline the experimentally realistic range of the parameters considered in our simulations. We consider the  $^{87}$Rb BECs with $N \sim 2\times 10^{4}$ number of atoms with FM interaction trapped under the axial trap frequency as $\omega_x = 2 \pi \times 50$ Hz and the transverse trap frequencies as $\omega_y = \omega_z = 2 \pi \times 500$ Hz. With these, the resultant characteristic lengths are $l_0 = 1.52 \, \mu$m and $l_\perp = 0.48 \, \mu$m. 
For the AFM interaction, one can consider the $^{23}$Na atoms with characteristic lengths as $l_{0} = 2.9 \mu m$, and $l_{\perp} = 0.9 \mu m$. The spin-dependent and spin-independent interactions can be achieved by controlling the $s$-wave scattering lengths via Feshbach resonance~\cite{Inouye1998, Marte2002, Chin2010}. The SO coupling strength $k_L = \{ 0.1 - 5 \}$ can be tuned by changing the laser wavelengths in the range of $\{68.86 \mu m$ - $1377.22 \mbox{nm}\}$. However, the dimensionless Rabi frequency interval $\Omega=[0,5]$ used in the simulation can be attained by tuning the Raman laser strength in the range of $2\pi \hbar\times \{5 - 250\}$ Hz.%
\section{Simulation results}
\label{sec:resdis}
We numerically solve the coupled GP equations (\ref{gpe03})-(\ref{gpe04}) using the split-time step Crank-Nicolson scheme. To determine the ground state, we use imaginary-time propagation, which relaxes the system to its lowest energy state but does not conserve normalization or magnetization; hence, renormalization is applied at each time step appropriately. The initial wave function is chosen as a Gaussian profile. We employ the real-time propagation method to investigate the dynamics of the condensate using the converged ground state obtained in the imaginary-time propagation method. The space and time steps used in the imaginary-time propagation method are $dx = 0.05$, $dt = 0.0002$, and in the real-time propagation method are $dx = 0.05$, $dt = 0.0001$. In the next section, we implement these and study the ground states and miscibility of the condensate in the presence of FM interactions.%
\subsection{Ground state phases for FM interactions}
\label{trapped_ferro}
In this section, we present the different ground state phases of quasi-1D SO coupled spin-1 FM BECs. We analyze the ground state phases in the absence of magnetization by examining key properties such as the spin-density vector (SDV) polarization, miscibility, and the root mean square size.%
\subsubsection{ Different ground state phases}
\begin{figure}[!htp]
\centering\includegraphics[width=0.99\linewidth]{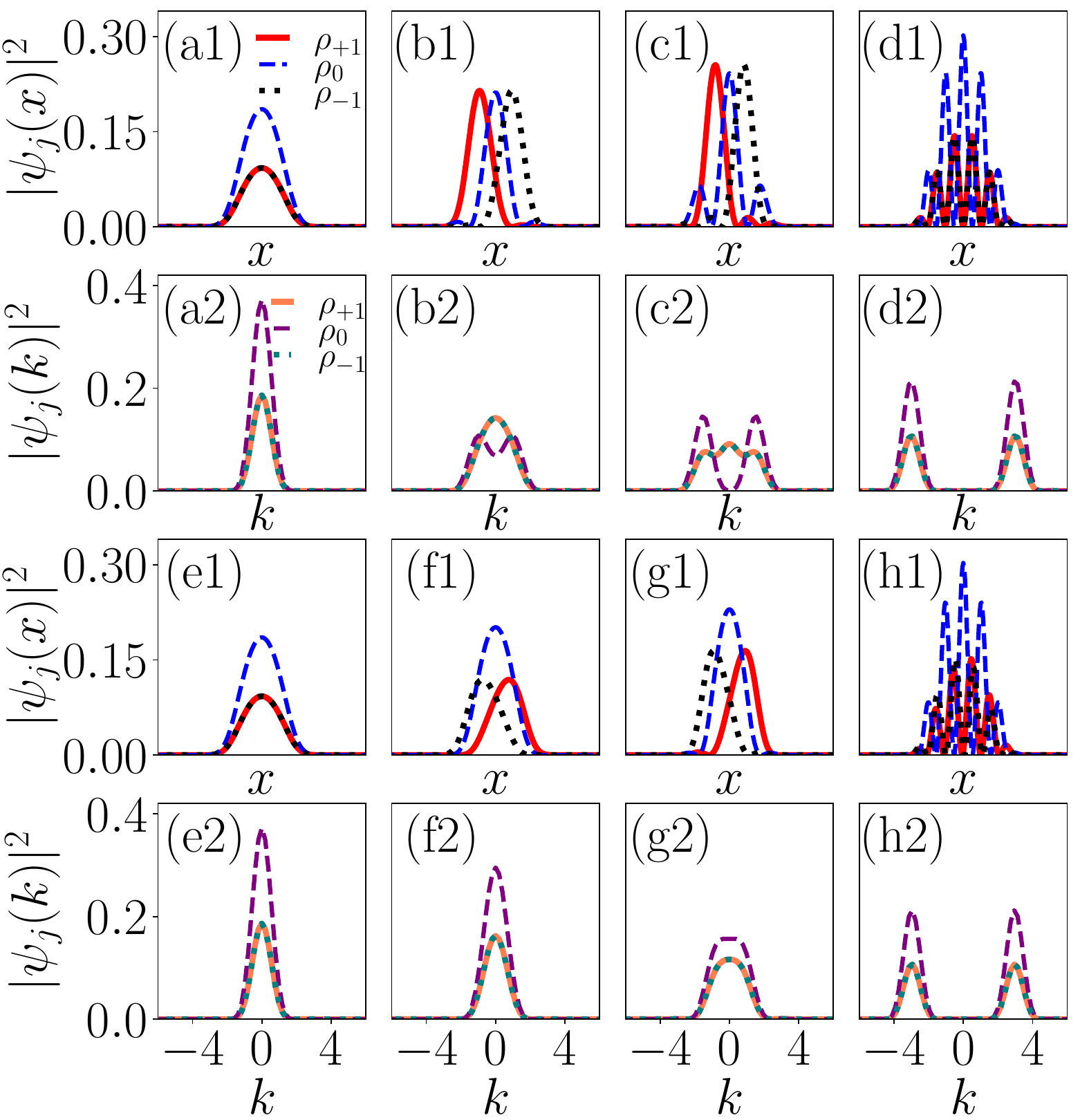}
\caption{Different ground state phases of the SO coupled spin-1 BECs with FM interaction. Real-space (top) and momentum-space (bottom) density profiles of spin components $\vert \psi_{+1}(x)\vert^{2}$ (solid red), $\vert \psi_{0}(x)\vert^{2}$ (dashed blue), and $\vert \psi_{-1}(x)\vert^{2}$ (dotted black), along with their corresponding momentum-space densities $\vert \psi_{+1}(k)\vert^{2}$ (solid coral), $\vert \psi_{0}(k)\vert^{2}$ (dashed purple), and $\vert \psi_{-1}(k)\vert^{2}$ (dotted green), under varying Rabi coupling $\Omega$ and spin-orbit coupling $k_L$. Interaction strengths are fixed at $c_0 = 10$ and $c_2 = -5.0$. Panels (a–d) correspond to $\Omega = 0$ and panels (e–h) to $\Omega = 1.0$, with $k_L = 0.0, 1.0, 1.5, 3.0$ increasing from left to right in each set. For $\Omega \approx 0$, increasing $k_L$ induces transitions through ZM, EZM~I, SW, and SSW ground state phases. At finite $\Omega$, the SSW phase is absent.
}
\label{fig01a} 
\end{figure}%
We begin our investigation with the ground state phases of the SO coupled spin-1 BECs in presence of FM interactions with the interaction strengths $c_{0} = 10$ and $c_{2} = -5.0$. Fig.~\ref{fig01a} illustrates the impact of SO ($k_{L}$) and Rabi ($\Omega$) couplings on various ground state phases. In Fig.~\ref{fig01a}(a1), we present the ground state density profile for $\Omega = k_{L} = 0.0$.  All condensate components exhibit a single density peak at the center with $\rho_{\pm 1}$ overlapping in the real space. To characterize the detailed structure, we show the corresponding momentum space distribution in Fig.~\ref{fig01a}(a2). The momentum state density distribution shows a single peak centered at $k = 0$, indicating that the condensate occupies the zero-momentum state, which we classify as a zero-momentum (ZM) phase. Upon increasing the SO coupling to $k_{L}=1.0$ leads to a clear spatial separation between the densities $\rho_{\pm 1}$ as seen in Fig.~\ref{fig01a}(b1). The densities $\rho_{+1}$ and $\rho_{-1}$ peak away from the trap center at $x = -0.95$ and $0.95$, respectively, with a similar amplitude $\rho_{\pm 1}(max) = 0.22$, while the zeroth component density starts developing the second hump in real space. The corresponding momentum distribution [Fig.~\ref{fig01a}(b2)] reveals two peaks for the zeroth component of the condensate, indicating a slight deviation from the ZM phase. We classify this as an elongated zero-momentum phase of the first type (EZM~I) phase due to the spatial deformation and partial separation of the spin components. 

Further, increasing $k_{L}$ from  $1.0$ to $1.5$ results in multi-peak structure in real space density profile [Fig.~\ref{fig01a}(c1)]. The corresponding density profile in the momentum space [Fig.~\ref{fig01a}(c2)] exhibits prominent peaks at finite $k$, especially the zeroth component, with $m_{F} = \pm 1$ also peaking up at finite $k$. These features signal the onset of the stripe wave (SW) phase, where interference between multiple plane wave components in momentum space leads to spatial modulation of the condensate density. Upon further increasing the SO coupling to $k_L=3.0$, the condensate develops a pronounced multi-peak density profile in the real space with overlap between the $\rho_{+1}$ and $\rho_{-1}$, as illustrated in Fig.~\ref{fig01a}(d1). The corresponding momentum density profile [Fig.~\ref{fig01a}(d2)] exhibits symmetric peaks at finite $k$ for all spin components.  This structure is consistent with the characteristics of the superstripe wave (SSW) phase, a subset of the SW phase characterized by complete overlap of $m_{F} = \pm 1$ components and unpolarized behaviour.  

To investigate the effect of SO coupling, we fix $\Omega = 1.0$. For $k_{L} = 0.0$, in real and momentum space densities [Fig.~\ref{fig01a}(e1,e2)] remain centered and single peaked across all components, indicating persistence of the ZM phase. For  $k_{L} = 1.0$  as shown in Fig.~\ref{fig01a}(f1), the second hump in the zeroth component density disappears and densities $\rho_{+1}$ and $\rho_{-1}$ posses peaks away from the trap center at $x = 0.75$, and $-0.75$ with $\rho_{\pm 1}(max) = 0.12$. The corresponding momentum density profile displays only a single peak across all components given in Fig.~\ref{fig01a}(f2). However, in real space $\rho_{\pm1}$ shows separation among themselves, resulting in an EZM~I phase. Upon increasing $k_{L} = 1.0$ to $1.5$, in Fig.~\ref{fig01a}(g1), we again obtain a single peak density across all components and $m_{F} = +1 (-1)$ components peaks at $x = 0.9 (-0.9)$ with a amplitude $\rho_{\pm 1}(max) = 0.17$. In the corresponding momentum distribution in Fig.~\ref{fig01a}(g2), a flat-top-like distribution emerges, peaks at $k \approx 0$, and the condensate remains in the EZM~I phase. Considering $\Omega = 1.0$ and $k_{L} = 3.0$ in Fig.~\ref{fig01a}(h1), we observe a multi-peak density profile across all components in real space, and the corresponding momentum space density profile in Fig.~\ref{fig01a}(h2) displays two peaks across all components. Here $m_{F} = \pm 1$ does not superpose on each other, which indicates that the condensate is in the SW phase. Although the polarization behaviour between $m_{F} = \pm 1$ component is not evident in the momentum space distributions, it becomes clearly discernible in the subsequent analysis of spin density vectors. Overall, we find that a finite Rabi coupling strength shifts the emergence of the SW phase to higher values of the SO coupling. The SSW phase appears only for $\Omega \sim 0 $ and for large $k_{L}$.  Otherwise, depending on the interplay between SO and Rabi coupling, the system predominantly exhibits the ZM, EZM~I, or SW phases.

\begin{figure}[!htp] 
\centering\includegraphics[width=0.99\linewidth]{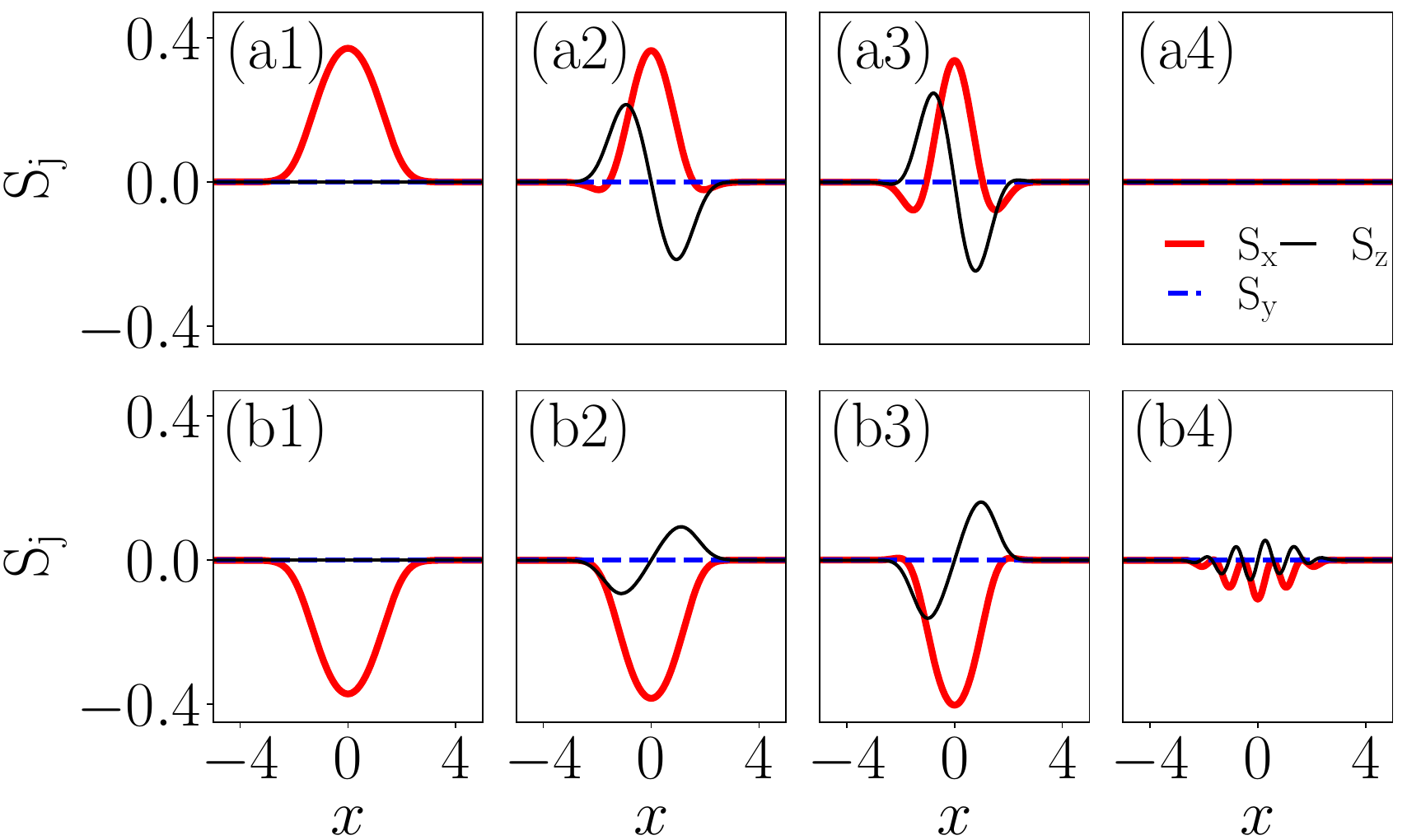}
\caption{Spin-density vector (SDV) components $\mathrm{S}_{x}$ (solid red line), $S_{y}$ (dashed blue line), and $\mathrm{S}_{z}$ (thin black line) shown in panels (a) and (b), corresponding to Fig.~\ref{fig01a}, using the same interaction and coupling parameters. The $\mathrm{S}_{y}$ component remains zero across all phases. In the ZM phase, $\mathrm{S}_{x}$ shows a single peak while $\mathrm{S}_{z}$ vanishes. The EZM~I phase features a single-peak $\mathrm{S}_{x}$ accompanied by finite $\mathrm{S}_{z}$. The SSW phase is characterized by vanishing SDV components, whereas the SW phase displays a multipeak structure in $\mathrm{S}_{x}$ and oscillations in $\mathrm{S}_{z}$, indicating density modulation.}
\label{fig01a-sdv} 
\end{figure}%

To characterize the different ground state phases, we compute the spin-density vector (SDV) polarization, which captures key features of the spin textures and their evolution. This makes it a valuable tool for distinguishing the phase transitions discussed above~\cite{Ravisankar_2021}. In Fig.\ref{fig01a-sdv}, we present the SDV components $\mathrm{S}_{x}$, $\mathrm{S}_{y}$ and $\mathrm{S}_{z}$ as defined in Eq.~(\ref{eqn:sdv}) corresponding to the different phases of Fig.~\ref{fig01a}. We begin with Fig.~\ref{fig01a-sdv}(a1), where $\Omega = k_{L} = 0.0$. In this case, $\mathrm{S}_{x}$ is non-zero and displays a single peak, consistent with the ZM phase of the condensate. Upon increasing the SO coupling to $1.0$ [Fig.~\ref{fig01a-sdv}(a2)], the single peak in $\mathrm{S}_{x}$ remains, while a non-zero $\mathrm{S}_{z}$ emerges, ranging from $-0.22$ to $0.22$. This non-zero $\mathrm{S}_{z}$ results from a distinct separation between the densities $\rho_{+1}$ and $\rho_{-1}$, confirming the transition to the EZM~I phase. The finite magnitude of $\mathrm{S}_{z}$ indicates that the $m_{F} = \pm 1$ are not superposed. 

For $k_{L} = 1.5$ with fixed $\Omega$ shown in Fig.~\ref{fig01a-sdv}(a3), $\mathrm{S}_{x}$ maintains a single peak but features a peculiar negative dip over a small range of $x$, forming a symmetric kink-like structure. Simultaneously the amplitude of $\mathrm{S}_{z}$ increases to the range $[-0.25, 0.25]$. This kink-like feature arises due to the multi-peak density profile, confirming a transition to the SW phase. At higher SO coupling $k_{L} = 3.0$, Fig.~\ref{fig01a-sdv}(a4) shows that all the SDV components vanish. Since non-zero $\mathrm{S}_{z}$ indicates a separation between the $m_{F} = \pm 1$ components, the condition $\mathrm{S}_{z} = 0$ confirms the unpolarized configuration, consistent with the SSW phase. The vanishing  $S_{x}$ results from the wavefunctions $\psi_{+1}$, and $\psi_{-1}$ being equal in magnitude but opposite in sign. In Fig.~\ref{fig01a-sdv}(b1), we explore $\Omega = 1.0$ while varying the SO coupling. Notably, for finite $\Omega$, spin flip effects appear in the SDV. For $k_{L} = 0.0$, $\mathrm{S}_{x}$ exhibits single peak and and $\mathrm{S}_{z} = 0$, indicating a complete overlap between the densities $\rho_{\pm 1}$), characteristics of the ZM phase. Upon $k_L$ is increased  to $1.0$ and then $1.5$ [Figs.~\ref{fig01a-sdv}(b2), and (b3)], the single peak structure in $\mathrm{S}_{x}$ persists and the amplitude of $\mathrm{S}_{z}$ increases within the ranges $[-0.1,0.1]$ and $[-0.16, 0.16]$, respectively, confirming the EZM~I phase. For $k_{L} = 3.0$ [Fig.~\ref{fig01a-sdv}(b4)], $\mathrm{S}_{x}$ develops a multi-peak structure, while, $\mathrm{S}_{z}$ displays oscillation, confirming the SW phase. Throughout all the configurations, $\mathrm{S}_{y}$ remains zero.

\subsubsection{Effect of SO coupling and Rabi coupling on the miscibility and ground state phase transformation}
Here, we examine the miscibility of the condensate and ground state phase transformation using the root mean square size of the condensate. Initially to characterize the overlap between $\rho_{+1}$ and $\rho_{-1}$, we define $\eta_{np}$ as given in Eq.(\ref{eqa:overlap}). We also define $\eta$, which calculates the total overlap between the densities $\rho_{+1}$, $\rho_{0}$, and $\rho_{-1}$ given in Eq.(\ref{eqb:overlap}). The total overlap between the condensate components is characterized as the miscibility of the condensate. If $\eta \approx 1$, the condensate is miscible, and if $\eta \approx 0$, it is immiscible; otherwise, it is partially miscible~\cite{Ravisankar_2021}.%
 
\begin{figure}[!htp] 
\centering\includegraphics[width=0.99\linewidth]{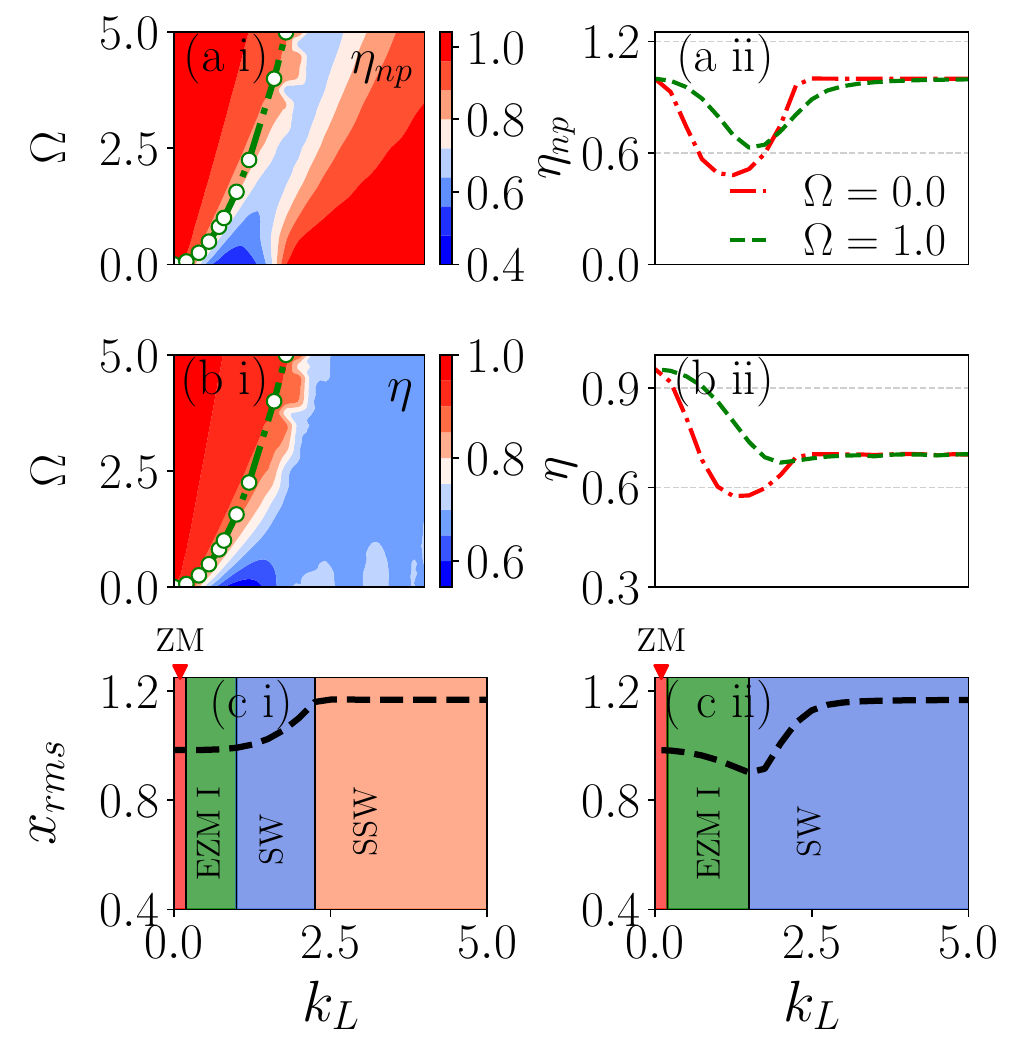}
\caption{Miscibility parameters (a) $\eta_{np}$ and (b) $\eta$ for interaction strengths $c_{0} = 10$ and $c_{2} =-5.0$. The top rows of (a) and (b) show pseudo color representation of $\eta_{np}$ and $\eta$ as a function of $k_{L}$ and $\Omega$ in the range [0,5]. The green line with white dots marks the critical relation $k_{L}^{2} \approx \Omega$, beyond which both $\eta_{np}$ and $\eta$ begin to decrease. $\eta_{np}$ shows decreasing trend for intermediate values of $k_{L}$ and then rises again. (a ii) and (b ii) show the $\eta_{np}$ and $\eta$ as a function of $k_{L}$ for $\Omega = 0.0$ (dashed-dotted red line), and $1.0$ (dashed green line). Panels (c i) and (c ii) show the root mean square size ($x_{\mathrm{rms}}$) as a function of $k_{L}$ for $\Omega = 0.0$ and $\Omega = 1.0$, respectively. Background color shading indicates distinct quantum phases: red (ZM), green (EZM~I), blue (SW), and coral (SSW).
}
\label{fig01f} 
\end{figure}%

In Fig.~\ref{fig01f}(a i), we report $\eta_{np}$ upon simultaneously varying $k_{L}$ and $\Omega$. Upon varying $k_{L}$, keeping $\Omega$ at some finite value. Initially, we obtain $\eta_{np} \approx 1.0$, representing the maximum overlap between densities $\rho_{\pm 1}$ corresponding to the ZM phase reported in the first column of Fig.~\ref{fig01a}. For intermediate values of $k_{L}$, the value of the overlap parameter $\eta_{np}$ decreases and reaches $\eta_{np} \approx 0.586$ at $\Omega = 0.5$, and $k_{L} = 1.50$, due to the separation between densities $\rho_{+1}$ and $\rho_{-1}$, characteristic of the EZM~I phase. Further increasing $k_{L}$ leads to a transition from the EZM~I phase to the SW phase, accompanied by an increasing trend in $\eta_{np}$, which reaches approximately $\eta = 0.999$. For $\Omega \sim 0$ and $k_{L} > 2.5$, we observe the maximum overlap between densities $\rho_{+1}$ and $\rho_{-1}$ components, resulting in an unpolarized SW phase (SSW phase), which is given in Fig.~\ref{fig01a}(a4). For a better description of the overlap between the $m_{F} = \pm 1$ components of the condensate in Fig.~\ref{fig01f}(a ii), we report the variation of $\eta_{np}$ for two different Rabi coupling strengths $\Omega = 0.0$ and $1.0$. Considering $\Omega = 0.0$, at $k_{L} = 0.0$, we observe $\eta_{np} \approx 1.0$ which is the maximum overlap between densities $\rho_{+1}$ and $\rho_{-1}$ components. Upon increasing the $k_{L}$, $\eta_{np}$ starts decreasing, gets a minimum at $k_{L} \approx 1.25$. After this point, it starts to increase and gets the maximum overlap ($\eta_{np} = 1.0$) at $k_{L} \approx 2.5$. For $\Omega = 1.0$, $\eta_{np}$ depicts the same behavior; however, the minimum shifts upward. At $k_{L} = 2.5$, it reaches $\eta_{np} \approx 0.889$, and for higher $k_{L}$, it approaches saturation with magnitude approximately equal to $0.999$.%

Next, in Fig.~\ref{fig01f}(b i), we report the variation of the miscibility parameter $\eta$, which varies in the range $0.58$ to $0.96$. The miscibility parameter gains the maximum amplitude $\eta \approx 0.96$ in the ZM phase. As $k_{L}$ increases, the strength of the miscibility parameter decreases, approaching a minimum value of $\eta \approx 0.58$ (in other phases, likely the EZM~I phase, the SW phase, and the SSW phase). The lowest values of both overlap parameters are observed for $\Omega \sim 0$, and $k_{L} \gtrsim 0.5$, and $\lesssim 2.5$. To understand miscibility of the condensate in a better way, in Fig.~\ref{fig01f}(b ii), we show the variation of the miscibility parameter $\eta$ with $k_{L}$ for two different Rabi coupling strengths $\Omega = 0.0$, and $1.0$. Considering $\Omega = 0.0$, $k_{L} = 0.0$, we obtain maximum overlap among the components ($\eta = 0.96$). Upon increasing $k_{L}$, the magnitude of the miscibility parameter decreases to $\eta = 0.58$ at $k_{L} \approx 1.30$. After this, it starts increasing and saturates at $k_{L} \approx 2.50$ with $\eta = 0.70$. We also calculate $\eta$ for $\Omega = 1.0$, which shows similar trends with increasing $k_{L}$; however, the minimum of $\eta$ shifts upward. The overlap parameter $\eta_{np}$ and miscibility order parameter $\eta$ begin to decrease $k_{L}^2 \approx \Omega$, which separates the ZM phase from other phases. However, $\eta_{np}$ decreases only for the intermediate values of $k_{L}$ and rises again to reach the near-maximum value. Both parameters show a minimum amplitude at $\Omega \sim 0$ and $k_{L}$ in the range [0.5, 2.5].%

In Figs.~\ref{fig01f}(c i), and (c ii), we analyze the $x_{rms}$ size of the condensate for Rabi coupling strengths $\Omega = 0.0$, and $1.0$, respectively upon varying the $k_{L}$ in the range [0, 5]. At first, in Fig.~\ref{fig01f}(c i), we consider $\Omega = 0.0$. For small $k_{L} \lesssim 0.20$, we obtain $x_{rms} = 0.984$, and the condensate is in the ZM phase. The $x_{rms}$ remains flat up to $k_{L} \approx 1.0$ due to no significant change in density profile except separation between the $m_{F} = \pm 1$ components, which corresponds to the EZM~I phase. Beyond $k_{L} > 1.0$, $x_{rms}$ starts to increase due to the density modulation across all components, indicating a transition from the EZM~I phase to the SW phase. The $x_{rms}$ reaches a maximum value of $1.169$ at $k_{L} = 2.50$ and then saturates, indicating the transition to the SSW phase with significant density modulations. Next, in Fig.~\ref{fig01f}(c ii), we consider $\Omega = 1.0$. For small $k_{L} \approx 0.20$, we obtain $x_{rms} = 0.982$, representing the ZM phase. Upon increasing $k_{L}$, the $x_{rms}$ size starts decreasing and forms a minimum $x_{rms} = 0.902$ at $k_{L} = 1.50$; in this range of $k_{L}$ condensate lies in the EZM~I phase, where the condensate experiences an interplay between SO and Rabi couplings. Furthermore, $x_{rms}$ increases and reaches $x_{rms} = 1.164$ at $k_{L} = 3.50$, after which it saturates, indicating the onset of the SW phase characterized by density modulations. For $\Omega \sim 0$, at high $k_{L}$, the SSW phase dominates, while for finite Rabi coupling, the SW phase dominates.%

\subsection{Ground state phases and spin texture for AFM interactions}
\label{trapped_antiferro}
So far, we have explored different types of ground-state phases arising from the FM interaction in SO-coupled spin-1 BECs. In this section, we extend the analysis to different ground-state phases with AFM interactions. Specifically, we consider the case of vanishing magnetization and characterize the phases using the SDV polarization, miscibility, and the root mean square size.
\subsubsection{Different ground state phases}
\begin{figure}[!htp]
\centering\includegraphics[width=0.99\linewidth]{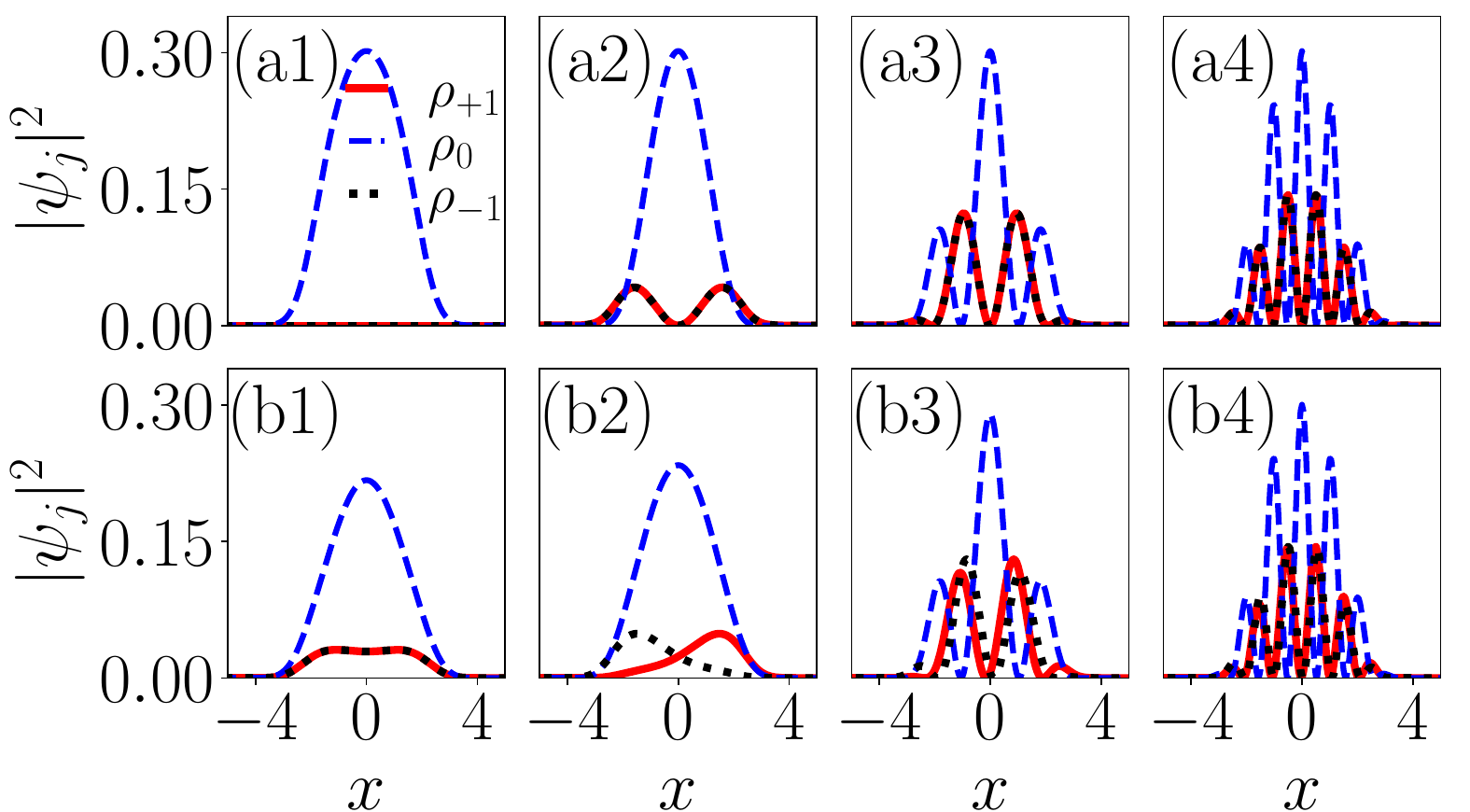}
\caption{Different ground state phases of the SO coupled spin-1 BECs with AFM interaction. Ground state density profile of the spin-component $\vert \psi_{+1} \vert^{2}$ (solid red line), $\vert \psi_{0} \vert^{2}$ (dashed blue line), and $\vert \psi_{-1} \vert^{2}$ (dotted black line) for (a1-a4) $\Omega = 0.0$ and (b1-b4) $\Omega = 1.0$, with interaction strengths $c_{0}= 10$ and $c_{2} = 5$. In each panel,  $k_{L}$ varies from left to right: $0.0$, $0.5$, $1.5$, and $3.0$. For $\Omega \sim 0$, the ZM phase, EZM~II phase, and SSW phase are observed. For finite $\Omega$, the SSW phase disappears, giving rise to the appearance of EZM~I and SW phases.}
\label{fig02a} 
\end{figure}%

We begin our discussion by demonstrating the different ground-state phases that arise under AFM interactions, with $c_{0} = 10$ and $c_{2} = 5$. Here, we vary the SO coupling row-wise and the Rabi coupling across columns.

We first consider zero SO and Rabi coupling. Fig.~\ref{fig02a}(a1) shows the condensate in the zero-momentum (ZM) polar phase, where all atoms occupy the $\psi_{0}$ component. Consequently, the densities $\rho_{+1}$ and $\rho_{-1}$ vanish identically. Upon increasing $k_{L} = 0.0 \rightarrow 0.5$ [Fig.\ref{fig02a}(a2)], the densities $\rho_{+1}$ and $\rho_{-1}$ begin to emerge with finite magnitude, while remaining fully spatially overlapping. These components peak away from the trap center at $x = 1.60$, with amplitude $\rho_{\pm 1}(\text{max}) = 0.42$. The behavior remains spatially symmetric, indicating a redistribution of population into the $m_{F} = \pm 1$ components. In momentum space, the $m_{F} = \pm 1$ components develop peaks at finite $k$, whereas the $\psi_0$ component remains centered at $k = 0$ (not shown here). These features---symmetric off-center density peaks in real space, overlapping $m_{F} = \pm 1$ components, and the absence of visible density modulation---are characteristic of the elongated zero-momentum phase of the second type (EZM~II). Increasing $k_{L} = 0.5 \rightarrow 1.5$ leads to a multipeak density profile, indicative of an unpolarized stripe wave (SW) or superstripe wave (SSW) phase, since the densities $\rho_{+1}$ and $\rho_{-1}$ remain fully overlapping [Fig.\ref{fig02a}(a3)]. Further increasing the SO coupling to $k_{L} = 3.0$ results in the same SSW phase, although with an increased number of stripes [Fig.\ref{fig02a}(a4)].

We now vary the Rabi coupling in the respective columns. Considering $\Omega = 1.0$ and $k_{L} = 0.0$, the condensate remains in the ZM phase, with finite and fully overlapping $\rho_{+1}$ and $\rho_{-1}$ densities [Fig.\ref{fig02a}(b1)]. The introduction of either SO or Rabi coupling induces the revival of the $m_{F} = \pm 1$ components under AFM interactions [see Figs.\ref{fig02a}(a2) and (b1)]. Upon increasing $k_{L}=0.0 \rightarrow 0.5$ in Fig.\ref{fig02a}(b2), we obtain the EZM~I phase with a single-peak density profile across all components. The $m_{F} = \pm 1$ components are centered away from the trap center at $x = \pm 1.50$, respectively, with $\rho_{\pm 1}(\text{max}) = 0.049$. The SW phase emerges as $k_{L}$ is increased to $1.5$ [Fig.\ref{fig02a}(b3)], and remains present at $k_{L} = 3.0$ with an enhanced number of stripes [Fig.\ref{fig02a}(b4)]. The polarization behavior between the $m_{F} = \pm 1$ components will become clearer in the subsequent discussion on spin-density vectors.

In the absence of SO and Rabi coupling, all atoms reside in the $\psi_{0}$ component. The introduction of SO coupling facilitates the revival of the $m_{F} = \pm 1$ components, enabling the formation of spin textures. As calculated in the FM case, we also compute the momentum distribution to verify the phase identity. We observe a single peak in the momentum distribution centered at $k = 0$, corresponding to the ZM phase. In contrast, the EZM~I and EZM~II phases show momentum-space peaks for the $m_{F} = \pm 1$ components at $k \neq 0$, while the $\psi_{0}$ component remains centered at $k = 0$; the SW and SSW phases exhibit multi-peak momentum distributions across all components (not shown here).

Similar to the FM interactions, the SSW phase appears only for $\Omega \sim 0$. However, in contrast to the FM case, we find that under AFM interactions, only the ZM, EZM~II, and SSW phases emerge at $\Omega \sim 0$.

\begin{figure}[!htp] 
\centering\includegraphics[width=0.99\linewidth]{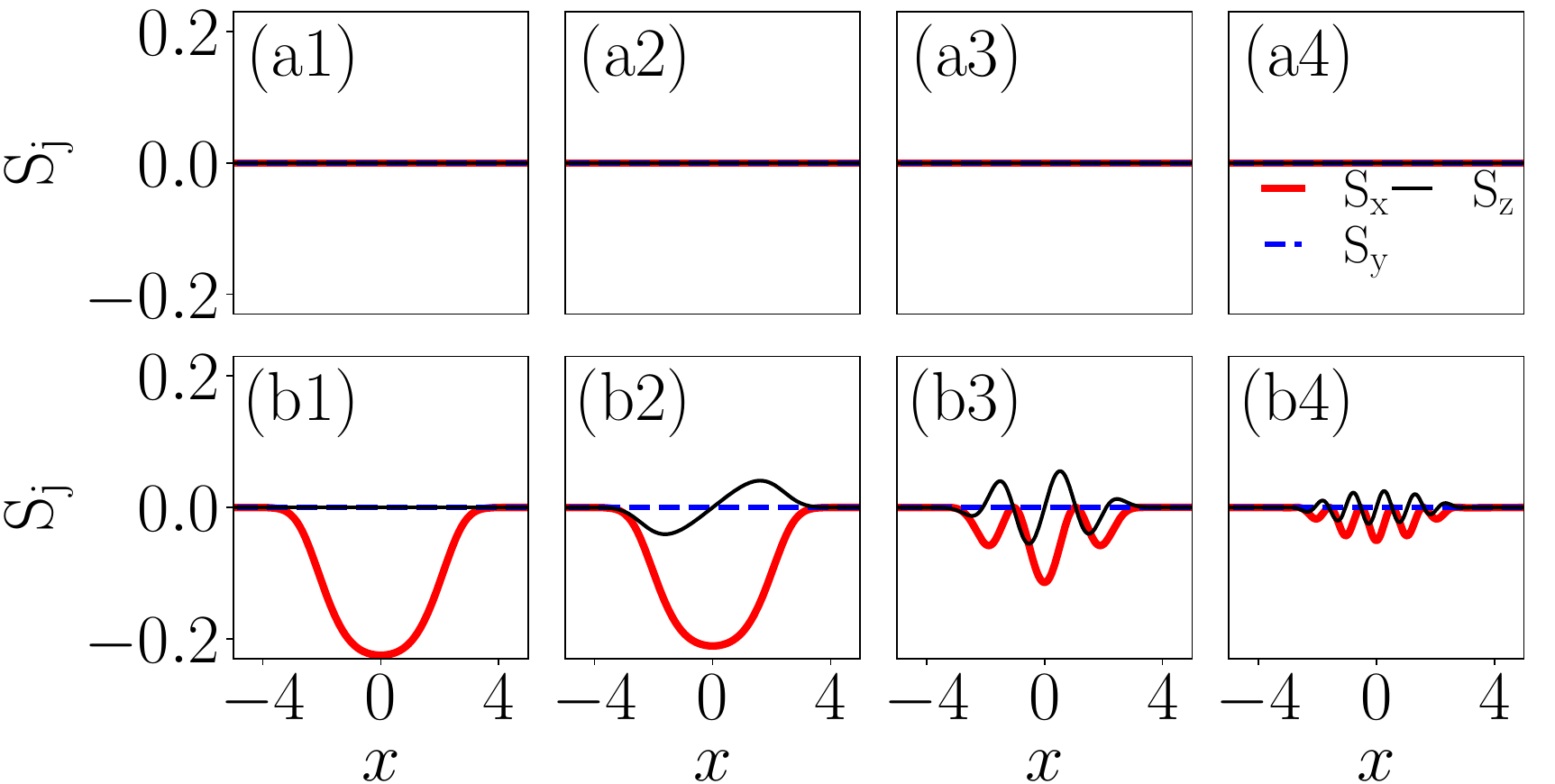}
\caption{SDV component $\mathrm{S}_{x}$, $\mathrm{S}_{y}$, and $\mathrm{S}_{z}$ in panels (a), and (b) corresponding to the Fig.~\ref{fig02a}. The lines representation is the same as in Fig.~\ref{fig01a-sdv}. Almost zero of $\mathrm{S}_{y}$ across all phases and a single peak $\mathrm{S}_{x}$ with zero (finite) $\mathrm{S}_{z}$, confirms the ZM (EZM~I) phase for finite $\Omega$. All SDVs remain zero in the ZM polar phase, EZM~II phase, and SSW phase for $\Omega \sim 0$. Multi-peak $\mathrm{S}_{x}$ along with oscillating $\mathrm{S}_{z}$ confirms the SW phase.
}
\label{fig02a-sdv} 
\end{figure}%
In Fig.~\ref{fig02a-sdv}, we report the SDV polarization corresponding to the density profile shown in Fig.~\ref{fig02a}. Initially, in Fig.~\ref{fig02a-sdv}(a1), we consider $\Omega = 0.0$, and $k_{L} = 0.0$. In this case, all atoms occupy state $\psi_{0}$, and densities $\rho_{+1}$ and $\rho_{-1}$ vanish, which leads to $\mathrm{S}_{x}$, $\mathrm{S}_{y}$, and $\mathrm{S}_{z}$ to be zero and the condensate remains in the ZM polar phase since the momentum distribution shows only a single peak, peaked at zero (not shown here). Therefore, in the case of AFM interaction, the ZM phase with $\Omega \sim 0.0$ differs from the FM interactions, where a single peak $\mathrm{S}_{x}$ was obtained. Upon increasing $k_{L}=0.0 \rightarrow 0.5$ in Fig.~\ref{fig02a-sdv}(a2), we again find all SDVs to be zero. The momentum space density shows only a single peak centered at zero for the zeroth component, and $m_{F} = \pm 1$ components exhibit peaks at $k \neq 0$ (not shown here), indicating the EZM~II phase. For a higher SO coupling strength $k_{L} = 1.5$ in Fig.~\ref{fig02a-sdv}(a3), we observe that all SDVs are zero with a multipeak density profile [see Fig.~\ref{fig02a}(a3)], confirming the presence of the unpolarized SW phase or SSW phase. This observation also holds for $k_{L} = 3.0$, as shown in Fig.~\ref{fig02a-sdv}(a4). The vanishing of all SDVs in the SSW phase is similar to the behaviour observed in the case of FM interactions. Next, in Fig.~\ref{fig02a-sdv}(b1), we consider $\Omega = 1.0$ and $k_{L} = 0.0$, in which $\mathrm{S}_{x}$ shows a single peak structure, confirming the presence of the ZM phase. Upon increasing $k_{L} = 0.0 \rightarrow 0.5$, we obtain a single peak $\mathrm{S}_{x}$ and non-zero $\mathrm{S}_{z}$, which is a signature of the EZM~I phase [see Fig.~\ref{fig02a-sdv}{b2}]. The non-zero value of $\mathrm{S}_{z}$ suggests the separation between the density of $m_{F} = \pm 1$ components of the condensates. Increasing $k_{L} = 0.5 \rightarrow 1.5$ in Fig.~\ref{fig02a-sdv}(b3), we obtain a multi-peak structure in $\mathrm{S}_{x}$, and $\mathrm{S}_{z}$, confirming the SW phase, where $\mathrm{S}_{z}$ lies between $-0.055$ to $0.055$. For $k_{L} = 3.0$, we also obtain similar behavior of $\mathrm{S}_{x}$, and $\mathrm{S}_{z}$, leading to the SW phase; however, the magnitude of both SDVs are less than the previous case [see Fig.~\ref{fig02a-sdv}(b4)]. Similar to the FM interactions, here also, spin flip occurs for finite $\Omega$. Due to the absence of the imaginary part in the wave function, $\mathrm{S}_{y}$ is zero throughout. For $\Omega \sim 0$, we obtain SDVs: $\mathrm{S}_{x}$, $\mathrm{S}_{y}$, and $\mathrm{S}_{z}$ are zero, in the ZM phase, which differs from the case of FM interactions.%

\begin{figure}[!htp] 
\centering\includegraphics[width=0.99\linewidth]{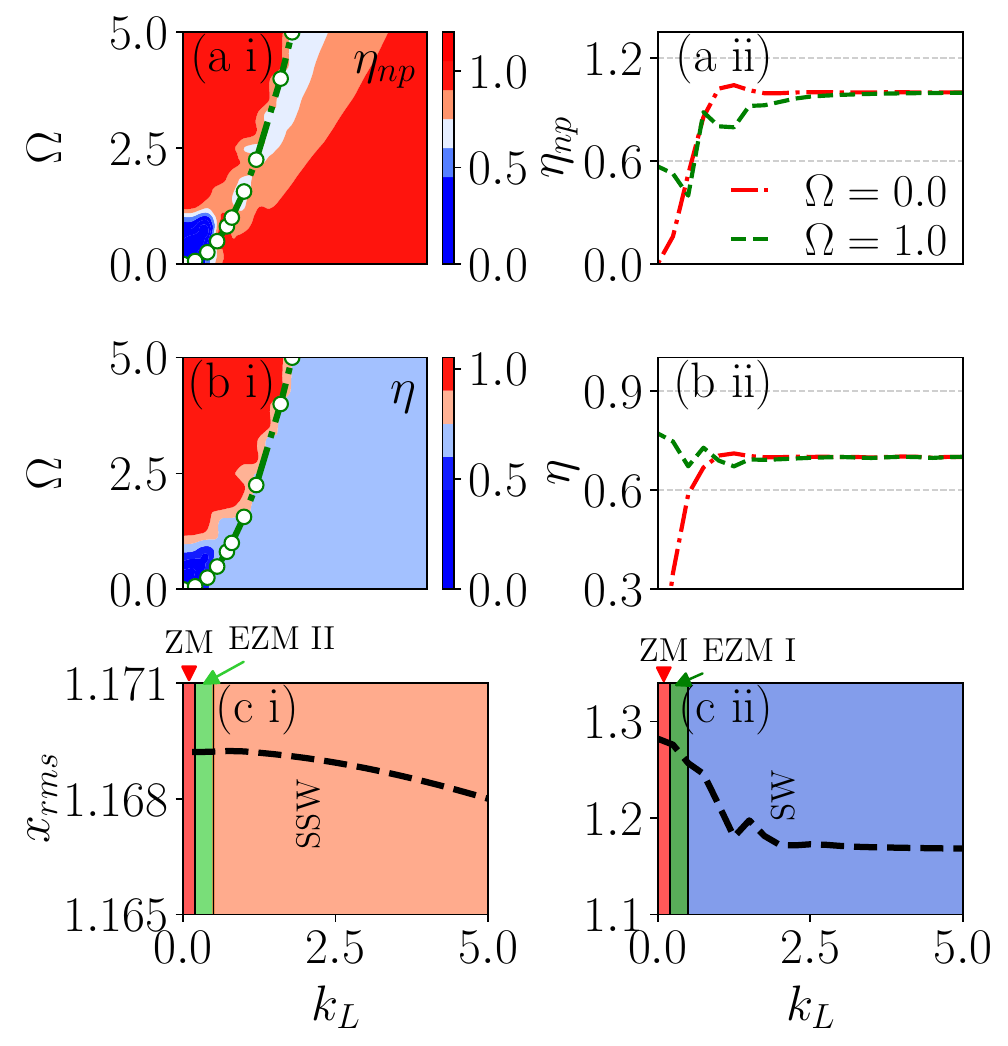}
\caption{
Miscibility parameters (a) $\eta_{np}$ and (b) $\eta$ for interaction strengths $c_0 = 10$ and $c_2 = 5.0$. (a i) and (b i) show $\eta_{np}$ and $\eta$ as functions of spin-orbit coupling strength $k_L$ and Rabi coupling $\Omega$, both varying in the range $[0, 5]$. The green line with white dots marks the critical relation $k_L^2 \approx \Omega$. Beyond this boundary ($k_L^2 > \Omega$), both $\eta_{np}$ and $\eta$ decrease. However, $\eta_{np}$ exhibits a non-monotonic behavior—decreasing initially and rising again to near-maximum values. (a ii) and (b ii) display $\eta_{np}$ and $\eta$ as functions of $k_L$ for fixed $\Omega = 0.0$ (dash-dotted red line) and $\Omega = 1.0$ (dashed green line). Panels (c i) and (c ii) show the root-mean-square condensate size $x_{\mathrm{rms}}$ versus $k_L$ for $\Omega = 0.0$ and $\Omega = 1.0$, respectively. The background color represents distinct quantum phases: red (ZM), green (EZM~I), limegreen (EZM~II), blue (SW), and coral (SSW).
}
\label{fig02e} 
\end{figure}%

\subsubsection{Effect of SO and Rabi coupling on miscibility and ground state phase transformation} 
Similar to FM interactions, we report $\eta_{np}$ defined in Eq.(\ref{eqa:overlap}), to characterize overlap between densities $\rho_{+1}$, and $\rho_{-1}$. In addition, we report the miscibility parameter $\eta$ defined in Eq.(\ref{eqb:overlap}), which quantifies the overlap between densities $\rho_{+1}$, $\rho_{0}$, and $\rho_{-1}$. 

In Fig.~\ref{fig02e}(a i), we show $\eta_{np}$ upon simultaneously varying the $\Omega$ and $k_{L}$ in the range [0,5] and keeping $c_{0} = 10.0$, and $c_{2} = 5.0$. Initially, for $\Omega \sim 0 $, and $k_{L} \sim 0$, the $\eta_{np} \sim 0$, which occurs due to the absence of densities of $\rho_{+1}$ and $\rho_{-1}$, as all atoms occupy the $\psi_{0}$ state. Upon considering the finite Rabi coupling ($\Omega > 0.5$) and gradually increase $k_{L}$ we find that the $\eta_{np} \approx 1.0$, which indicates a complete overlap between densities $\rho_{+1}$, and $\rho_{-1}$ and condensate resides in the ZM phase. For the intermediate range of $k_{L}$, $\eta_{np}$ decreases to $0.40$ for $\Omega =1.0$ and $k_{L} = 0.5$, which shows the separation between densities $\rho_{+1}$ and $\rho_{-1}$ a typical characteristics of the EZM~I phase. Upon further increasing of $k_{L}$, $\eta_{np}$ further increases and reaches as $\eta_{np} \approx 0.982$, which is the polarized SW phase. For a better description of the overlap between $m_{F} = \pm 1$ components of the condensate, in Fig.~\ref{fig02e}(a ii), we report variation in $\eta_{np}$ for two different Rabi coupling strengths $\Omega = 0.0$, and $1.0$. Initially considering $\Omega = 0.0$, and $k_{L} = 0.0$, where we observe that $\eta_{np} \approx 0$; increasing $k_{L}$ leads to the higher magnitude, and saturates further at $\eta_{np} \approx 1.0$. 
Upon increasing the Rabi coupling to $\Omega = 1.0$ while keeping $k_{L} = 0.0$, we find the revival of the densities of the $\rho_{+1}$ and $\rho_{-1}$ components which reflects in $\eta_{np}$ showing a finite magnitude $\eta_{np} \approx 0.568$. Upon increasing $k_{L}$, the $\eta_{np}$ initially decreases to $\eta_{np} \approx 0.40$ at $k_{L} = 0.5$ which increases upon increasing the SO coupling to $k_{L} =0.75$ which yields $\eta_{np} \approx 0.885$. Further increasing $k_{L}$, another minimum occurs at $k_{L} = 1.25$ with $\eta \approx 0.796$. Beyond this point, it increases and saturates at $\eta_{np} \approx 0.999$.%

To better characterize the different phases in Fig.~\ref{fig02e}(b i), we show the miscibility $\eta$ upon simultaneously varying $\Omega$, and $k_{L}$. Initially at $k_{L} \sim 0$, and $\Omega \sim 0$, $\eta \sim 0$, due to the absence of $m_{F} = \pm 1$ components. While considering finite $\Omega > 0.5$ and increasing $k_{L}$ gradually, we find the maximum value of $\eta \approx 0.96$ in the ZM phase, which starts to decrease upon increasing $k_{L}$ and attains $\eta \approx 0.70$ in the EZM~I, EZM~II, SW, and SSW phases. For a detailed understanding of miscibility $\eta$, in Fig.~\ref{fig02e}(b ii), we also show the variation of $\eta$ upon varying $k_{L}$ in the range [0,5] for two different $\Omega = 0.0$ and $1.0$. For $\Omega = k_{L} = 0.0$, as all atoms occupy the state $\psi_{0}$, $\eta$ appears to be zero. Upon increasing $k_{L}$, it starts to increase and reaches $\eta \approx 0.70$, which eventually gets saturated. Next, we consider $\Omega = 1.0$, and $k_{L} = 0.0$. Initially, upon increasing $k_{L}$, $\eta$ starts from a higher magnitude $\eta \approx 0.772$, further increasing $k_{L}$, reduces its magnitude and forms a minimum at $k_{L} = 0.50$ with magnitude $\eta \approx 0.68$. Increasing $k_{L}$ beyond this point, $\eta$ reaches to $\eta \approx 0.70$ and saturates. %

Similar to FM interactions, here we also obtain $\eta_{np}$ decreases beyond the $k_{L}^{2} \approx \Omega$, in the $k_{L}-\Omega$ plane; however, the $\eta_{np}$ show decreasing behaviour only for medium range of $k_{L}$, and rises again. For higher $k_{L}$, we obtain a value closer to the maximum of $\eta_{np}$, which arises from the increase of overlap between condensate components. The miscibility parameter $\eta$ is maximum for the region $k_{L}^{2} < \Omega$ (ZM phase), and decreases in the $k_{L}^{2} > \Omega$ region (EZM~I phase, EZM~II phase, SW phase, and SSW phase). However, for small values of $k_{L}$, and $\Omega$, we obtain minimum of $\eta_{np}$, and $\eta$ in the $k_{L}^{2} < \Omega$ region (ZM phase).%

Next, in Fig.~\ref{fig02e}(c i), and (c ii), we report the $x_{rms}$ size of the condensate keeping $\Omega = 0.0$, and $1.0$, varying $k_{L}$ in the range [0,5]. For $\Omega = 0.0$ and $k_{L} = 0.0$, we obtain $x_{rms} \approx 1.169$ and the condensate resides in the ZM phase. The $x_{rms}$ remains flat up to $k_{L} \lesssim 0.70$ in the EZM~II phase. Upon increasing $k_{L}$, $x_{rms}$ starts to decrease, as a result of the modulation of the density of the condensate components, which reduces $x_{rms} \approx 1.165$. This multi-peak density profile shows the presence of the unpolarized SW phase, which is the SSW phase. The difference is visible in the third decimal place, which is given in Fig.~\ref{fig02e}(c i). Further, in Fig.~\ref{fig02e}(c ii), considering $\Omega = 1.0$, and $k_{L} = 0.0$, we obtain $x_{rms} \approx 1.282$, which remains flat up to $k_{L} \lesssim 0.20$ in the ZM phase. Upon increasing $k_{L}$, $x_{rms}$ starts decreasing up to $k_{L} \approx 0.70$ in the EZM~I phase. The $x_{rms}$ drops further to $x_{rms} = 1.170$ as a result of density modulation of the condensate components, and then saturates. The multipeak density profile is the polarized SW phase. The root mean square ($x_{rms}$) size of the condensate decreases with increasing SO coupling in the case of AFM interactions, whereas it increases in the case of FM interactions. For $\Omega \sim 0$, at high $k_{L}$, the SSW phase dominates, while for finite Rabi coupling, the SW phase dominates.%

\begin{figure*}[!htp]
\centering\includegraphics[width=0.9\linewidth]{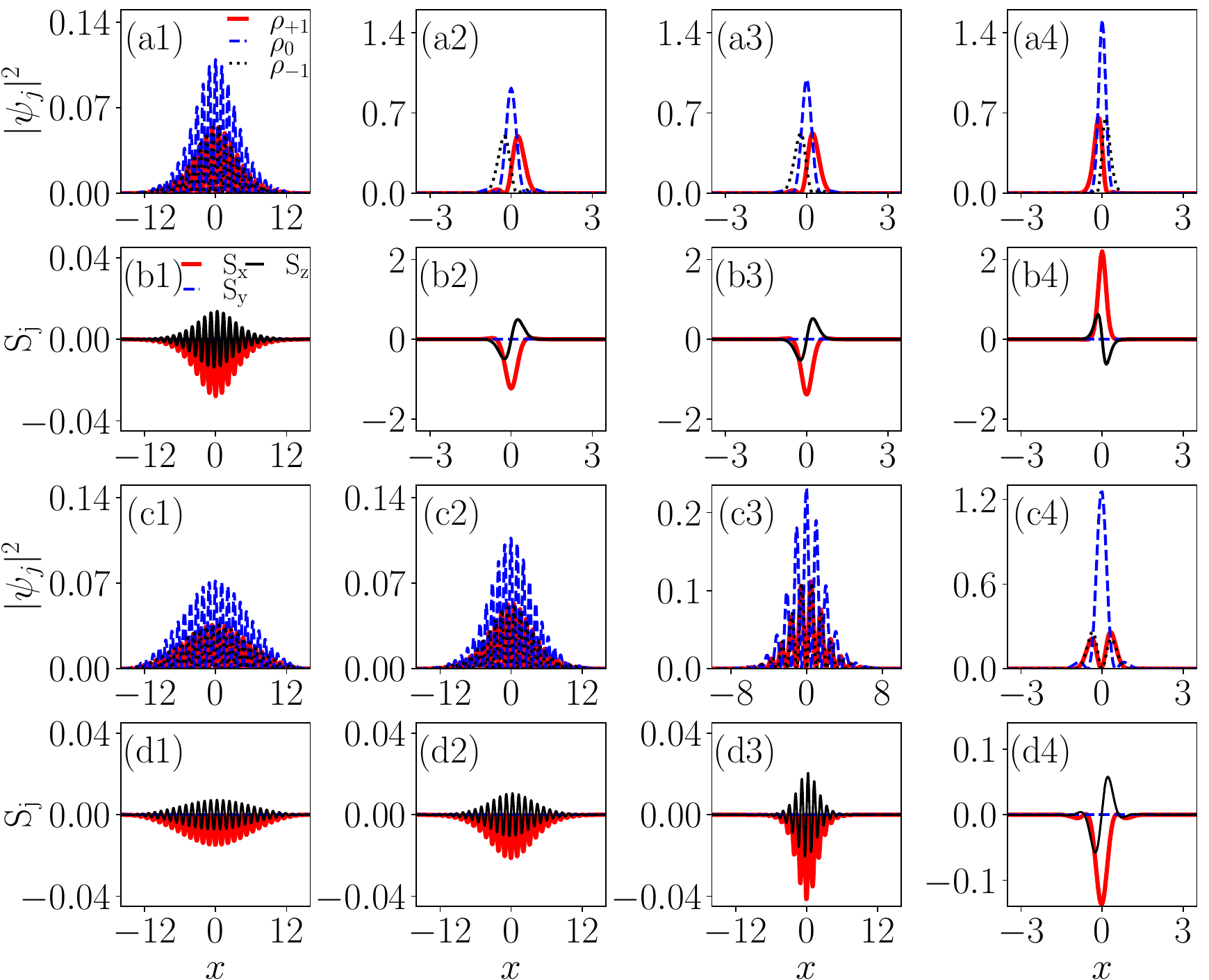}
\caption{Ground state density profiles of $\vert \psi_{+1} \vert^{2}$ (solid red line), $\vert \psi_{0} \vert^{2}$ (dashed blue line), and $\vert \psi_{-1} \vert^{2}$ (dotted black line) in panels (a) and (c), and corresponding spin-density vector components $\mathrm{S}_{x}$ (solid red line), $\mathrm{S}_{y}$ (dashed blue line), and $\mathrm{S}_{z}$ (thin black line) in panels (b) and (d), for $k_{L} = 3.0$ and $\Omega = 1.0$. Panels (a, b) correspond to $c_{2} = -5.0$, and panels (c, d) to $c_{2} = 5.0$. In each case, the density-density interaction $c_{0}$ varies as $-0.25$, $-0.50$, $-1.0$, and $-5.0$ from left to right. The appearance of multipeak structures in $\mathrm{S}_{x}$ and oscillations in $\mathrm{S}_{z}$ indicates the SW phase, while a single peak in $\mathrm{S}_{x}$ with finite $\mathrm{S}_{z}$ signifies the EZM~I phase. A transition from the SW to EZM~I phase occurs with increasing $|c_{0}|$, for both $c_{2} < 0$ and $c_{2} > 0$. As $c_{0} < 0$ in all cases, no external trapping potential is applied.
}

\label{fig05a} 
\end{figure*}%
\subsection{ Effect of spin-independent ($c_{0}$) and spin-exchange ($c_{2}$) interactions on the ground state phases} 
\label{grndintr}

In the above parts of the study, we have reported different ground state phases (ZM phase, EZM~I phase, EZM~II phase, SW phase, and SSW phase), their miscibility, and the transformation induced by varying SO coupling and Rabi coupling strength in FM and AFM interactions. Here, we fix the SO and Rabi coupling strength corresponding to (i) ZM phase, (ii) SW phase, and (iii) SSW phase, and vary the interaction parameters; density-density interaction ($c_{0}$) and spin-exchange interaction ($c_{2}$) to analyze miscibility and phase transitions. The EZM phases are excluded from the analysis, as it is a trivial extension of the ZM phase. We begin our discussion with the ZM phase.%

\subsubsection{Impact on the ZM phase}
In this section, we focus on the ZM phase with $\Omega = 1.0$ in the absence of SO coupling strength ($k_{L} = 0$), as illustrated in Figs.~\ref{fig01a}(c1) and~\ref{fig02a}(c1). We consider two values of the spin-dependent interaction strength $c_{2} = -5.0, \; 5.0$, while varying $c_{0}$ in the range [-5,5]. For attractive density-density interactions ($c_0 < 0$), the condensate naturally localizes due to self-attraction, forming a self-bound state without the need for an external trapping potential. In contrast, when $c_{0} > 0$, the condensate is not naturally localized and requires an external trap for confinement~\cite{Gautam2015Mob, Adhikari2020stable}. Based on this, we organize the discussion into two parts: (i) without external trapping, and (ii) with a trap potential of strength ($\lambda = 1.0$). Considering the case without a trap and with $c_{0} < 0$, we observe that the condensate remains in the ZM phase for both $c_{2} = -5.0$ and $c_{2} = 5.0$, exhibiting a single peak density profile across all components. Upon increasing the magnitude of $c_{0}$, the condensate shrinks spatially, and its amplitude increases significantly (not shown here)~\cite{Adhikari2020stable}. This behaviour is consistent regardless of whether the $c_{2} < 0$ or $c_{2} > 0$. 

Importantly, within the ZM phase, varying $c_{0}$ and $c_{2}$ does not lead to a phase transition. However, varying $c_{0}$ can significantly affect the condensate's spatial extent and amplitude~\cite{Adhikari2020stable}. For repulsive interactions ($c_{0} \gtrsim 0$), the nature of the ground state has already been discussed in Sec .~\ref {trapped_ferro} and~\ref{trapped_antiferro}, corresponding to $c_{2} < 0$ and $c_{2} > 0$ regimes, respectively.

\subsubsection{Impact on the SW phase}
In this section, we present the effect of the variation of $c_{0}$ and $c_{2}$ on the ground state (SW phase), keeping $k_{L}= 3.0$, and $\Omega = 1.0$. We consider $c_{2} < 0$ and $ > 0$, and varies $c_{0}$ in the range [-5,5]. For $c_{0} < 0$, the condensate localizes naturally, and no external trapping is required; however, for $c_{0} \gtrsim 0$, a trap is introduced with strength $\lambda = 1.0$.%

\paragraph{Different ground state phases:} We consider $k_{L} = 3.0$, and $\Omega = 1.0$, which belong to the SW phase as shown in Figs.~\ref{fig01a}(c4) and ~\ref{fig02a}(c4), and vary the interaction strengths $c_{0}$, and $c_{2}$. Initially, considering $c_{2} = -5.0$ and $c_{0} = -0.25$ in Fig.~\ref{fig05a}(a1), we obtain the SW phase with 24 maxima and minima in the density profile, which extends over a large spatial region. The corresponding SDV polarization in Fig.~\ref{fig05a}(b1) exhibits a multi-peak structure in $\mathrm{S}_{x}$, and $\mathrm{S}_{z}$ shows a variation of amplitude with maximum range $-0.014$ to $0.014$, which confirms the SW phase. Next, increasing the attractive density-density interaction $c_{0} = -0.50$ in Fig.~\ref{fig05a}(a2), we obtain a single peak structure across all components, among them the densities $\rho_{+1}$ and $\rho_{-1}$ peaks away from the trap center at $ x = 0.20$ and $x = -0.20$, respectively, with $\rho_{\pm 1}(max) = 0.52$, resulting in an elongated zero-momentum phase of first type (EZM~I phase). The corresponding SDV polarization in Fig.~\ref{fig05a}(b2) shows a single peak structure in $\mathrm{S}_{x}$ and finite $\mathrm{S}_{z}$, validating the EZM~I phase of the condensate, which supports a phase transition from the SW phase to the EZM~I phase. The SW phase transitions to the EZM~I phase at $c_{0}(SW \rightarrow EZM~I) = -0.50$.  Further, increasing the attractive strength of $c_{0}$, we observe similar trends in the density profile, as well as in the corresponding SDVs: $\mathrm{S}_{x}$, and $\mathrm{S}_{z}$ in Figs.~\ref{fig05a}(a3, b3), and (a4, b4); therefore, the condensate remains in the EZM~I phase. However a spin flip occurs in SDVs ($\mathrm{S}_{x}$, and $\mathrm{S}_{z}$) in Figs.~\ref{fig05a}(b4). Next, considering $c_{2} = 5.0$ and $c_{0} = -0.25$ in Fig.~\ref{fig05a}(c1), we obtain a multi-peak structure in the density profile with 24 maxima and minima, which is the SW phase. Correspondingly, we obtain the multi-peak $\mathrm{S}_{x}$ and finite $\mathrm{S}_{z}$ varies with maximum range $-0.021$ to $0.021$, which is the result of separation between densities $\rho_{+1}$, and $\rho_{-1}$, and confirms the SW phase in Fig.~\ref{fig05a}(d1). As the attractive strength of $c_{0}$ increases, the number of stripes in the density profile continuously decreases, as shown in Figs.~\ref{fig05a}(c2,c3); however, the condensate remains in the SW phase. We obtain similar features in $\mathrm{S}_{x}$ and $\mathrm{S}_{z}$ in Figs.~\ref{fig05a}(d2,d3) with the reduced number of maxima and minima; however, the magnitude of $\mathrm{S}_{x}$, and $\mathrm{S}_{z}$ increases. Finally, in Fig.~\ref{fig05a}(c4), we consider a higher attractive strength of $c_{0} = -5.0$, where the condensate transitions from the SW wave phase to the EZM~I phase. This transition takes place at $c_{0}(SW \rightarrow EZM~I) = -3.75$. The corresponding SDVs show that $\mathrm{S}_{x}$ has a single peak and $\mathrm{S}_{z}$ exhibits finite magnitude, which confirms that the condensate is in the EZM~I phase. We do not include the discussion of the case $c_{0} \gtrsim 0$ since it is given above in sec.~\ref{trapped_ferro}, and sec.~\ref{trapped_antiferro} for $c_{2} < 0.0$, and $c_{2} > 0.0$, respectively. For attractive interactions $c_{0} < 0$ and $c_{2} < 0$, a transition from the SW phase to EZM~I phases takes place at $c_{0}(SW \rightarrow EZM~I) = -0.50$, and for $c_{2} >0$, it takes place at $c_{0}(SW \rightarrow EZM~I) = -3.75$.%

\begin{figure}[!htp]
\centering\includegraphics[width=0.95\linewidth]{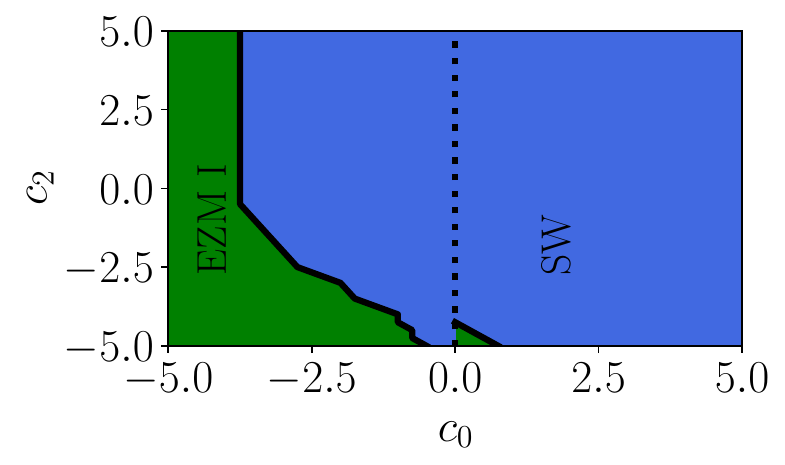}
\caption{{Ground state phase diagram in the $c_{0}$-$c_{2}$ plane for fixed $k_{L} = 3.0$ and $\Omega = 1.0$. The black dotted line separates the trapless ($c_{0} < 0$) and trapped ($c_{0} \gtrsim 0$) regimes. Different quantum phases are indicated by color: green denotes the EZM~I phase and blue denotes the SW phase. For considered parameter ($k_{L}, \Omega$) = (3.0, 1.0), the system resides in the SW phase under weak or repulsive interaction strengths, while the higher attractive strength of both $c_{0}$ and $c_{2}$ drives a transition into the EZM~I phase.}}
\label{fig05a} 
\end{figure}%

In Fig.~\ref{fig05a}, we present the ground state phase diagram as a function of $c_{0}$ and $c_{2}$, with $k_{L} = 3.0$ and $\Omega = 1.0$ held fixed. The black dashed line divides the diagram into two regimes: the untrapped region ($c_{0} < 0$) and the trapped region ($c_{0} \gtrsim 0$), where a harmonic potential confines the condensate.

We begin by analyzing the case $c_{2} = -5.0$, varying $c_{0}$. For strong attractive interaction ($c_{0} = -5.0$), the condensate is in the EZM~I phase. As $c_{0}$ increases, weakening the attraction, a phase transition to the SW phase occurs around $c_{0} = -0.50$, which continues throughout the remaining untrapped region ($c_{0} < 0$). Entering the trapped regime ($c_{0} \gtrsim 0$), the condensate briefly re-enters the EZM~I phase for small repulsive values of $c_{0}$, up to $c_{0} = 0.75$. Beyond this, the system transitions back to the SW phase.

With increasing $c_{2}$, the region occupied by the EZM~I phase gradually shrinks in the $c_{0}$--$c_{2}$ plane. At $c_{2} = -0.5$, the EZM~I--SW transition shifts to $c_{0} = -3.75$, remaining unchanged for larger $c_{2}$. For $c_{2} = 5.0$, the condensate starts in the EZM~I phase at $c_{0} = -5.0$ and transitions to the SW phase at $c_{0} = -3.75$, remaining in this phase for higher $c_{0}$, including the trapped regime.

Overall, we find that for weakly attractive or repulsive interactions, the system with parameters $(k_{L}, \Omega) = (3.0, 1.0)$ resides in the SW phase. Increasing the attractive strengths of $c_{0}$ and $c_{2}$ drives a transition from the SW phase to the EZM~I phase.

\begin{figure}[!htp] 
\centering\includegraphics[width=0.99\linewidth]{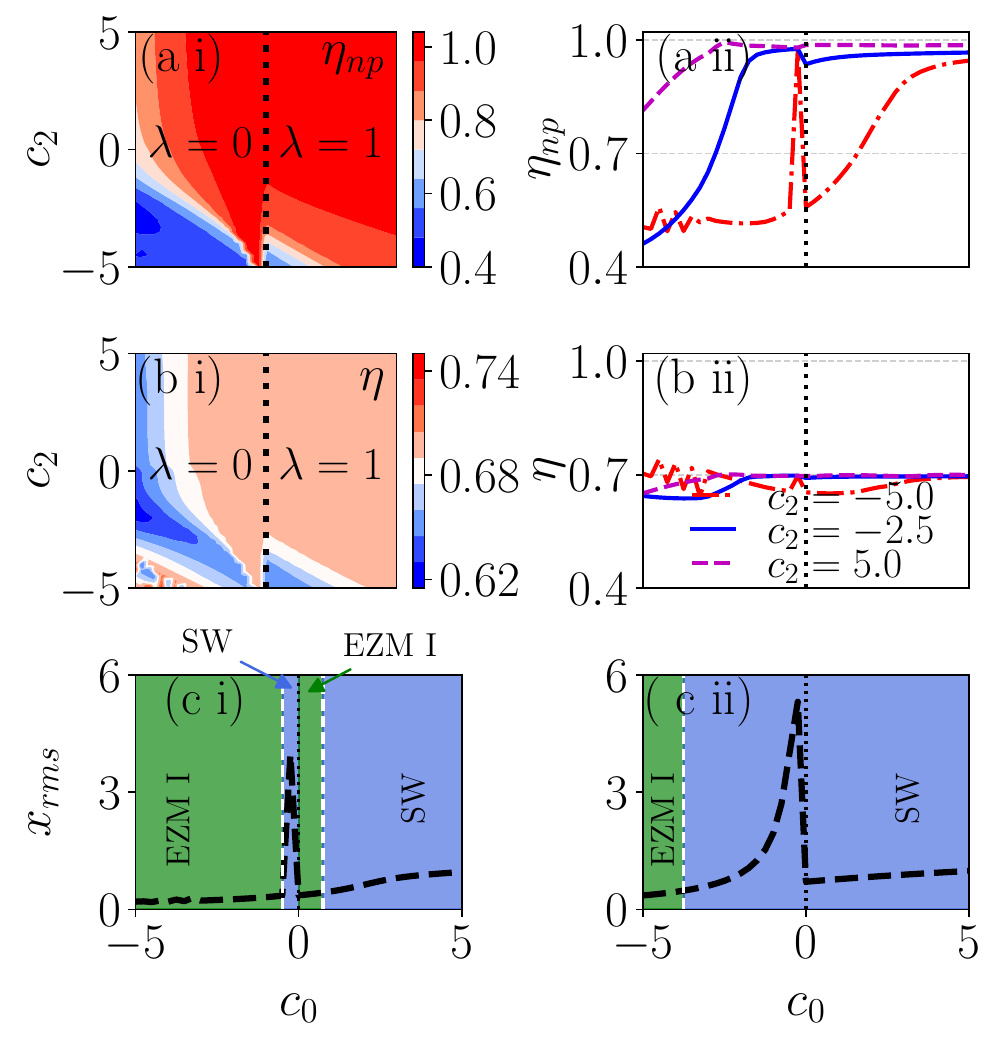}
\caption{
 Miscibility parameters (a) $\eta_{np}$ and (b) $\eta$ as functions of interaction strengths $c_0$ and $c_2$, for fixed  $k_L = 3.0$ and $\Omega = 1.0$. In (a i), and (b i), both $c_0$ and $c_2$ are varied within the range $[-5, 5]$. (a ii), and (b ii) show $\eta_{np}$ and $\eta$ as functions of $c_0$ for fixed values of $c_2 = -5.0$ (dash-dotted red line), $-2.5$ (solid blue line), and $5.0$ (dashed magenta line). Panels (c i) and (c ii) depict the root-mean-square size $x_{\mathrm{rms}}$ of the condensate for $c_2 = -5.0$ and $c_2 = 5.0$, respectively, as $c_0$ is varied. The background color shading represents distinct quantum phases: green indicates the EZM~I phase and blue indicates the SW phase, highlighting the phase transition of the condensate. The black dotted line indicates the boundary separating trapped regime ($c_0 \gtrsim 0$) and untrapped regime ($c_0 < 0$) condensates.}

\label{fig05c} 
\end{figure}%

\paragraph{Effect of $c_{0}$ and $c_{2}$ on miscibility and ground state phase transformation:}
In the above parts, we reported ground state density profiles such as the SW and EZM~I phases, and the corresponding SDVs specifically for the $c_{0} < 0$ case. Here, we report $\eta_{np}$, overlap between densities $\rho_{+1}$, and $\rho_{-1}$ given in Eq.(\ref{eqa:overlap}). We also report the miscibility parameter $\eta$, which is the overlap among the densities of $\rho_{+1}$, $\rho_{0}$, and $\rho_{-1}$ given in Eq.(\ref{eqb:overlap}). 

Initially, we examine the variation of $\eta_{np}$ in Fig.~\ref{fig05c}(a i) as both $c_{0}$ and $c_{2}$ vary simultaneously within the range $[-5, 5]$, while keeping $k_{L} = 3.0$ and $\Omega = 1.0$ fixed. The dotted black line divides the phase plot into two regions: (i) the left half, corresponding to the absence of trap strength, and (ii) the right half, where the trap is present.

First, we focus on $c_{2} < 0$, setting $c_{2} = -5.0$ and varying $c_{0}$. At $c_{0} = -5.0$, we find $\eta_{np} = 0.505$, arising from the separation between densities $\rho_{+1}$ and $\rho_{-1}$. In this regime, the condensate is in the EZM~I phase. As $c_{0}$ increases, $\eta_{np}$ also increases, reaching $\eta_{np} = 0.974$ at $c_{0} = -0.25$, where the condensate enters the SW phase. For $c_{0} \gtrsim 0$, a harmonic trap confines the condensate. Increasing $c_{0}$ beyond zero causes $\eta_{np}$ to decrease to $\eta_{np} = 0.573$ at $c_{0} = 0.25$, signaling a return to the EZM~I phase. Further increases in $c_{0}$ result in $\eta_{np}$ rising again, reaching $\eta_{np} = 0.957$ at $c_{0} = 5.0$, indicating a transition back to the SW phase. Next, considering $c_{2} > 0$ with $c_{2} = 5.0$, at $c_{0} = -5.0$ we observe $\eta_{np} = 0.813$, and the condensate is in the EZM~I phase. As $c_{0}$ increases, $\eta_{np}$ grows, reaching $\eta_{np} = 0.980$ at $c_{0} = -0.25$, where the condensate transitions to the SW phase, which persists throughout the trapped regime as well.

To describe the overlap parameter $\eta_{np}$ in a better way, in Fig.~\ref{fig05c}(a ii), we present it by choosing three different values of $c_{2} = -5.0, -2.5$, and $5.0$, varying $c_{0}$ in the range [-5,5] at fixed $k_{L} = 3.0$, and $\Omega = 1.0$. For $c_{2} = -5.0$ and $c_{0} = -5.0$, we obtain $\eta_{np} \approx 0.505$; increasing $c_{0}$, $\eta_{np}$ remains almost flat up to $c_{0} \lesssim -2.5$. Upon further increasing $c_{0}$, $\eta_{np}$ peaks at $c_{0} = -0.25$ with $\eta_{np} \approx 0.974$ and drops to $\eta_{np} \approx 0.558$ at $c_{0} = 0$. For $c_{0} > 0$ case, upon increasing the $c_{0}$ the $\eta_{np}$ increases and reaches $\eta_{np} \approx 0.945$ at $c_{0} = 5.0$. Next, considering $c_{2} = -2.5$ and $c_{0} = -5.0$, we obtain $\eta_{np} \approx 0.461$, which increases upon increasing the $c_{0}$, and reaches $\eta_{np} \approx 0.976$ at $c_{0} = -0.25$. Beyond this point, it decreases slightly to $\eta_{np} \approx 0.936$ at $c_{0}=0.0$. For $c_{0} > 0$, it remains almost constant. Now, considering $c_{2} = 5.0$, and $c_{0} = -5.0$, we obtain $\eta_{np} \approx 0.813$, which is higher than for $c_{2} < 0$. Upon increasing $c_{0}$, $\eta_{np}$ peaks at $c_{0} = -2.5$ with magnitude $\eta_{np} \approx 0.994$; further, it shows a slight decrease and reaches $\eta_{np} \approx 0.978$ at $c_{0} = -0.25$. For $c_{0} =0.0$, it rises again to the value $\eta_{np} \approx 0.986$, and after which it saturates. 

In Fig.\ref{fig05c}(b i), we show the variation of the miscibility order parameter $\eta$ of the condensate in the $c_{0}$--$c_{2}$ plane, with $k_{L} = 3.0$ and $\Omega = 1.0$ fixed. We first consider $c_{2} < 0$ by setting $c_{2} = -5.0$ and varying $c_{0}$. At $c_{0} = -5.0$, we find $\eta \approx 0.703$, corresponding to the EZM~I phase. As $c_{0}$ increases, $\eta$ decreases to $\eta = 0.656$ at $c_{0} = -0.50$, where the condensate transitions into the SW phase. For $c_{0} \gtrsim 0$, a harmonic trap is introduced to confine the condensate. Increasing $c_{0}$ further causes $\eta$ to rise slightly, reaching $\eta = 0.654$ at $c_{0} = 5.0$. Next, considering $c_{2} > 0$ with $c_{2} = 5.0$, we vary $c_{0}$. At $c_{0} = -5.0$, $\eta = 0.651$, and the condensate resides in the EZM~I phase. As $c_{0}$ increases, $\eta$ rises to $\eta = 0.697$ at $c_{0} = -0.25$, indicating the SW phase. Similar to the previous case, the introduction of a harmonic trap occurs for $c_{0} \gtrsim 0$, beyond which $\eta$ remains nearly constant. In particular, we find that the miscibility parameter varies between approximately $0.62$ and $0.75$, indicating that both the EZM~I and SW phases are partially miscible.
 
Now, to explore it more systematically, we compute the miscibility of the condensate upon varying $c_{0}$ for different values of $c_{2} = -5.0, -2.5$, and $5.0$ with fixed $k_{L} = 3.0$ and $\Omega = 1.0$ in Fig.~\ref{fig05c}(b ii). We consider $c_{2} =-5.0$, and $c_{0} = -5.0$, we obtain $\eta = 0.705$. Beyond this point, $\eta$ decreases upon increasing $c_{0}$, showing fluctuating behaviour. At $c_{0} = -0.25$, we obtain a peak with $\eta \approx 0.696$, which reduces at $c_{0} = 0.0$, and reaches $\eta \approx 0.653$. For $c_{0} > 0$, the $\eta$ increases with increasing $c_{0}$ and reaches to $\eta\approx 0.693$ at $c_{0} = 5.0$. Next, for $c_{2} = -2.5$, and $c_{0} = -5.0$, we obtain $\eta \approx 0.643$, beyond this $\eta$ increases upon increasing $c_{0}$ and falls slightly at $c_{0} = 0.0$. For $c_{0} > 0$, it rises again and reaches $\eta = 0.690$, after which it saturates. Next, considering $c_{2} = 5.0$, and $c_{0} = -5.0$, we obtain $\eta \approx 0.653$. It increases upon increasing $c_{0}$ and reaches to $\eta \approx 0.702$ at $c_{0} = -2.5$, and further it saturate.

Finally, we report $x_{rms}$ of the condensate with respect to $c_{0}$, and $c_{2}$. Considering $c_{2} = -5.0$, and $c_{0} = -5.0$ in Fig.~\ref{fig05c}(c i), we obtain $x_{rms} \approx 0.201$ which is the EZM~I phase. Upon decreasing the attractive strength of $c_{0}$, the $x_{rms}$ increases gradually; however, the condensate remains in the EZM~I phase. At $c_{0} = -0.25$, the $x_{rms}$ peaks with a higher magnitude of $x_{rms} = 3.965$, which arises due to the density modulation and condensate transitions into the SW phase. For $c_{0} \gtrsim 0$ (in the presence of a trap), the $x_{rms}$ falls to $x_{rms} \approx 0.363$, which transforms the condensate to the EZM~I phase again and exists up to $c_{0} \approx 0.75$. Beyond this point, the $x_{rms}$ increases again due to the density modulation, and the condensate transforms to the SW phase. Next, in Fig.~\ref{fig05c}(c ii) considering $c_{2} = 5.0$, and $c_{0} = -5.0$, we obtain $x_{rms} \approx 0.360$, which lies in the EZM~I phase. The condensate remains in the EZM~I phase up to $c_{0} = -3.75$. Beyond this point, the $x_{rms}$ increases with increasing $c_{0}$ and the condensate transitions to the SW phase. The $x_{rms}$ peaks to $5.314$ at $c_{0} = -0.25$ arise as a result of the large number of maxima and minima of the SW phase. For $c_{0} = 0.0$, the $x_{rms}$ falls to $0.701$; however, the condensate remains in the SW phase with a lesser number of maxima and minima in component density. Upon further increasing the $c_{0}$, the $x_{rms}$ gradually increases with condensate to be the SW phase, and the number of maxima and minima increases with increasing $c_{0}$.

\subsubsection{Impact on the SSW phase}
\label{ImpSSW}
In this section, we present the impact of the variation of $c_{0}$ and $c_{2}$ on the ground state (SSW phase) for $k_{L}= 3.0$, and in the absence of $\Omega$. We consider $c_{2} < 0$ and $ > 0$, and varies $c_{0}$ in the range [-5,5]. For $c_{0} < 0$, the condensate localizes naturally, and no external trapping is required; however, for $c_{0} > 0$, a trap is introduced with trap aspect ratio $\lambda = 1.0$~\cite{Gautam2015Mob, Adhikari2020stable}.%
\begin{figure}[!htp]
\centering\includegraphics[width=0.9\linewidth]{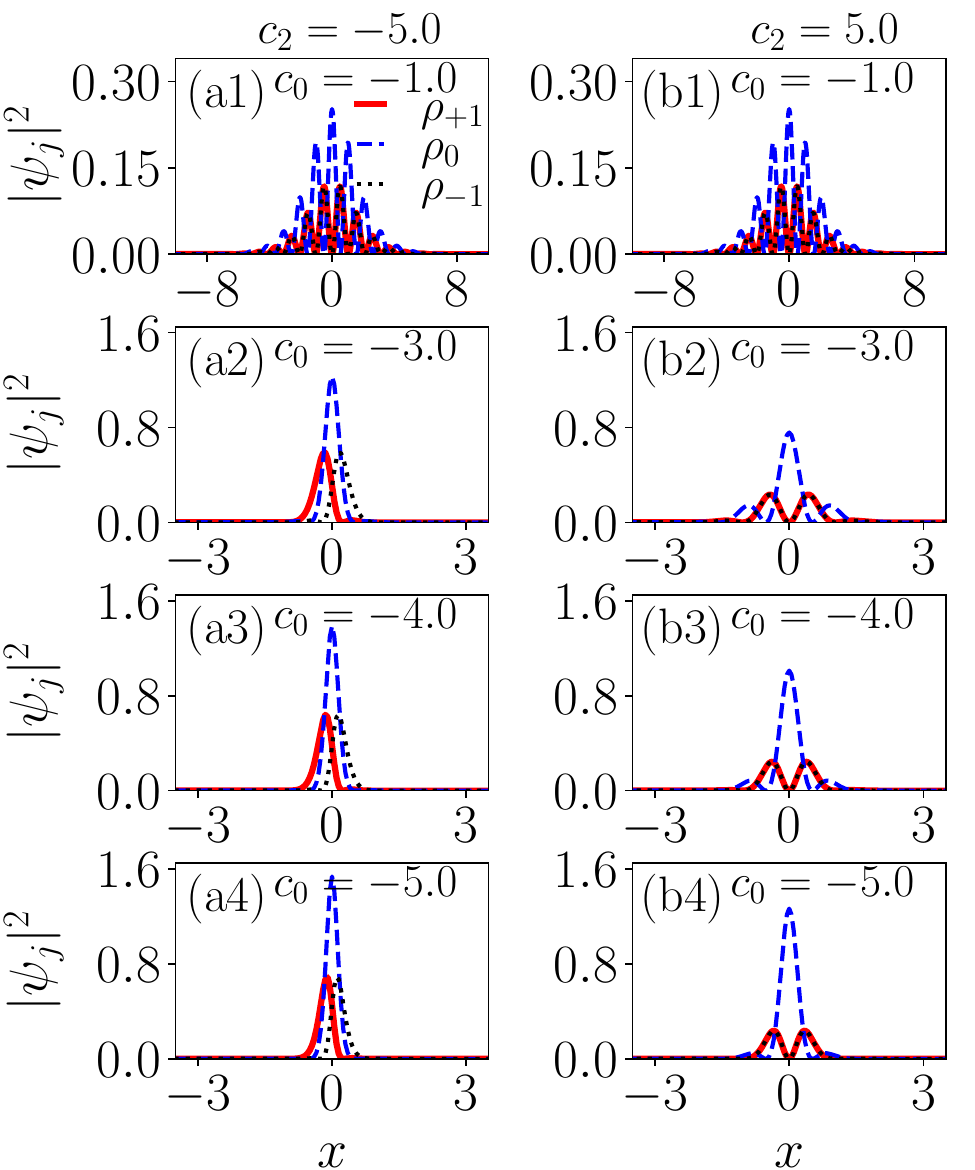}
\caption{Ground state density profile $\vert\psi_{+1}\vert^{2}$ (solid red line), $\vert\psi_{0}\vert^{2}$ (dashed blue line), and $\vert\psi_{-1}\vert^{2}$ (dotted black line) upon varying $c_{0}$, keeping $k_{L} = 3.0$ and $\Omega = 0.0$ with (a) $c_{2} = -5.0$, and (b) $c_{2} = 5.0$. The density-density interaction strength varies as $c_{0} = -1.0, -3.0, -4.0$, and $-5.0$ in columns from top to bottom. A phase transition takes place from the SSW phase to the EZM~I phase and EZM~II phase upon increasing the magnitude of $c_{0}$ for $c_{2} < 0$ and $ c_{2} > 0$, respectively. Since $c_{0} < 0$, no external trapping is required.}
\label{fig07a} 
\end{figure}

\paragraph{Different ground state phases:}
We vary the interaction strength $c_{0}$ and $c_{2}$ by considering $k_{L} = 3.0$ in the absence of Rabi coupling strength, which corresponds to the SSW phase given in Figs.~\ref{fig01a}(a4) and ~\ref{fig02a}(a4). Initially, considering $c_{2} = -5.0$ and $c_{0} = -1.0$, in Fig.~\ref{fig07a}(a1) we obtain the multi-peak structure in the density profile across all components, in which the $m_{F} = \pm 1$ components superpose completely, leading to unpolarized stripe wave phase (SSW phase). Increasing the attractive strength in Fig.~\ref{fig07a}(a2) to a higher magnitude $c_{0} = -3.0$, the SSW phase transitions into the elongated zero-momentum phase of the first type (EZM~I phase). This transition is characterized by the spatial separation of the $\rho_{+1}$ and $\rho_{-1}$ densities, resulting in a single-peak density profile for all three components.

The densities $\rho_{+1}$ and $\rho_{-1}$ components peaks away from the trap center at $x = -0.15$ and $x= 0.15$, respectively with amplitude $\rho_{\pm 1}(max) = 0.64$. This transition takes place at an earlier strength of $c_{0}(SSW \rightarrow EZM~I) = -1.25$. Further, increasing the attractive strength of $c_{0}$, condensate remains in the EZM~I phase; however, it shrinks spatially, and the amplitude increases given in Figs.~\ref{fig07a}(a3) and (a4). Next, considering $c_{2} = 5.0$ and $c_{0} = -1.0$, we obtain the multi-peak structure, which is the SSW phase and is given in Fig.~\ref{fig07a}(b1). Furthermore,  increasing the attractive strength of $c_{0} = -3.0$ in Fig.~\ref{fig07a}(b2), we find that the condensate remains in the SSW phase, with two peaks in the density profile across all components. Upon further increasing the magnitude of $c_{0}$ in Fig.~\ref{fig07a}(b3), at $c_{0}(SSW \rightarrow EZM~II)= -3.75$ the system makes a transition into the EZM~II phase, where the densities $\rho_{+1}$, and $\rho_{-1}$ overlap. This EZM~II phase differs from the EZM~I phase, which exhibits polarization between the $m_{F} = \pm 1$ components of the condensate. Beyond this point, condensate remains in the EZM~II phase as shown in the Fig.~\ref{fig07a}(b4). We do not discuss the case $c_{0} \gtrsim 0$, as it is already elaborated extensively above in sec.~\ref{trapped_ferro} and in sec.~\ref{trapped_antiferro} for $c_{2} < 0.0$, and $c_{2} > 0.0$, respectively. For attractive interactions with $c_{0} < 0$ and $c_{2} < 0$, a transition from the SSW phase to EZM~I phase occurs at $c_{0}(SSW \rightarrow EZM~I) = -1.25$, while for $c_{2} >0$, it makes a transition to EZM~II phase at $c_{0}(SSW \rightarrow EZM~II)= -3.75$. The effects of $c_{0}$ and $c_{2}$ on the overlap parameter $\eta_{np}$, miscibility $\eta$, and ground state phase transformation in the SSW phase are discussed in Appendix~\ref{appdxa}.

\begin{figure}[!htp]
\centering\includegraphics[width=0.95\linewidth]{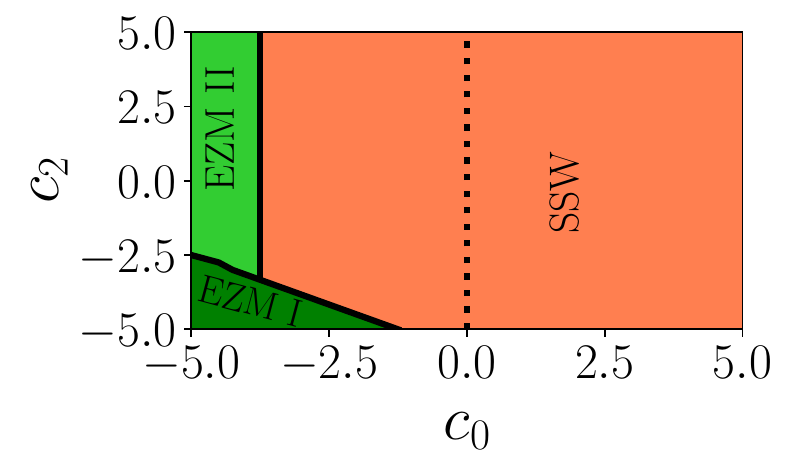}
\caption{{Ground state phase diagram in the $c_{0}$-$c_{2}$ plane for fixed $k_{L} = 3.0$ in absence of Rabi coupling ($\Omega$). The black dotted line separates the trapless ($c_{0} < 0$) and trapped ($c_{0} \gtrsim 0$) regimes. Different quantum phases are indicated by color: green denotes the EZM~I phase, lime green denotes EZM~II, and coral denotes the SSW phase. For considered parameter ($k_{L}, \Omega$) = (3.0, 0.0), the system resides in the SSW phase under weak or repulsive interaction strengths, while the higher attractive strength of both $c_{0}$ and $c_{2}$ drives a transition into the EZM's phase.}}
\label{fig07b} 
\end{figure}%
In Fig.~\ref{fig07b}, we present the ground state phase diagram in the $c_{0}$ - $c_{2}$ plane. The black dashed line separates the phase diagram into two distinct regimes: (i) in the absence of a trap ($c_{0} < 0$), where no external confinement is applied, and (ii) the trapped regime ($c_{0} \gtrsim 0$), where a harmonic potential is applied to confine the condensate. Starting with $c_{2} = -5.0$ we vary $c_{0}$. For strong attractive interaction $c_{0} = -5.0$, the condensate remains in the EZM~I phase, which persists up to $c_{0} = -1.25$. Beyond this value, the condensate undergoes a phase transition into the SSW phase. Notably, the SSW phase extends into the trapped regime, demonstrating the robustness against the external confinement. 

As $c_{2}$ increases, the region occupied by the EZM~I phase in the $c_{0}-c_{2}$ plane progressively shrinks. For $c_{2} > -2.5$, a different EZM phase (EZM~II phase) emerges, which persists up to $c_{0} = -3.75$. The EZM~II is characterized by overlapping densities $\rho_{+1}$ and $\rho_{-1}$, with no observable density modulation in real space, indicating a fully mixed spin-state with suppressed stripes. Finally, for strongly repulsive spin-dependent interactions ($c_{2} = 5.0$), a transition from the EZM~II phase to the SSW phase occurs at $c_{0} = -3.75$, beyond which the condensate remains in the SSW phase across the entire range of $c_{0}$.

\begin{figure*}[!htp] 
\begin{centering}
\centering\includegraphics[width=0.99\linewidth]{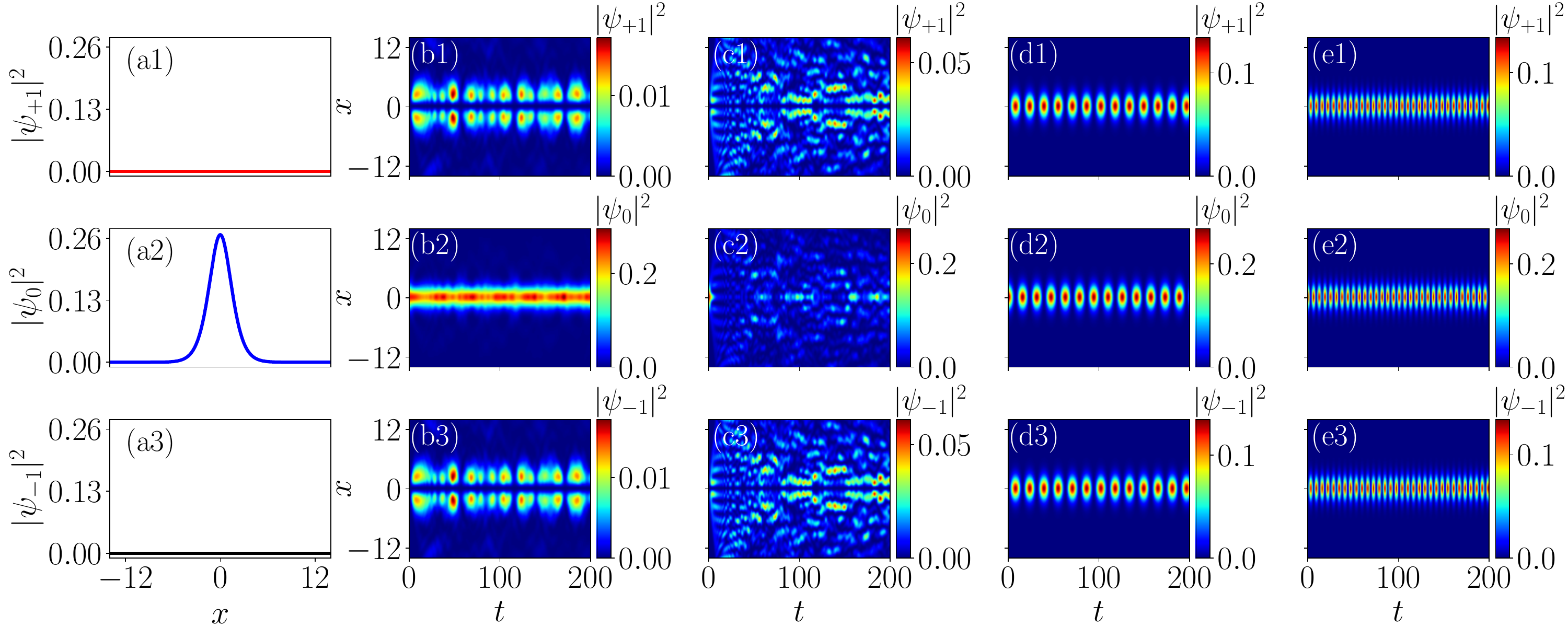}
 \caption{Ground state density profile (a1-a3) for $m_{F} = +1, 0, -1$, respectively, with $c_{0} =-1.0$, and $c_{2} =5.0$, in absence of the $k_{L}$, and $\Omega$. The condensate is in the ZM phase. Panels (b1-b3) and (c1-c3) depict the time evolution of the ground state profile upon quenching the SO coupling from $k_{L} \sim 0 \rightarrow 0.20$ and $k_{L} \sim 0.0 \rightarrow 1.0$, respectively. For a small quench of the SO coupling strength, periodic high-density peaks appear along the propagation direction, whereas a complex pattern emerges for stronger quench. Panels (d1-d3) and (e1-e3) exhibit the dynamics of the ground state profile upon quenching the Rabi coupling $\Omega = 0 \rightarrow 0.20$ and  $\Omega = 0.0 \rightarrow 0.50$, respectively. In both cases, the quench of the Rabi coupling leads to the stabilization of the condensate. The SO and Rabi coupling is quenched at $t = 0$ units.}
 \label{figq-01a}
 \end{centering}
 \end{figure*}%

The zero-momentum (ZM) phase and the stripe wave (SW) phase discussed above can be characterized using the single-particle spectrum. To obtain the single particle dispersion, we consider $V(x) = 0$, in the absence of interaction strengths, that is, $ c_{0} =  c_{2} = 0$. Substituting $\psi_{0,\pm 1} = \phi_{0,\pm 1} e^{\mathrm{i}({q_{x} x -\omega t})}$ into the Eqs.(\ref{gpe03})-(\ref{gpe04}), we obtain the following dispersion relations~\cite{Rajaswathi2023, Gangwar2024},
\begin{subequations}\label{eqn7}
\begin{align}%
\omega_{0}= &\frac{q_{x}^{2}}{2}, \\
\omega_{\pm}=&\frac{1}{2}(q_{x}^{2}\pm 2 \sqrt{\Omega^{2}+ k_{L}^{2} q_{x}^{2}}).
\end{align}
\end{subequations}
Based on this dispersion relation, the low-lying branch $\omega_{-}$ displays the single minimum in the $k_{L}^{2} < \Omega$ regime, while a double minimum appears for the $k_{L}^{2} > \Omega$ regime. The presence of a single minimum in the quasi-momentum direction leads to the zero-momentum phase (plane wave phase with zero momentum). In contrast, the appearance of the double minimum corresponds to the SW phase, characterized as the superposition of two momentum states, which leads to the density modulation in real space.

The conditions mentioned above, i.e., $k_{L}^{2}<\Omega$ (ZM phase) and $k_{L}^{2}>\Omega$ (SW phase), worked well in FM interaction and AFM interaction with repulsive density-density interaction. However, for highly attractive density-density interaction and in the presence of FM interaction and AFM interaction, we obtained the elongated zero-momentum phases (EZM~I phase, and EZM~II phase) in the $k_{L}^{2}>\Omega$ regime given in Figs.~\ref{fig05a} and ~\ref{fig07b}. The EZM~I phase in $k_{L}^{2}>\Omega$ regime also appeared for weak repulsive density-density interaction in the presence of FM interactions [see Fig.~\ref{fig05a}].

In the above sections, we reported various ground state phases: the ZM phase, EZM~I phase, EZM~II phase, SW phase, and the SSW phase with respect to ($k_{L}, \Omega$) in the presence of FM interaction and AFM interaction. We also studied the effect of interaction strengths $c_{0}$ and $c_{2}$ in the ZM phase, SW phase, and the SSW phase. We extensively studied the miscibility of the condensate and ground state transformation from one phase to another with respect to coupling strengths ($k_{L}, \Omega$) and interaction strengths ($c_{0}, c_{2}$). In the next section, we investigate the quench dynamics across these ground state phases by quenching the SO and Rabi coupling strengths, excluding the EZM phases, since these phases are trivial extensions of the ZM phase.

\subsection{Quench dynamics of the ground state phases} 
\label{quench}
After analyzing the detailed nature of the ground state phases of the spin-1 SO-coupled BECs for FM and AFM interactions, we now turn to their dynamics using the quenched dynamics.  First, we generate the ground state profiles, namely, the ZM phase, the SW phase, and the SSW phase of the condensate with interaction strengths $c_{0} = -1.0$ and $c_{2} = 5.0$ (as discussed above), in the presence of a harmonic trap with the trap strength $\lambda = 0.05$. By choosing a weak trap to minimize its influence on the system and to better reveal the underlying dynamics of the condensate following a quench, we study the quench dynamics by abruptly changing the SO ($k_{L}$) and Rabi coupling strength ($\Omega$).%
\subsubsection{ZM phase: $k_{L} \sim 0.0$, and $\Omega = 0.0$} 

Initially, we generate the ground state profile given in Fig.~\ref{figq-01a}(a) corresponding to $k_{L} \sim 0.0$, and $\Omega = 0.0$, with interactions $c_{0} = -1.0$, $c_{2} = 5.0$, which is a ZM phase, where all the atoms occupy the state $\psi_{0}$. A similar type of ground state profile is shown above in Fig.~\ref{fig02a}(a1). To start with, in Figs.~\ref{figq-01a}(b1)-(b3), we quench the SO coupling strength $k_{L} \sim 0.0 \rightarrow 0.20$ in the absence of $\Omega$ at $t = 0$, and study the time evolution of the ground state profile. During time evolution, periodic high-density peaks emerge along the propagation direction in the $m_{F} = \pm 1$ component of the condensate, which shows the revival of the $m_{F} = \pm 1$ components. This behavior is symmetric in the spatial direction. The zeroth component shows breather-like dynamics. Here, in real-time propagation, we observe all three components present~\cite{Ma2019}. Next, in Figs.~\ref{figq-01a}(c1)-(c3), we quench $k_{L} \sim 0.0 \rightarrow 1.0$. We observe the appearance of the complex pattern of $m_{F} = \pm 1$ components along the direction of propagation, which is symmetric in spatial direction. The zeroth component also exhibits similar behavior, which is centered at $x \approx 0$. The density profile is spatially extended, showing the spread of the condensate. The fragmentation of the density profile in spatio-temporal evaluation indicates the onset of dynamical instability~\cite{Mithun_2019}.%

Now, we quench the Rabi coupling strength, keeping the SO coupling strength fixed. At first, in Figs.~\ref{figq-01a}(d1)-(d3), quenching $\Omega = 0.0 \rightarrow 0.20$ results in breather-like dynamics across all three components of the condensate, with the zeroth component exhibiting the highest density magnitude. Similar to SO coupling quenching, here, also, the quenching dynamics induces the revival of the $m_{F} = \pm 1$ components of the condensate; however, it shows stable behaviour throughout. Next, we quench $\Omega = 0.0 \rightarrow 0.50$ in Figs.~\ref{fig01a}(e1)-(e3). Here, we also obtain stable breather dynamics in all three components of the condensate; the zeroth component remains more dominant. However, the number of breathers is more than the $\Omega = 0.20$ case~\cite{Ravisankar_2020}.%

\subsubsection{SW phase: $k_{L} = 3.0$, $\Omega = 1.0$}

\begin{figure}[!htp] 
\begin{centering}
\centering\includegraphics[width=0.99\linewidth]{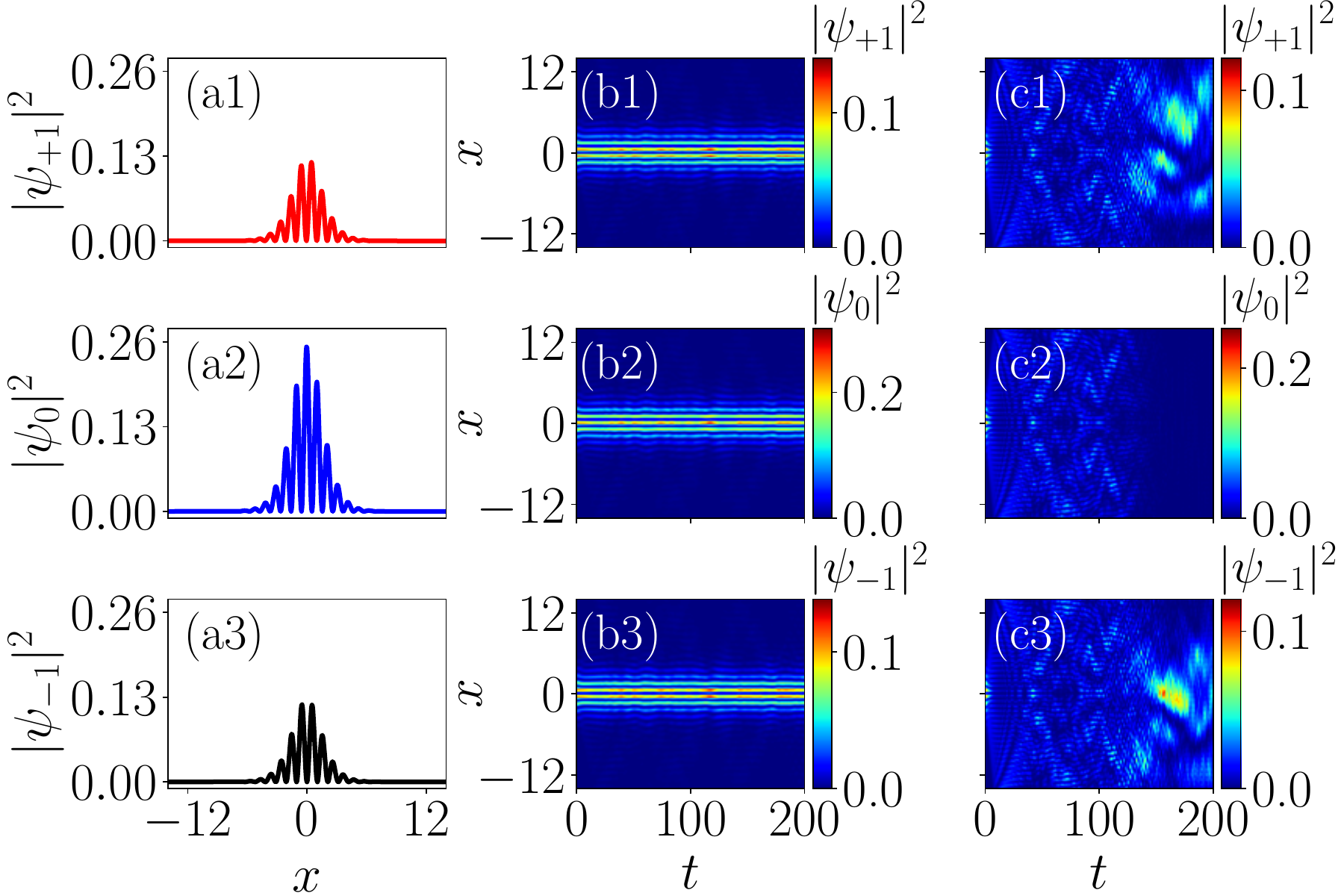}
\caption{Ground state density profile (a1-a3) for $m_{F} = +1, 0, -1$, respectively, with $(k_{L}, \Omega) = (3.0, 1.0)$. The interaction strengths are $c_{0} =-1.0$, and $c_{2} =5.0$. The condensate is in the SW phase. Panels (b1-b3) and (c1-c3) depict the time evolution of the ground state profile in the $x-t$ plane upon quenching the SO coupling strength $k_{L} = 3.0 \rightarrow 3.20$ and $k_{L} = 3.0 \rightarrow 4.0$, respectively. The SO coupling is quenched at $t = 0$ units. For a stronger quench of the Rabi coupling strength, the condensate fragments into small domains, forming a complex pattern, and the zeroth component disappears at a later time unit.}
 \label{figq-02a}
 \end{centering}
 \end{figure}%
 
In this section, we generate the ground state corresponding to $k_{L} = 3.0$, $\Omega = 1.0$, with interaction strengths $c_{0} = -1.0$, and $c_{2} = 5.0$, which is the stripe wave (SW) phase given in Fig.~\ref{figq-02a}(a). A similar type of ground state profile is shown in Fig.~\ref{fig05a}(b1). To initiate the quench dynamics, we first increase $k_{L} = 3.0 \rightarrow 3.20$ suddenly, while keeping $\Omega$ fixed and evolving it in time. At first, small fluctuations occur in the density profile in the $x$-$t$ plane; further, beyond $t = 180$ units, the density profile shows higher fluctuations, as shown in Figs.~\ref {figq-02a}(b1)-(b3). Next, in Figs.~\ref {figq-02a}(c1)-(c3), we quench $k_{L} = 3.0 \rightarrow 4.0$, which breaks the density profile initially into several small parts; further, it shows formation of complex patterns in the $x$-$t$ plane. Beyond $t > 120$ units, the density profile of the zeroth component starts to disappear, and the $m_{F} = \pm1$ components gain amplitude. The fluctuation behaviour and formation of complex patterns during time evolution in the density profile indicate the onset of dynamical instability~\cite{Mithun_2019, Gangwar2024}.%

\subsubsection{SSW phase: $k_{L} = 3.0$, $\Omega = 0.0$}
\begin{figure}[!htp] 
\begin{centering}
\centering\includegraphics[width=0.99\linewidth]{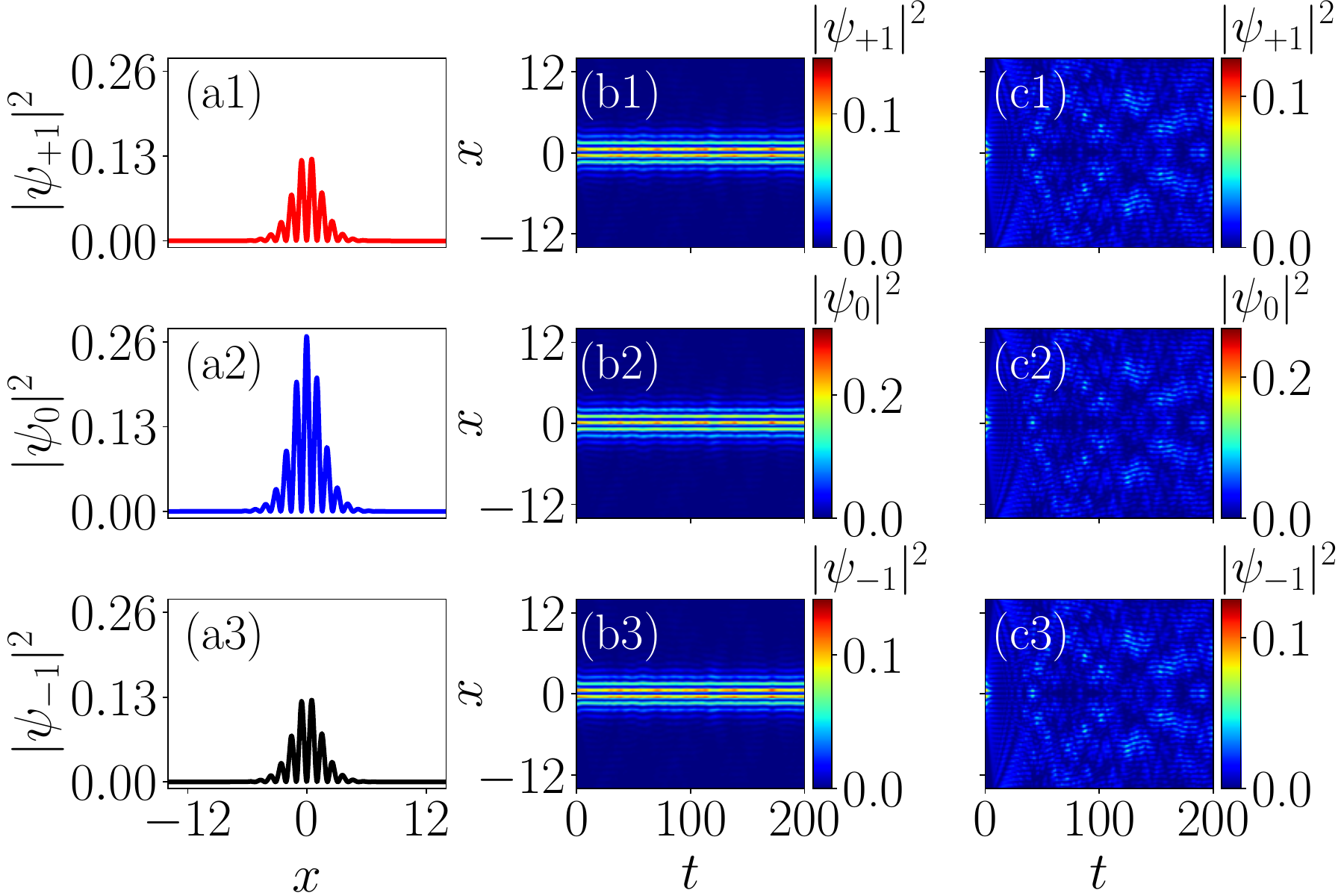}
\caption{Ground state density profile (a1-a3) for $m_{F} = +1, 0, -1$, respectively, with $(k_{L}, \Omega) = (3.0, 0.0)$. The interaction strengths are $c_{0} =-1.0$, and $c_{2} =5.0$. The condensate is in the SSW phase. Panels (b1-b3) and (c1-c3) exhibit dynamics of the ground state profile in the $x-t$ plane upon quenching the SO coupling strength $k_{L} = 3.0 \rightarrow 3.20$ and $k_{L} = 3.0 \rightarrow 4.0$, respectively. The SO coupling is quenched at $t = 0$ units. The strong quench of the SO coupling strength leads to the emergence of complex patterns.}
 \label{figq-03a}
 \end{centering}
 \end{figure}%

In this part of the study, we generate a ground state considering $k_{L} = 3.0$ in the absence of $\Omega$ with interaction strengths $c_{0}= -1.0$, and $c_{2} = 5.0$, which is an unpolarized SW phase (SSW phase) given in Fig.~\ref{figq-03a}(a). A similar type of density profile is given in Fig.~\ref{fig07a}(b1). To initiate the quench dynamics, firstly, in Figs.~\ref{figq-03a}(b1)-(b3) at $t = 0$, we quench the SO coupling to a smaller value $k_{L} = 3.0 \rightarrow 3.20$. During time evolution, the density profile exhibits minute fluctuations in the $x$–$t$ plane, and breather-like dynamics appear across all components; however, they are more pronounced in the zeroth component of the density profile. Next, in Figs.~\ref {figq-03a}(c1)-(c3) we sudden increase the SO coupling $k_{L} = 3.0 \rightarrow 4.0$. The density profile splits into several parts and further fragments into small domains across all components. These fluctuations and the formation of complex patterns during time evolution indicate the dynamical instability of the condensate~\cite{Mithun_2019, Gangwar2024}.%

\section{Summary and Conclusions}
\label{sec:sumcon}
In this work, we have numerically investigated the ground state phases and quench dynamics of an SO coupled spin-1 BEC with both FM and AFM interactions. The interplay between interaction strengths and coupling parameters gives rise to a rich variety of quantum phases. For FM interactions, at $\Omega \sim 0$, we identify four distinct phases: the zero-momentum (ZM) phase, the elongated zero-momentum of first type (EZM~I) phase, the stripe wave (SW) phase, and the superstripe wave (SSW) phase. Notably, the SSW phase is absent when the Rabi coupling is finite. In contrast, with the AFM interactions for $\Omega \sim 0$, only the ZM, EZM~II, and the SSW phases are observed. As the Rabi coupling increases, the condensate with AFM interaction exhibits the ZM, EZM~I, and SW phases. In the absence of SO and Rabi couplings, the AFM ZM phase is characterized by full population of the $\psi_{0}$ component, while the $\psi_{\pm 1}$ components are unpopulated. However, the introduction of either SO or Rabi coupling strength revives the $\psi_{\pm 1}$ components of the condensate, signaling a transition from the pure to the ZM phase. 

We have further explored the influence of the density-density interaction $c_{0}$ and the spin-density interaction $c_{2}$ on these ground state phases. The ZM phase remains unchanged under variations of $c_{0}$ and $c_{2}$. However, increasing the attractive strength of $c_{0}$ leads to spatial compression and a higher peak amplitude of the condensate density profile, while repulsive interactions have the opposite effect. Both the SW and SSW phases can transition into the EZM (EZM~I or EZM~II) phase under sufficiently strong attractive  ($c_{0}$). Specific phase transition points under different interaction strengths are summarized in the table below.


\begin{center}
\begin{tabular}{|c|c|c|c|}
\hline
$k_L = 3.0$, $\Omega = 0$ & $c_2 = -5.0$ & $c_0 = -1.25$ & SSW-EZM~I \\
\cline{2-4}
& $c_2 = 5.0$ & $c_0 = -3.75$ & SSW-EZM~II \\
\hline
$k_L = 3.0$, $\Omega = 1.0$ & $c_2 = -5.0$ & $c_0 = -0.5$ & SW-EZM~I \\
\cline{2-4}
& $c_2 = 5.0$ & $c_0 = -3.75$ & SW-EZM~I \\
\hline
\end{tabular}
\end{center}

To characterize the spin texture of the condensate, we computed the spin density vector (SDV) components: $\mathrm{S}_{x}$, $\mathrm{S}_{y}$, and $\mathrm{S}_{z}$. We find that $\mathrm{S}_{y}$ is identically zero across all observed phases. A single peak profile in $\mathrm{S}_{x}$ accompanied by the vanishing $\mathrm{S}_{z}$ identifies the ZM phase, whereas the presence of finite $\mathrm{S}_z$ with a single-peak $S_x$ corresponds to the elongated zero-momentum phase of first type (EZM~I) phase. The stripe wave (SW) phase is marked by a multipeak structure in $\mathrm{S}_x$ and an oscillatory behavior in $\mathrm{S}_z$. In contrast, all SDV components vanish in the superstripe wave (SSW) phase, indicating a suppressed spin texture. Similar to the SSW phase, all the SDV components vanish for the EZM~II phase, which, in general, lacks the spatial density modulation. 

We also calculated the overlap parameter $\eta_{np}$ between densities $\rho_{+1}$, and $\rho_{-1}$ to quantify the phase separation. Spatial separation between these components emerges at intermediate SO coupling strengths for both FM and AFM interactions. In the ZM phase, the condensate remains fully miscible, as reflected by a high total overlap parameter $\eta$. Other phases exhibit partial miscibility, corresponding to reduced overlap values. The root mean square size ($x_{\text{rms}}$) serves as another diagnostic of phase transitions. For FM interactions, $x_{\text{rms}}$ increases during the transition from the EZM~I phase to the SW or SSW phases. Conversely, in the AFM regime, the same transitions exhibit a decreasing trend in $x_{\text{rms}}$.

Finally, we have investigated the quench dynamics across these ground state phases by abruptly changing the SO and Rabi coupling strength. In the ZM phase, quenching of these coupling parameters induces a revival of the $\psi_{\pm 1}$ components. A sudden change in the Rabi coupling tends to stabilize the condensate, while the SO coupling yields a more prominent effect: small quenches generate spatial-temporal ($x$-$t$) density fluctuation and large quenches result in the formation of complex domain structure during time evolution. For sufficiently strong SO quenches, the density of the $\psi_0$ component diminishes and eventually vanishes, particularly in the SW phase.

Overall, our results provide a comprehensive picture of the interplay between spin-orbit coupling, Rabi coupling, and interaction strengths in shaping the ground-state properties and dynamics of the SO coupled spin-1 BECs. These insights could be useful for future efforts to engineer and control quantum phases in the multicomponent ultracold atomic systems.%

\acknowledgments
We gratefully acknowledge our Param-Ishan and Param Kamrupa supercomputing facility (IITG), where all numerical simulations were performed. S.K.G gratefully acknowledges a research fellowship from MoE, Government of India.

\twocolumngrid
\appendix
\counterwithin{figure}{section}

\section{Miscibility and Ground state phase transformation in the SSW phase}
\label{appdxa}

In Sec .~\ref {ImpSSW}, we reported density profiles of the ground state, where we obtained the phase transition from the SSW phase to the EZM~I phase and EZM~II phase, specifically for the $c_{0} < 0$ case. Here, we report miscibility and ground state phase transformation upon varying the $c_{0}$ and $c_{2}$ in the SSW phase. 

\begin{figure}[!htp] 
\centering\includegraphics[width=0.99\linewidth]{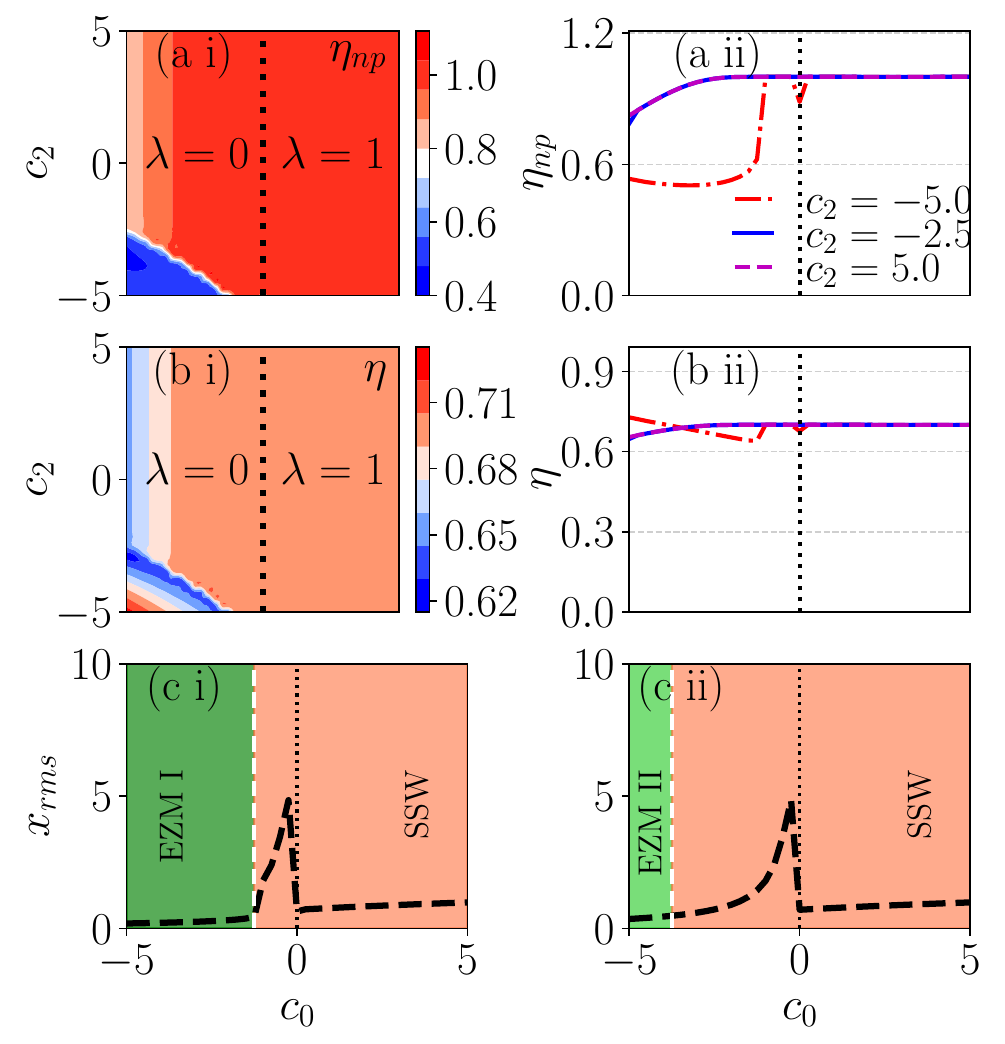}
\caption{Miscibility parameters (a) $\eta_{np}$, and (b) $\eta$ upon varying $c_{0}$ and keeping $k_{L} = 3.0$ in absence of $\Omega$. In (a i) and (b i), $c_{0}$ and $c_{2}$ vary within the range [-5,5]. (a ii) and (b ii) exhibits the $\eta_{np}$, and $\eta$ upon varying $c_{0}$ for $c_{2} = -5.0$ (dashed-dotted rd line), $-2.5$ (solid blue line), and $5.0$ (dashed magenta line). In the bottom row, panels (c i) and (c ii) represent the root mean square ($x_{rms}$) size of the condensate for $c_{2}=-5.0$ and $5.0$, respectively, which indicate different phases of the condensate. The color code represents different quantum phases: green represents the EZM~I phase, limegreen represents the EZM~II phase, and coral represents the SSW phase. The black dotted line separates untrapped ($c_{0} < 0$) and trapped ($c_{0} \gtrsim 0$) regimes of the condensate.}
\label{fig07c} 
\end{figure}

We report the variation of $\eta_{np}$ in Fig.~\ref{fig07c}(a i) by simultaneously varying the $c_{0}$, and $c_{2}$ in the range [-5,5] for $k_{L} = 3.0$ and $\Omega=0$. The black dotted line separates the phase plot into two parts: (i) the left half, in the absence of trap strength, and (ii) the right half, where the trap is present. For $c_{2} = -5.0$  and $c_{0} = -5.0$, we obtain the $\eta_{np} = 0.534$ which indicates a separation between the densities components $\rho_{+1}$ and $\rho_{-1}$ and condensate is in EZM~I phase. The $\eta_{np}$ shows an increasing trend upon increase in $c_{0}$, which finally attains as $\eta_{np} = 1.0$ for $c_{0} = -1.0$, which also results in making the transition to the SSW phase. Note that for $c_{0} \gtrsim 0$, the condensate is trapped under the harmonic potential. For $c_{0} > 0$, $\eta_{np}$ remains flat and therefore condensate retains the SSW phase. However, for $c_{2} = 5.0$ and $c_{0} = -5.0$, we obtain $\eta_{np} = 0.824$ that indicates a separation between density component $\rho_{+1}$, and $\rho_{-1}$, and the condensate is in the EZM~II phase. With increasing $c_{0}$, $\eta_{np}$ keeps increasing and reaches to $\eta_{np} = 1.0$ at $c_{0} = -1.0$, where the condensate transitions into the SSW phase. Similar to the previous case, for $c_{0} \gtrsim 0$, a harmonic trap is introduced. Beyond this point, the $\eta_{np}$ remains flat and the condensate remains in the SSW phase. 

For a better description of the overlap parameters $\eta_{np}$, in Fig.~\ref{fig07c}(a ii), we report it by choosing three different values of $c_{2} = -5.0, -2.5$, and $5.0$, varying $c_{0}$ in the range [-5,5] at fixed $k_{L} = 3.0$, in absence of $\Omega$. At first, considering $c_{2}  = -5.0$, and $c_{0} = -5.0$, we obtain $\eta_{np} = 0.534$. Increasing $c_{0}$, it remains constant up to $c_{0} = -1.50$. Further, increasing $c_{0}$, $\eta_{np}$ increases to a higher value at $c_{0} = -1.0$ with magnitude $\eta_{np} \approx 1.0$ and saturates; however, a kink to lower amplitude appears at $c_{0} = 0.0$ with magnitude $\eta_{np} = 0.885$. Next, considering $c_{2} = -2.5$, we obtain $\eta_{np} = 0.787$ at $c_{0} = -5.0$. Upon increasing $c_{0}$, $\eta_{np}$ increases gradually and saturates at $c_{0} = -1.0$ with magnitude $\eta_{np} \approx 1.0$. For $c_{2} = 5.0$, $\eta_{np}$ varies similar to $c_{2} = -2.5$.%

In Fig.~\ref{fig07c}(b i), we report the variation of miscibility parameter $\eta$ in the $c_{0}$- $c_{2}$ plane for $k_{L} = 3.0$ and in the absence of $\Omega$. We begin with $c_{2} < 0$ with $c_{2} = -5.0$, and varying $c_{0}$. At $c_{0} = -5.0$, we find $\eta = 0.728$, corresponding to the condensate residing in the EZM~I phase. As $c_{0}$ increases, $\eta$ decreases until $c_{0} = -1.25$, reaching a minimum value of $\eta = 0.641$. Beyond this point, the $\eta$ increases again and the condensate makes a transition to the SSW phase. For $c_{0} \gtrsim 0$, a harmonic trap is introduced to confine the condensate, and $\eta$ remains flat while the condensate stays in the SSW phase. For $c_{2} > 0$ we fix $c_{2} = 5.0$ and vary $c_{0}$. At $c_{0} = -5.0$, indicating the condensate is in the  EZM~II phase. As $c_{0}$ increases $\eta$  rises continuously, reaching $\eta = 0.70$  at $c_{0} = -1.0$, where the condensate transitions to the SSW phase.  For $c_{0} \gtrsim 0$, with the harmonic trap applied, the $\eta$ remains flat and the condensate retains the SSW phase. Overall in the $c_{0}-c_{2}$ plane, the $\eta$ varies between $0.60$ and $0.75$, indicating that the EZM~I, EZM~II, and SSW phases of the condensate are partially miscible.

To probe the impact of density-density interaction and spin-exchange interaction is more quantitative way, in Fig.~\ref{fig07c}(b ii), we show the miscibility parameter $\eta$ by varying $c_{0}$ in the range [-5.0,5.0] for $c_{2} = -5.0$, $-2.5$, and $5.0$, keeping $k_{L} = 3.0$ in absence of $\Omega $. At first, considering $c_{2} = -5.0$, and $c_{0} = -5.0$, we obtain $\eta = 0.728$, which decreases upon increasing $c_{0}$ reaching to minimum at $c_{0} = -1.25$ with magnitude $\eta = 0.641$. Upon further increasing $c_{0}$, $\eta$ increases to a higher value and saturates. Next, we consider $c_{2} = -2.5$, and $c_{0} = -5.0$, we obtain $\eta = 0.648$. Increasing $c_{0}$, $\eta$ gradually increases to $\eta = 0.70$ at $c_{0} = -2.25$, and saturates further. We obtain a similar behaviour of $\eta$ throughout for $c_{2} = 5.0$ as we obtained for $c_{2} = -2.5$.%

Finally, we present the $x_{rms}$ of the condensate with respect to $c_{0}$, and $c_{2}$. Taking into account $c_{2} = -5.0$, and $c_{0} = -5.0$, in Fig.~\ref{fig07c}(c i), we obtain $x_{rms} \approx 0.188$ where condensate is in the EZM~I phase. Upon increasing $c_{0}$, the $x_{rms}$ increases gradually; however, the condensate remains in the EZM~I phase. At $c_{0}(SSW \rightarrow EZM~I) = -1.25$, the $x_{rms}$ starts to increase and peaks with a higher magnitude of $x_{rms} = 4.847$ at $c_{0} = -0.25$, which arises due to the density modulation and condensate transitions to the SSW phase. For $c_{0} \gtrsim 0$ (in the presence of a trap), the $x_{rms}$ falls to $x_{rms} \approx 0.640$ and the condensate remains in the SSW phase. Beyond this point, the $x_{rms}$ increases gradually, leading to an increase in the number of stripes, and the condensate remains in the SSW phase. However, at $c_{0} = 0$, we obtain the SW phase. In Fig.~\ref{fig07c}(c ii), considering $c_{2} = 5.0$, and $c_{0} = -5.0$, we obtain $x_{rms} \approx 0.359$, corresponding to the EZM~II phase. Upon increasing $c_{0}$, the $x_{rms}$ increases gradually, and condensate transitions to SSW phase at $c_{0}(SSW \rightarrow EZM~II) = -3.75$. Increasing $c_{0}$ beyond this point, the condensate is in the SSW phase. Further, increasing $c_{0}$, the $x_{rms}$ peaks at $c_{0} = -0.25$ with magnitude $x_{rms} = 4.847$. The increase in $x_{rms}$ arises as the SSW phase consists of a large number of maxima and minima. For $c_{0} = 0.0$, the $x_{rms}$ falls to $0.706$; however, the condensate remains in the SSW phase with fewer maxima and minima across all density components. Upon further increasing the $c_{0}$, the $x_{rms}$ gradually increases with condensate remaining in the SSW phase, and the number of maxima and minima increases with increasing $c_{0}$.%

\twocolumngrid
\bibliography{reference.bib}

\begin{thebibliography}{47}%
\makeatletter
\providecommand \@ifxundefined [1]{%
 \@ifx{#1\undefined}
}%
\providecommand \@ifnum [1]{%
 \ifnum #1\expandafter \@firstoftwo
 \else \expandafter \@secondoftwo
 \fi
}%
\providecommand \@ifx [1]{%
 \ifx #1\expandafter \@firstoftwo
 \else \expandafter \@secondoftwo
 \fi
}%
\providecommand \natexlab [1]{#1}%
\providecommand \enquote  [1]{``#1''}%
\providecommand \bibnamefont  [1]{#1}%
\providecommand \bibfnamefont [1]{#1}%
\providecommand \citenamefont [1]{#1}%
\providecommand \href@noop [0]{\@secondoftwo}%
\providecommand \href [0]{\begingroup \@sanitize@url \@href}%
\providecommand \@href[1]{\@@startlink{#1}\@@href}%
\providecommand \@@href[1]{\endgroup#1\@@endlink}%
\providecommand \@sanitize@url [0]{\catcode `\\12\catcode `\$12\catcode
  `\&12\catcode `\#12\catcode `\^12\catcode `\_12\catcode `\%12\relax}%
\providecommand \@@startlink[1]{}%
\providecommand \@@endlink[0]{}%
\providecommand \url  [0]{\begingroup\@sanitize@url \@url }%
\providecommand \@url [1]{\endgroup\@href {#1}{\urlprefix }}%
\providecommand \urlprefix  [0]{URL }%
\providecommand \Eprint [0]{\href }%
\providecommand \doibase [0]{https://doi.org/}%
\providecommand \selectlanguage [0]{\@gobble}%
\providecommand \bibinfo  [0]{\@secondoftwo}%
\providecommand \bibfield  [0]{\@secondoftwo}%
\providecommand \translation [1]{[#1]}%
\providecommand \BibitemOpen [0]{}%
\providecommand \bibitemStop [0]{}%
\providecommand \bibitemNoStop [0]{.\EOS\space}%
\providecommand \EOS [0]{\spacefactor3000\relax}%
\providecommand \BibitemShut  [1]{\csname bibitem#1\endcsname}%
\let\auto@bib@innerbib\@empty
\bibitem [{\citenamefont {Hasan}\ and\ \citenamefont {Kane}(2010)}]{Hasan2010}%
  \BibitemOpen
  \bibfield  {author} {\bibinfo {author} {\bibfnamefont {M.~Z.}\ \bibnamefont
  {Hasan}}\ and\ \bibinfo {author} {\bibfnamefont {C.~L.}\ \bibnamefont
  {Kane}},\ }\bibfield  {title} {\bibinfo {title} {{Colloquium: Topological
  insulators}},\ }\href {https://doi.org/10.1103/RevModPhys.82.3045} {\bibfield
   {journal} {\bibinfo  {journal} {Rev. Mod. Phys.}\ }\textbf {\bibinfo
  {volume} {82}},\ \bibinfo {pages} {3045} (\bibinfo {year}
  {2010})}\BibitemShut {NoStop}%
\bibitem [{\citenamefont {Qi}\ and\ \citenamefont {Zhang}(2011)}]{Liang2011}%
  \BibitemOpen
  \bibfield  {author} {\bibinfo {author} {\bibfnamefont {X.-L.}\ \bibnamefont
  {Qi}}\ and\ \bibinfo {author} {\bibfnamefont {S.-C.}\ \bibnamefont {Zhang}},\
  }\bibfield  {title} {\bibinfo {title} {{Topological insulators and
  superconductors}},\ }\href {https://doi.org/10.1103/RevModPhys.83.1057}
  {\bibfield  {journal} {\bibinfo  {journal} {Rev. Mod. Phys.}\ }\textbf
  {\bibinfo {volume} {83}},\ \bibinfo {pages} {1057} (\bibinfo {year}
  {2011})}\BibitemShut {NoStop}%
\bibitem [{\citenamefont {Kato}\ \emph {et~al.}(2004)\citenamefont {Kato},
  \citenamefont {Myers}, \citenamefont {Gossard},\ and\ \citenamefont
  {Awschalom}}]{Kato2004}%
  \BibitemOpen
  \bibfield  {author} {\bibinfo {author} {\bibfnamefont {Y.~K.}\ \bibnamefont
  {Kato}}, \bibinfo {author} {\bibfnamefont {R.~C.}\ \bibnamefont {Myers}},
  \bibinfo {author} {\bibfnamefont {A.~C.}\ \bibnamefont {Gossard}},\ and\
  \bibinfo {author} {\bibfnamefont {D.~D.}\ \bibnamefont {Awschalom}},\
  }\bibfield  {title} {\bibinfo {title} {{Observation of the spin Hall effect
  in semiconductors}},\ }\href {https://doi.org/10.1126/science.1105514}
  {\bibfield  {journal} {\bibinfo  {journal} {Science}\ }\textbf {\bibinfo
  {volume} {306}},\ \bibinfo {pages} {1910} (\bibinfo {year}
  {2004})}\BibitemShut {NoStop}%
\bibitem [{\citenamefont {Konig}\ \emph {et~al.}(2007)\citenamefont {Konig},
  \citenamefont {Wiedmann}, \citenamefont {Brune}, \citenamefont {Roth},
  \citenamefont {Buhmann}, \citenamefont {Molenkamp}, \citenamefont {Qi},\ and\
  \citenamefont {Zhang}}]{Konig2007}%
  \BibitemOpen
  \bibfield  {author} {\bibinfo {author} {\bibfnamefont {M.}~\bibnamefont
  {Konig}}, \bibinfo {author} {\bibfnamefont {S.}~\bibnamefont {Wiedmann}},
  \bibinfo {author} {\bibfnamefont {C.}~\bibnamefont {Brune}}, \bibinfo
  {author} {\bibfnamefont {A.}~\bibnamefont {Roth}}, \bibinfo {author}
  {\bibfnamefont {H.}~\bibnamefont {Buhmann}}, \bibinfo {author} {\bibfnamefont
  {L.~W.}\ \bibnamefont {Molenkamp}}, \bibinfo {author} {\bibfnamefont {X.-L.}\
  \bibnamefont {Qi}},\ and\ \bibinfo {author} {\bibfnamefont {S.-C.}\
  \bibnamefont {Zhang}},\ }\bibfield  {title} {\bibinfo {title} {{Quantum spin
  Hall insulator state in HgTe quantum wells}},\ }\href
  {https://doi.org/10.1126/science.1148047} {\bibfield  {journal} {\bibinfo
  {journal} {Science}\ }\textbf {\bibinfo {volume} {318}},\ \bibinfo {pages}
  {766} (\bibinfo {year} {2007})}\BibitemShut {NoStop}%
\bibitem [{\citenamefont {\ifmmode \check{Z}\else
  \v{Z}\fi{}uti\ifmmode~\acute{c}\else \'{c}\fi{}}\ \emph
  {et~al.}(2004)\citenamefont {\ifmmode \check{Z}\else
  \v{Z}\fi{}uti\ifmmode~\acute{c}\else \'{c}\fi{}}, \citenamefont {Fabian},\
  and\ \citenamefont {Das~Sarma}}]{Igor2004}%
  \BibitemOpen
  \bibfield  {author} {\bibinfo {author} {\bibfnamefont {I.}~\bibnamefont
  {\ifmmode \check{Z}\else \v{Z}\fi{}uti\ifmmode~\acute{c}\else \'{c}\fi{}}},
  \bibinfo {author} {\bibfnamefont {J.}~\bibnamefont {Fabian}},\ and\ \bibinfo
  {author} {\bibfnamefont {S.}~\bibnamefont {Das~Sarma}},\ }\bibfield  {title}
  {\bibinfo {title} {{Spintronics: Fundamentals and applications}},\ }\href
  {https://doi.org/10.1103/RevModPhys.76.323} {\bibfield  {journal} {\bibinfo
  {journal} {Rev. Mod. Phys.}\ }\textbf {\bibinfo {volume} {76}},\ \bibinfo
  {pages} {323} (\bibinfo {year} {2004})}\BibitemShut {NoStop}%
\bibitem [{\citenamefont {Lin}\ \emph {et~al.}(2011)\citenamefont {Lin},
  \citenamefont {Jim{\'e}nez-Garc{\'\i}a},\ and\ \citenamefont
  {Spielman}}]{Lin2011}%
  \BibitemOpen
  \bibfield  {author} {\bibinfo {author} {\bibfnamefont {Y.-J.}\ \bibnamefont
  {Lin}}, \bibinfo {author} {\bibfnamefont {K.}~\bibnamefont
  {Jim{\'e}nez-Garc{\'\i}a}},\ and\ \bibinfo {author} {\bibfnamefont {I.~B.}\
  \bibnamefont {Spielman}},\ }\bibfield  {title} {\bibinfo {title}
  {{Spin--orbit-coupled Bose--Einstein condensates}},\ }\href
  {https://doi.org/10.1038/nature09887} {\bibfield  {journal} {\bibinfo
  {journal} {Nature (London)}\ }\textbf {\bibinfo {volume} {471}},\ \bibinfo
  {pages} {83} (\bibinfo {year} {2011})}\BibitemShut {NoStop}%
\bibitem [{\citenamefont {Bychkov}\ and\ \citenamefont
  {Rashba}(1984)}]{Rashba1984}%
  \BibitemOpen
  \bibfield  {author} {\bibinfo {author} {\bibfnamefont {Y.~A.}\ \bibnamefont
  {Bychkov}}\ and\ \bibinfo {author} {\bibfnamefont {E.~I.}\ \bibnamefont
  {Rashba}},\ }\bibfield  {title} {\bibinfo {title} {{Oscillatory effects and
  the magnetic susceptibility of carriers in inversion layers}},\ }\href
  {https://doi.org/10.1088/0022-3719/17/33/015} {\bibfield  {journal} {\bibinfo
   {journal} {Jour. Phys. C}\ }\textbf {\bibinfo {volume} {17}},\ \bibinfo
  {pages} {6039} (\bibinfo {year} {1984})}\BibitemShut {NoStop}%
\bibitem [{\citenamefont {Dresselhaus}(1955)}]{Dresselhaus1955}%
  \BibitemOpen
  \bibfield  {author} {\bibinfo {author} {\bibfnamefont {G.}~\bibnamefont
  {Dresselhaus}},\ }\bibfield  {title} {\bibinfo {title} {Spin-orbit coupling
  effects in zinc blende structures},\ }\href
  {https://doi.org/10.1103/PhysRev.100.580} {\bibfield  {journal} {\bibinfo
  {journal} {Phys. Rev.}\ }\textbf {\bibinfo {volume} {100}},\ \bibinfo {pages}
  {580} (\bibinfo {year} {1955})}\BibitemShut {NoStop}%
\bibitem [{\citenamefont {Hamner}\ \emph {et~al.}(2014)\citenamefont {Hamner},
  \citenamefont {Qu}, \citenamefont {Zhang}, \citenamefont {Chang},
  \citenamefont {Gong}, \citenamefont {Zhang},\ and\ \citenamefont
  {Engels}}]{Hamner2014}%
  \BibitemOpen
  \bibfield  {author} {\bibinfo {author} {\bibfnamefont {C.}~\bibnamefont
  {Hamner}}, \bibinfo {author} {\bibfnamefont {C.}~\bibnamefont {Qu}}, \bibinfo
  {author} {\bibfnamefont {Y.}~\bibnamefont {Zhang}}, \bibinfo {author}
  {\bibfnamefont {J.}~\bibnamefont {Chang}}, \bibinfo {author} {\bibfnamefont
  {M.}~\bibnamefont {Gong}}, \bibinfo {author} {\bibfnamefont {C.}~\bibnamefont
  {Zhang}},\ and\ \bibinfo {author} {\bibfnamefont {P.}~\bibnamefont
  {Engels}},\ }\bibfield  {title} {\bibinfo {title} {{Dicke-type phase
  transition in a spin-orbit-coupled Bose--Einstein condensate}},\ }\href
  {https://doi.org/10.1038/ncomms5023} {\bibfield  {journal} {\bibinfo
  {journal} {Nat. Commun.}\ }\textbf {\bibinfo {volume} {5}},\ \bibinfo {pages}
  {4023} (\bibinfo {year} {2014})}\BibitemShut {NoStop}%
\bibitem [{\citenamefont {Qu}\ \emph {et~al.}(2013)\citenamefont {Qu},
  \citenamefont {Hamner}, \citenamefont {Gong}, \citenamefont {Zhang},\ and\
  \citenamefont {Engels}}]{Qu2013}%
  \BibitemOpen
  \bibfield  {author} {\bibinfo {author} {\bibfnamefont {C.}~\bibnamefont
  {Qu}}, \bibinfo {author} {\bibfnamefont {C.}~\bibnamefont {Hamner}}, \bibinfo
  {author} {\bibfnamefont {M.}~\bibnamefont {Gong}}, \bibinfo {author}
  {\bibfnamefont {C.}~\bibnamefont {Zhang}},\ and\ \bibinfo {author}
  {\bibfnamefont {P.}~\bibnamefont {Engels}},\ }\bibfield  {title} {\bibinfo
  {title} {{Observation of Zitterbewegung in a spin-orbit-coupled Bose-Einstein
  condensate}},\ }\href {https://doi.org/10.1103/PhysRevA.88.021604} {\bibfield
   {journal} {\bibinfo  {journal} {Phys. Rev. A}\ }\textbf {\bibinfo {volume}
  {88}},\ \bibinfo {pages} {021604} (\bibinfo {year} {2013})}\BibitemShut
  {NoStop}%
\bibitem [{\citenamefont {Zhang}\ \emph {et~al.}(2022)\citenamefont {Zhang},
  \citenamefont {Liu},\ and\ \citenamefont {Zhang}}]{Zhang2022}%
  \BibitemOpen
  \bibfield  {author} {\bibinfo {author} {\bibfnamefont {H.}~\bibnamefont
  {Zhang}}, \bibinfo {author} {\bibfnamefont {S.}~\bibnamefont {Liu}},\ and\
  \bibinfo {author} {\bibfnamefont {Y.-S.}\ \bibnamefont {Zhang}},\ }\bibfield
  {title} {\bibinfo {title} {{Faraday patterns in spin-orbit-coupled
  Bose-Einstein condensates}},\ }\href
  {https://doi.org/10.1103/PhysRevA.105.063319} {\bibfield  {journal} {\bibinfo
   {journal} {Phys. Rev. A}\ }\textbf {\bibinfo {volume} {105}},\ \bibinfo
  {pages} {063319} (\bibinfo {year} {2022})}\BibitemShut {NoStop}%
\bibitem [{\citenamefont {Martone}\ and\ \citenamefont
  {Stringari}(2021)}]{Martone2021}%
  \BibitemOpen
  \bibfield  {author} {\bibinfo {author} {\bibfnamefont {G.~I.}\ \bibnamefont
  {Martone}}\ and\ \bibinfo {author} {\bibfnamefont {S.}~\bibnamefont
  {Stringari}},\ }\bibfield  {title} {\bibinfo {title} {{Supersolid phase of a
  spin-orbit-coupled Bose-Einstein condensate: A perturbation approach}},\
  }\href {https://doi.org/10.21468/SciPostPhys.11.5.092} {\bibfield  {journal}
  {\bibinfo  {journal} {SciPost Phys.}\ }\textbf {\bibinfo {volume} {11}},\
  \bibinfo {pages} {092} (\bibinfo {year} {2021})}\BibitemShut {NoStop}%
\bibitem [{\citenamefont {Chen}\ \emph {et~al.}(2022)\citenamefont {Chen},
  \citenamefont {Lyu}, \citenamefont {Xu},\ and\ \citenamefont
  {Zhang}}]{Chen2022}%
  \BibitemOpen
  \bibfield  {author} {\bibinfo {author} {\bibfnamefont {Y.}~\bibnamefont
  {Chen}}, \bibinfo {author} {\bibfnamefont {H.}~\bibnamefont {Lyu}}, \bibinfo
  {author} {\bibfnamefont {Y.}~\bibnamefont {Xu}},\ and\ \bibinfo {author}
  {\bibfnamefont {Y.}~\bibnamefont {Zhang}},\ }\bibfield  {title} {\bibinfo
  {title} {{Elementary excitations in a spin--orbit-coupled spin-1
  Bose--Einstein condensate}},\ }\href
  {https://doi.org/10.1088/1367-2630/ac7fb1} {\bibfield  {journal} {\bibinfo
  {journal} {New J. Phys.}\ }\textbf {\bibinfo {volume} {24}},\ \bibinfo
  {pages} {073041} (\bibinfo {year} {2022})}\BibitemShut {NoStop}%
\bibitem [{\citenamefont {Campbell}\ \emph {et~al.}(2016)\citenamefont
  {Campbell}, \citenamefont {Price}, \citenamefont {Putra}, \citenamefont
  {Vald{\'e}s-Curiel}, \citenamefont {Trypogeorgos},\ and\ \citenamefont
  {Spielman}}]{Campbell2016}%
  \BibitemOpen
  \bibfield  {author} {\bibinfo {author} {\bibfnamefont {D.}~\bibnamefont
  {Campbell}}, \bibinfo {author} {\bibfnamefont {R.}~\bibnamefont {Price}},
  \bibinfo {author} {\bibfnamefont {A.}~\bibnamefont {Putra}}, \bibinfo
  {author} {\bibfnamefont {A.}~\bibnamefont {Vald{\'e}s-Curiel}}, \bibinfo
  {author} {\bibfnamefont {D.}~\bibnamefont {Trypogeorgos}},\ and\ \bibinfo
  {author} {\bibfnamefont {I.}~\bibnamefont {Spielman}},\ }\bibfield  {title}
  {\bibinfo {title} {{Magnetic phases of spin-1 spin--orbit-coupled Bose
  gases}},\ }\href {https://doi.org/10.1038/ncomms10897} {\bibfield  {journal}
  {\bibinfo  {journal} {Nat. Commun.}\ }\textbf {\bibinfo {volume} {7}},\
  \bibinfo {pages} {10897} (\bibinfo {year} {2016})}\BibitemShut {NoStop}%
\bibitem [{\citenamefont {Luo}\ \emph {et~al.}(2016)\citenamefont {Luo},
  \citenamefont {Wu}, \citenamefont {Chen}, \citenamefont {Guan}, \citenamefont
  {Gao}, \citenamefont {Xu}, \citenamefont {You},\ and\ \citenamefont
  {Wang}}]{Luo2016}%
  \BibitemOpen
  \bibfield  {author} {\bibinfo {author} {\bibfnamefont {X.}~\bibnamefont
  {Luo}}, \bibinfo {author} {\bibfnamefont {L.}~\bibnamefont {Wu}}, \bibinfo
  {author} {\bibfnamefont {J.}~\bibnamefont {Chen}}, \bibinfo {author}
  {\bibfnamefont {Q.}~\bibnamefont {Guan}}, \bibinfo {author} {\bibfnamefont
  {K.}~\bibnamefont {Gao}}, \bibinfo {author} {\bibfnamefont {Z.-F.}\
  \bibnamefont {Xu}}, \bibinfo {author} {\bibfnamefont {L.}~\bibnamefont
  {You}},\ and\ \bibinfo {author} {\bibfnamefont {R.}~\bibnamefont {Wang}},\
  }\bibfield  {title} {\bibinfo {title} {{Tunable atomic spin-orbit coupling
  synthesized with a modulating gradient magnetic field}},\ }\href
  {https://doi.org/10.1038/srep18983} {\bibfield  {journal} {\bibinfo
  {journal} {Scient. Rep.}\ }\textbf {\bibinfo {volume} {6}},\ \bibinfo {pages}
  {18983} (\bibinfo {year} {2016})}\BibitemShut {NoStop}%
\bibitem [{\citenamefont {Bulgakov}\ and\ \citenamefont
  {Sadreev}(2003)}]{Bulgakov2003}%
  \BibitemOpen
  \bibfield  {author} {\bibinfo {author} {\bibfnamefont {E.~N.}\ \bibnamefont
  {Bulgakov}}\ and\ \bibinfo {author} {\bibfnamefont {A.~F.}\ \bibnamefont
  {Sadreev}},\ }\bibfield  {title} {\bibinfo {title} {{Vortex Phase Diagram of
  $F=1$ Spinor Bose-Einstein Condensates}},\ }\href
  {https://doi.org/10.1103/PhysRevLett.90.200401} {\bibfield  {journal}
  {\bibinfo  {journal} {Phys. Rev. Lett.}\ }\textbf {\bibinfo {volume} {90}},\
  \bibinfo {pages} {200401} (\bibinfo {year} {2003})}\BibitemShut {NoStop}%
\bibitem [{\citenamefont {Kato}\ \emph {et~al.}(2011)\citenamefont {Kato},
  \citenamefont {Nakano}, \citenamefont {Kasamatsu},\ and\ \citenamefont
  {Matsui}}]{Kato2011}%
  \BibitemOpen
  \bibfield  {author} {\bibinfo {author} {\bibfnamefont {A.}~\bibnamefont
  {Kato}}, \bibinfo {author} {\bibfnamefont {Y.}~\bibnamefont {Nakano}},
  \bibinfo {author} {\bibfnamefont {K.}~\bibnamefont {Kasamatsu}},\ and\
  \bibinfo {author} {\bibfnamefont {T.}~\bibnamefont {Matsui}},\ }\bibfield
  {title} {\bibinfo {title} {{Vortex formation of a Bose-Einstein condensate in
  a rotating deep optical lattice}},\ }\href
  {https://doi.org/10.1103/PhysRevA.84.053623} {\bibfield  {journal} {\bibinfo
  {journal} {Phys. Rev. A}\ }\textbf {\bibinfo {volume} {84}},\ \bibinfo
  {pages} {053623} (\bibinfo {year} {2011})}\BibitemShut {NoStop}%
\bibitem [{\citenamefont {Adhikari}(2020{\natexlab{a}})}]{Adhikari2021}%
  \BibitemOpen
  \bibfield  {author} {\bibinfo {author} {\bibfnamefont {S.~K.}\ \bibnamefont
  {Adhikari}},\ }\bibfield  {title} {\bibinfo {title} {{Vortex-lattice
  formation in a spin–orbit coupled rotating spin-1 condensate}},\ }\href
  {https://doi.org/10.1088/1361-648X/abc5d7} {\bibfield  {journal} {\bibinfo
  {journal} {J. Phys.: Conden. Matt.}\ }\textbf {\bibinfo {volume} {33}},\
  \bibinfo {pages} {065404} (\bibinfo {year} {2020}{\natexlab{a}})}\BibitemShut
  {NoStop}%
\bibitem [{\citenamefont {Katsimiga}\ \emph {et~al.}(2021)\citenamefont
  {Katsimiga}, \citenamefont {Mistakidis}, \citenamefont {Schmelcher},\ and\
  \citenamefont {Kevrekidis}}]{Katsimiga2021}%
  \BibitemOpen
  \bibfield  {author} {\bibinfo {author} {\bibfnamefont {G.~C.}\ \bibnamefont
  {Katsimiga}}, \bibinfo {author} {\bibfnamefont {S.~I.}\ \bibnamefont
  {Mistakidis}}, \bibinfo {author} {\bibfnamefont {P.}~\bibnamefont
  {Schmelcher}},\ and\ \bibinfo {author} {\bibfnamefont {P.~G.}\ \bibnamefont
  {Kevrekidis}},\ }\bibfield  {title} {\bibinfo {title} {{Phase diagram,
  stability and magnetic properties of nonlinear excitations in spinor
  Bose–Einstein condensates}},\ }\href
  {https://doi.org/10.1088/1367-2630/abd27c} {\bibfield  {journal} {\bibinfo
  {journal} {New J. Phys.}\ }\textbf {\bibinfo {volume} {23}},\ \bibinfo
  {pages} {013015} (\bibinfo {year} {2021})}\BibitemShut {NoStop}%
\bibitem [{\citenamefont {Adhikari}(2021)}]{Adhikari2021sym}%
  \BibitemOpen
  \bibfield  {author} {\bibinfo {author} {\bibfnamefont {S.~K.}\ \bibnamefont
  {Adhikari}},\ }\bibfield  {title} {\bibinfo {title} {{Symbiotic solitons in
  quasi-one- and quasi-two-dimensional spin-1 condensates}},\ }\href
  {https://doi.org/10.1103/PhysRevE.104.024207} {\bibfield  {journal} {\bibinfo
   {journal} {Phys. Rev. E}\ }\textbf {\bibinfo {volume} {104}},\ \bibinfo
  {pages} {024207} (\bibinfo {year} {2021})}\BibitemShut {NoStop}%
\bibitem [{\citenamefont {Yang}\ and\ \citenamefont {Zhang}(2023)}]{Zhang2023}%
  \BibitemOpen
  \bibfield  {author} {\bibinfo {author} {\bibfnamefont {J.}~\bibnamefont
  {Yang}}\ and\ \bibinfo {author} {\bibfnamefont {Y.}~\bibnamefont {Zhang}},\
  }\bibfield  {title} {\bibinfo {title} {{Spin-orbit-coupled spinor gap
  solitons in Bose-Einstein condensates}},\ }\href
  {https://doi.org/10.1103/PhysRevA.107.023316} {\bibfield  {journal} {\bibinfo
   {journal} {Phys. Rev. A}\ }\textbf {\bibinfo {volume} {107}},\ \bibinfo
  {pages} {023316} (\bibinfo {year} {2023})}\BibitemShut {NoStop}%
\bibitem [{\citenamefont {Sun}\ \emph {et~al.}(2016)\citenamefont {Sun},
  \citenamefont {Qu}, \citenamefont {Xu}, \citenamefont {Zhang},\ and\
  \citenamefont {Zhang}}]{Sun2016}%
  \BibitemOpen
  \bibfield  {author} {\bibinfo {author} {\bibfnamefont {K.}~\bibnamefont
  {Sun}}, \bibinfo {author} {\bibfnamefont {C.}~\bibnamefont {Qu}}, \bibinfo
  {author} {\bibfnamefont {Y.}~\bibnamefont {Xu}}, \bibinfo {author}
  {\bibfnamefont {Y.}~\bibnamefont {Zhang}},\ and\ \bibinfo {author}
  {\bibfnamefont {C.}~\bibnamefont {Zhang}},\ }\bibfield  {title} {\bibinfo
  {title} {{Interacting spin-orbit-coupled spin-1 Bose-Einstein condensates}},\
  }\href {https://doi.org/10.1103/PhysRevA.93.023615} {\bibfield  {journal}
  {\bibinfo  {journal} {Phys. Rev. A}\ }\textbf {\bibinfo {volume} {93}},\
  \bibinfo {pages} {023615} (\bibinfo {year} {2016})}\BibitemShut {NoStop}%
\bibitem [{\citenamefont {Gautam}\ and\ \citenamefont
  {Adhikari}(2015{\natexlab{a}})}]{Gautam2015Mob}%
  \BibitemOpen
  \bibfield  {author} {\bibinfo {author} {\bibfnamefont {S.}~\bibnamefont
  {Gautam}}\ and\ \bibinfo {author} {\bibfnamefont {S.~K.}\ \bibnamefont
  {Adhikari}},\ }\bibfield  {title} {\bibinfo {title} {{Mobile vector soliton
  in a spin--orbit coupled spin-1 condensate}},\ }\href
  {https://doi.org/10.1088/1612-2011/12/4/045501} {\bibfield  {journal}
  {\bibinfo  {journal} {Laser Phys. Lett.}\ }\textbf {\bibinfo {volume} {12}},\
  \bibinfo {pages} {045501} (\bibinfo {year} {2015}{\natexlab{a}})}\BibitemShut
  {NoStop}%
\bibitem [{\citenamefont {Adhikari}(2020{\natexlab{b}})}]{Adhikari2020stable}%
  \BibitemOpen
  \bibfield  {author} {\bibinfo {author} {\bibfnamefont {S.~K.}\ \bibnamefont
  {Adhikari}},\ }\bibfield  {title} {\bibinfo {title} {{Stable multi-peak
  vector solitons in spin–orbit coupled spin-1 polar condensates}},\ }\href
  {https://doi.org/https://doi.org/10.1016/j.physe.2019.113892} {\bibfield
  {journal} {\bibinfo  {journal} {Phys. E: Low-Dimens. Syst. Nanostructures}\
  }\textbf {\bibinfo {volume} {118}},\ \bibinfo {pages} {113892} (\bibinfo
  {year} {2020}{\natexlab{b}})}\BibitemShut {NoStop}%
\bibitem [{\citenamefont {Gautam}\ and\ \citenamefont
  {Adhikari}(2014)}]{Gautam2014}%
  \BibitemOpen
  \bibfield  {author} {\bibinfo {author} {\bibfnamefont {S.}~\bibnamefont
  {Gautam}}\ and\ \bibinfo {author} {\bibfnamefont {S.~K.}\ \bibnamefont
  {Adhikari}},\ }\bibfield  {title} {\bibinfo {title} {{Phase separation in a
  spin-orbit-coupled Bose-Einstein condensate}},\ }\href
  {https://doi.org/10.1103/PhysRevA.90.043619} {\bibfield  {journal} {\bibinfo
  {journal} {Phys. Rev. A}\ }\textbf {\bibinfo {volume} {90}},\ \bibinfo
  {pages} {043619} (\bibinfo {year} {2014})}\BibitemShut {NoStop}%
\bibitem [{\citenamefont {Gautam}\ and\ \citenamefont
  {Adhikari}(2015{\natexlab{b}})}]{Gautam2015S2}%
  \BibitemOpen
  \bibfield  {author} {\bibinfo {author} {\bibfnamefont {S.}~\bibnamefont
  {Gautam}}\ and\ \bibinfo {author} {\bibfnamefont {S.~K.}\ \bibnamefont
  {Adhikari}},\ }\bibfield  {title} {\bibinfo {title} {{Spontaneous symmetry
  breaking in a spin-orbit-coupled $f=2$ spinor condensate}},\ }\href
  {https://doi.org/10.1103/PhysRevA.91.013624} {\bibfield  {journal} {\bibinfo
  {journal} {Phys. Rev. A}\ }\textbf {\bibinfo {volume} {91}},\ \bibinfo
  {pages} {013624} (\bibinfo {year} {2015}{\natexlab{b}})}\BibitemShut
  {NoStop}%
\bibitem [{\citenamefont {Bookjans}\ \emph {et~al.}(2011)\citenamefont
  {Bookjans}, \citenamefont {Vinit},\ and\ \citenamefont
  {Raman}}]{Bookjans2011}%
  \BibitemOpen
  \bibfield  {author} {\bibinfo {author} {\bibfnamefont {E.~M.}\ \bibnamefont
  {Bookjans}}, \bibinfo {author} {\bibfnamefont {A.}~\bibnamefont {Vinit}},\
  and\ \bibinfo {author} {\bibfnamefont {C.}~\bibnamefont {Raman}},\ }\bibfield
   {title} {\bibinfo {title} {{Quantum Phase Transition in an Antiferromagnetic
  Spinor Bose-Einstein Condensate}},\ }\href
  {https://doi.org/10.1103/PhysRevLett.107.195306} {\bibfield  {journal}
  {\bibinfo  {journal} {Phys. Rev. Lett.}\ }\textbf {\bibinfo {volume} {107}},\
  \bibinfo {pages} {195306} (\bibinfo {year} {2011})}\BibitemShut {NoStop}%
\bibitem [{\citenamefont {Yang}\ \emph {et~al.}(2019)\citenamefont {Yang},
  \citenamefont {Tian}, \citenamefont {Yang}, \citenamefont {Qiu},
  \citenamefont {Liang}, \citenamefont {Chu}, \citenamefont
  {Da\ifmmode~\breve{g}\else \u{g}\fi{}}, \citenamefont {Xu}, \citenamefont
  {Liu},\ and\ \citenamefont {Duan}}]{Yang2019}%
  \BibitemOpen
  \bibfield  {author} {\bibinfo {author} {\bibfnamefont {H.-X.}\ \bibnamefont
  {Yang}}, \bibinfo {author} {\bibfnamefont {T.}~\bibnamefont {Tian}}, \bibinfo
  {author} {\bibfnamefont {Y.-B.}\ \bibnamefont {Yang}}, \bibinfo {author}
  {\bibfnamefont {L.-Y.}\ \bibnamefont {Qiu}}, \bibinfo {author} {\bibfnamefont
  {H.-Y.}\ \bibnamefont {Liang}}, \bibinfo {author} {\bibfnamefont {A.-J.}\
  \bibnamefont {Chu}}, \bibinfo {author} {\bibfnamefont {C.~B.}\ \bibnamefont
  {Da\ifmmode~\breve{g}\else \u{g}\fi{}}}, \bibinfo {author} {\bibfnamefont
  {Y.}~\bibnamefont {Xu}}, \bibinfo {author} {\bibfnamefont {Y.}~\bibnamefont
  {Liu}},\ and\ \bibinfo {author} {\bibfnamefont {L.-M.}\ \bibnamefont
  {Duan}},\ }\bibfield  {title} {\bibinfo {title} {{Observation of dynamical
  quantum phase transitions in a spinor condensate}},\ }\href
  {https://doi.org/10.1103/PhysRevA.100.013622} {\bibfield  {journal} {\bibinfo
   {journal} {Phys. Rev. A}\ }\textbf {\bibinfo {volume} {100}},\ \bibinfo
  {pages} {013622} (\bibinfo {year} {2019})}\BibitemShut {NoStop}%
\bibitem [{\citenamefont {Saito}\ \emph {et~al.}(2007)\citenamefont {Saito},
  \citenamefont {Kawaguchi},\ and\ \citenamefont {Ueda}}]{Saito2007}%
  \BibitemOpen
  \bibfield  {author} {\bibinfo {author} {\bibfnamefont {H.}~\bibnamefont
  {Saito}}, \bibinfo {author} {\bibfnamefont {Y.}~\bibnamefont {Kawaguchi}},\
  and\ \bibinfo {author} {\bibfnamefont {M.}~\bibnamefont {Ueda}},\ }\bibfield
  {title} {\bibinfo {title} {{Topological defect formation in a quenched
  ferromagnetic Bose-Einstein condensates}},\ }\href
  {https://doi.org/10.1103/PhysRevA.75.013621} {\bibfield  {journal} {\bibinfo
  {journal} {Phys. Rev. A}\ }\textbf {\bibinfo {volume} {75}},\ \bibinfo
  {pages} {013621} (\bibinfo {year} {2007})}\BibitemShut {NoStop}%
\bibitem [{\citenamefont {Ravisankar}\ \emph
  {et~al.}(2021{\natexlab{a}})\citenamefont {Ravisankar}, \citenamefont
  {Fabrelli}, \citenamefont {Gammal}, \citenamefont {Muruganandam},\ and\
  \citenamefont {Mishra}}]{Mishra2021}%
  \BibitemOpen
  \bibfield  {author} {\bibinfo {author} {\bibfnamefont {R.}~\bibnamefont
  {Ravisankar}}, \bibinfo {author} {\bibfnamefont {H.}~\bibnamefont
  {Fabrelli}}, \bibinfo {author} {\bibfnamefont {A.}~\bibnamefont {Gammal}},
  \bibinfo {author} {\bibfnamefont {P.}~\bibnamefont {Muruganandam}},\ and\
  \bibinfo {author} {\bibfnamefont {P.~K.}\ \bibnamefont {Mishra}},\ }\bibfield
   {title} {\bibinfo {title} {{Effect of Rashba spin-orbit and Rabi couplings
  on the excitation spectrum of binary Bose-Einstein condensates}},\ }\href
  {https://doi.org/10.1103/PhysRevA.104.053315} {\bibfield  {journal} {\bibinfo
   {journal} {Phys. Rev. A}\ }\textbf {\bibinfo {volume} {104}},\ \bibinfo
  {pages} {053315} (\bibinfo {year} {2021}{\natexlab{a}})}\BibitemShut
  {NoStop}%
\bibitem [{\citenamefont {Gangwar}\ \emph {et~al.}(2024)\citenamefont
  {Gangwar}, \citenamefont {Ravisankar}, \citenamefont {Fabrelli},
  \citenamefont {Muruganandam},\ and\ \citenamefont {Mishra}}]{Gangwar2024}%
  \BibitemOpen
  \bibfield  {author} {\bibinfo {author} {\bibfnamefont {S.~K.}\ \bibnamefont
  {Gangwar}}, \bibinfo {author} {\bibfnamefont {R.}~\bibnamefont {Ravisankar}},
  \bibinfo {author} {\bibfnamefont {H.}~\bibnamefont {Fabrelli}}, \bibinfo
  {author} {\bibfnamefont {P.}~\bibnamefont {Muruganandam}},\ and\ \bibinfo
  {author} {\bibfnamefont {P.~K.}\ \bibnamefont {Mishra}},\ }\bibfield  {title}
  {\bibinfo {title} {{Emergence of unstable avoided crossing in the collective
  excitations of spin-1 spin-orbit-coupled Bose-Einstein condensates}},\ }\href
  {https://doi.org/10.1103/PhysRevA.109.043306} {\bibfield  {journal} {\bibinfo
   {journal} {Phys. Rev. A}\ }\textbf {\bibinfo {volume} {109}},\ \bibinfo
  {pages} {043306} (\bibinfo {year} {2024})}\BibitemShut {NoStop}%
\bibitem [{\citenamefont {Liu}\ and\ \citenamefont {Zhang}(2019)}]{Liu2019}%
  \BibitemOpen
  \bibfield  {author} {\bibinfo {author} {\bibfnamefont {S.}~\bibnamefont
  {Liu}}\ and\ \bibinfo {author} {\bibfnamefont {Y.}~\bibnamefont {Zhang}},\
  }\bibfield  {title} {\bibinfo {title} {{Quench dynamics in a trapped
  Bose-Einstein condensate with spin-orbit coupling}},\ }\href
  {https://doi.org/10.1103/PhysRevA.99.053609} {\bibfield  {journal} {\bibinfo
  {journal} {Phys. Rev. A}\ }\textbf {\bibinfo {volume} {99}},\ \bibinfo
  {pages} {053609} (\bibinfo {year} {2019})}\BibitemShut {NoStop}%
\bibitem [{\citenamefont {Deng}\ \emph {et~al.}(2016)\citenamefont {Deng},
  \citenamefont {Zhang}, \citenamefont {Yi},\ and\ \citenamefont
  {Guo}}]{Deng2016}%
  \BibitemOpen
  \bibfield  {author} {\bibinfo {author} {\bibfnamefont {T.-S.}\ \bibnamefont
  {Deng}}, \bibinfo {author} {\bibfnamefont {W.}~\bibnamefont {Zhang}},
  \bibinfo {author} {\bibfnamefont {W.}~\bibnamefont {Yi}},\ and\ \bibinfo
  {author} {\bibfnamefont {G.-C.}\ \bibnamefont {Guo}},\ }\bibfield  {title}
  {\bibinfo {title} {{Quench dynamics of a Bose-Einstein condensate under
  synthetic spin-orbit coupling}},\ }\href
  {https://doi.org/10.1103/PhysRevA.93.053621} {\bibfield  {journal} {\bibinfo
  {journal} {Phys. Rev. A}\ }\textbf {\bibinfo {volume} {93}},\ \bibinfo
  {pages} {053621} (\bibinfo {year} {2016})}\BibitemShut {NoStop}%
\bibitem [{\citenamefont {Ravisankar}\ \emph {et~al.}(2020)\citenamefont
  {Ravisankar}, \citenamefont {Sriraman}, \citenamefont {Salasnich},\ and\
  \citenamefont {Muruganandam}}]{Ravisankar_2020}%
  \BibitemOpen
  \bibfield  {author} {\bibinfo {author} {\bibfnamefont {R.}~\bibnamefont
  {Ravisankar}}, \bibinfo {author} {\bibfnamefont {T.}~\bibnamefont
  {Sriraman}}, \bibinfo {author} {\bibfnamefont {L.}~\bibnamefont
  {Salasnich}},\ and\ \bibinfo {author} {\bibfnamefont {P.}~\bibnamefont
  {Muruganandam}},\ }\bibfield  {title} {\bibinfo {title} {{Quenching dynamics
  of the bright solitons and other localized states in spin–orbit coupled
  Bose–Einstein condensates}},\ }\href
  {https://doi.org/10.1088/1361-6455/aba661} {\bibfield  {journal} {\bibinfo
  {journal} {J. Phys. B: At. Mol. Opt. Phys.}\ }\textbf {\bibinfo {volume}
  {53}},\ \bibinfo {pages} {195301} (\bibinfo {year} {2020})}\BibitemShut
  {NoStop}%
\bibitem [{\citenamefont {Ho}\ and\ \citenamefont {Zhang}(2011)}]{Ho2011}%
  \BibitemOpen
  \bibfield  {author} {\bibinfo {author} {\bibfnamefont {T.-L.}\ \bibnamefont
  {Ho}}\ and\ \bibinfo {author} {\bibfnamefont {S.}~\bibnamefont {Zhang}},\
  }\bibfield  {title} {\bibinfo {title} {{Bose-Einstein Condensates with
  Spin-Orbit Interaction}},\ }\href
  {https://doi.org/10.1103/PhysRevLett.107.150403} {\bibfield  {journal}
  {\bibinfo  {journal} {Phys. Rev. Lett.}\ }\textbf {\bibinfo {volume} {107}},\
  \bibinfo {pages} {150403} (\bibinfo {year} {2011})}\BibitemShut {NoStop}%
\bibitem [{\citenamefont {Yu}(2016)}]{Yu2016}%
  \BibitemOpen
  \bibfield  {author} {\bibinfo {author} {\bibfnamefont {Z.-Q.}\ \bibnamefont
  {Yu}},\ }\bibfield  {title} {\bibinfo {title} {Phase transitions and
  elementary excitations in spin-1 bose gases with raman-induced spin-orbit
  coupling},\ }\href {https://doi.org/10.1103/PhysRevA.93.033648} {\bibfield
  {journal} {\bibinfo  {journal} {Phys. Rev. A}\ }\textbf {\bibinfo {volume}
  {93}},\ \bibinfo {pages} {033648} (\bibinfo {year} {2016})}\BibitemShut
  {NoStop}%
\bibitem [{\citenamefont {Gangwar}\ \emph {et~al.}(2025)\citenamefont
  {Gangwar}, \citenamefont {Ravisankar}, \citenamefont {Fabrelli},
  \citenamefont {Muruganandam},\ and\ \citenamefont {Mishra}}]{Gangwar2025}%
  \BibitemOpen
  \bibfield  {author} {\bibinfo {author} {\bibfnamefont {S.~K.}\ \bibnamefont
  {Gangwar}}, \bibinfo {author} {\bibfnamefont {R.}~\bibnamefont {Ravisankar}},
  \bibinfo {author} {\bibfnamefont {H.}~\bibnamefont {Fabrelli}}, \bibinfo
  {author} {\bibfnamefont {P.}~\bibnamefont {Muruganandam}},\ and\ \bibinfo
  {author} {\bibfnamefont {P.~K.}\ \bibnamefont {Mishra}},\ }\bibfield  {title}
  {\bibinfo {title} {Double unstable avoided crossings and
  complex-domain-pattern formation in spin-orbit-coupled spin-1 condensates},\
  }\href {https://doi.org/10.1103/PhysRevA.111.063303} {\bibfield  {journal}
  {\bibinfo  {journal} {Phys. Rev. A}\ }\textbf {\bibinfo {volume} {111}},\
  \bibinfo {pages} {063303} (\bibinfo {year} {2025})}\BibitemShut {NoStop}%
\bibitem [{\citenamefont {Kawaguchi}\ and\ \citenamefont
  {Ueda}(2012)}]{Ueda2012}%
  \BibitemOpen
  \bibfield  {author} {\bibinfo {author} {\bibfnamefont {Y.}~\bibnamefont
  {Kawaguchi}}\ and\ \bibinfo {author} {\bibfnamefont {M.}~\bibnamefont
  {Ueda}},\ }\bibfield  {title} {\bibinfo {title} {{Spinor bose--einstein
  condensates}},\ }\href {https://doi.org/10.1016/j.physrep.2012.07.005}
  {\bibfield  {journal} {\bibinfo  {journal} {Phys. Rep.}\ }\textbf {\bibinfo
  {volume} {520}},\ \bibinfo {pages} {253} (\bibinfo {year}
  {2012})}\BibitemShut {NoStop}%
\bibitem [{\citenamefont {Stamper-Kurn}\ and\ \citenamefont
  {Ueda}(2013)}]{Stamper2013}%
  \BibitemOpen
  \bibfield  {author} {\bibinfo {author} {\bibfnamefont {D.~M.}\ \bibnamefont
  {Stamper-Kurn}}\ and\ \bibinfo {author} {\bibfnamefont {M.}~\bibnamefont
  {Ueda}},\ }\bibfield  {title} {\bibinfo {title} {{Spinor Bose gases:
  Symmetries, magnetism, and quantum dynamics}},\ }\href
  {https://doi.org/10.1103/RevModPhys.85.1191} {\bibfield  {journal} {\bibinfo
  {journal} {Rev. Mod. Phys.}\ }\textbf {\bibinfo {volume} {85}},\ \bibinfo
  {pages} {1191} (\bibinfo {year} {2013})}\BibitemShut {NoStop}%
\bibitem [{\citenamefont {Ravisankar}\ \emph
  {et~al.}(2021{\natexlab{b}})\citenamefont {Ravisankar}, \citenamefont
  {Sriraman}, \citenamefont {Kumar}, \citenamefont {Muruganandam},\ and\
  \citenamefont {Mishra}}]{Ravisankar_2021}%
  \BibitemOpen
  \bibfield  {author} {\bibinfo {author} {\bibfnamefont {R.}~\bibnamefont
  {Ravisankar}}, \bibinfo {author} {\bibfnamefont {T.}~\bibnamefont
  {Sriraman}}, \bibinfo {author} {\bibfnamefont {R.~K.}\ \bibnamefont {Kumar}},
  \bibinfo {author} {\bibfnamefont {P.}~\bibnamefont {Muruganandam}},\ and\
  \bibinfo {author} {\bibfnamefont {P.~K.}\ \bibnamefont {Mishra}},\ }\bibfield
   {title} {\bibinfo {title} {{Influence of Rashba spin–orbit and Rabi
  couplings on the spin-mixing and ground state phases of binary
  Bose–Einstein condensates}},\ }\href
  {https://doi.org/10.1088/1361-6455/ac41b2} {\bibfield  {journal} {\bibinfo
  {journal} {J. Phys. B: At. Mol. Opt. Phys.}\ }\textbf {\bibinfo {volume}
  {54}},\ \bibinfo {pages} {225301} (\bibinfo {year}
  {2021}{\natexlab{b}})}\BibitemShut {NoStop}%
\bibitem [{\citenamefont {Sarkar}\ \emph {et~al.}(2024)\citenamefont {Sarkar},
  \citenamefont {Mardonov}, \citenamefont {Sherman}, \citenamefont
  {Muruganandam},\ and\ \citenamefont {Mishra}}]{SKSarkar2024}%
  \BibitemOpen
  \bibfield  {author} {\bibinfo {author} {\bibfnamefont {S.~K.}\ \bibnamefont
  {Sarkar}}, \bibinfo {author} {\bibfnamefont {S.}~\bibnamefont {Mardonov}},
  \bibinfo {author} {\bibfnamefont {E.~Y.}\ \bibnamefont {Sherman}}, \bibinfo
  {author} {\bibfnamefont {P.}~\bibnamefont {Muruganandam}},\ and\ \bibinfo
  {author} {\bibfnamefont {P.~K.}\ \bibnamefont {Mishra}},\ }\bibfield  {title}
  {\bibinfo {title} {{Spin-dependent localization of spin-orbit and
  Rabi-coupled Bose-Einstein condensates in a random potential}},\ }\href
  {https://doi.org/10.1088/1367-2630/adafd8} {\bibfield  {journal} {\bibinfo
  {journal} {New J. Phys.}\ }\textbf {\bibinfo {volume} {27}},\ \bibinfo
  {pages} {023018} (\bibinfo {year} {2024})}\BibitemShut {NoStop}%
\bibitem [{\citenamefont {Inouye}\ \emph {et~al.}(1998)\citenamefont {Inouye},
  \citenamefont {Andrews}, \citenamefont {Stenger}, \citenamefont {Miesner},
  \citenamefont {Stamper-Kurn},\ and\ \citenamefont {Ketterle}}]{Inouye1998}%
  \BibitemOpen
  \bibfield  {author} {\bibinfo {author} {\bibfnamefont {S.}~\bibnamefont
  {Inouye}}, \bibinfo {author} {\bibfnamefont {M.}~\bibnamefont {Andrews}},
  \bibinfo {author} {\bibfnamefont {J.}~\bibnamefont {Stenger}}, \bibinfo
  {author} {\bibfnamefont {H.-J.}\ \bibnamefont {Miesner}}, \bibinfo {author}
  {\bibfnamefont {D.~M.}\ \bibnamefont {Stamper-Kurn}},\ and\ \bibinfo {author}
  {\bibfnamefont {W.}~\bibnamefont {Ketterle}},\ }\bibfield  {title} {\bibinfo
  {title} {{Observation of Feshbach resonances in a Bose--Einstein
  condensate}},\ }\href {https://doi.org/10.1038/32354} {\bibfield  {journal}
  {\bibinfo  {journal} {Nature (London)}\ }\textbf {\bibinfo {volume} {392}},\
  \bibinfo {pages} {151} (\bibinfo {year} {1998})}\BibitemShut {NoStop}%
\bibitem [{\citenamefont {Marte}\ \emph {et~al.}(2002)\citenamefont {Marte},
  \citenamefont {Volz}, \citenamefont {Schuster}, \citenamefont {D\"urr},
  \citenamefont {Rempe}, \citenamefont {van Kempen},\ and\ \citenamefont
  {Verhaar}}]{Marte2002}%
  \BibitemOpen
  \bibfield  {author} {\bibinfo {author} {\bibfnamefont {A.}~\bibnamefont
  {Marte}}, \bibinfo {author} {\bibfnamefont {T.}~\bibnamefont {Volz}},
  \bibinfo {author} {\bibfnamefont {J.}~\bibnamefont {Schuster}}, \bibinfo
  {author} {\bibfnamefont {S.}~\bibnamefont {D\"urr}}, \bibinfo {author}
  {\bibfnamefont {G.}~\bibnamefont {Rempe}}, \bibinfo {author} {\bibfnamefont
  {E.~G.~M.}\ \bibnamefont {van Kempen}},\ and\ \bibinfo {author}
  {\bibfnamefont {B.~J.}\ \bibnamefont {Verhaar}},\ }\bibfield  {title}
  {\bibinfo {title} {{Feshbach Resonances in Rubidium 87: Precision Measurement
  and Analysis}},\ }\href {https://doi.org/10.1103/PhysRevLett.89.283202}
  {\bibfield  {journal} {\bibinfo  {journal} {Phys. Rev. Lett.}\ }\textbf
  {\bibinfo {volume} {89}},\ \bibinfo {pages} {283202} (\bibinfo {year}
  {2002})}\BibitemShut {NoStop}%
\bibitem [{\citenamefont {Chin}\ \emph {et~al.}(2010)\citenamefont {Chin},
  \citenamefont {Grimm}, \citenamefont {Julienne},\ and\ \citenamefont
  {Tiesinga}}]{Chin2010}%
  \BibitemOpen
  \bibfield  {author} {\bibinfo {author} {\bibfnamefont {C.}~\bibnamefont
  {Chin}}, \bibinfo {author} {\bibfnamefont {R.}~\bibnamefont {Grimm}},
  \bibinfo {author} {\bibfnamefont {P.}~\bibnamefont {Julienne}},\ and\
  \bibinfo {author} {\bibfnamefont {E.}~\bibnamefont {Tiesinga}},\ }\bibfield
  {title} {\bibinfo {title} {{Feshbach resonances in ultracold gases}},\ }\href
  {https://doi.org/10.1103/RevModPhys.82.1225} {\bibfield  {journal} {\bibinfo
  {journal} {Rev. Mod. Phys.}\ }\textbf {\bibinfo {volume} {82}},\ \bibinfo
  {pages} {1225} (\bibinfo {year} {2010})}\BibitemShut {NoStop}%
\bibitem [{\citenamefont {Rajaswathi}\ \emph {et~al.}(2023)\citenamefont
  {Rajaswathi}, \citenamefont {Bhuvaneswari}, \citenamefont {Radha},\ and\
  \citenamefont {Muruganandam}}]{Rajaswathi2023}%
  \BibitemOpen
  \bibfield  {author} {\bibinfo {author} {\bibfnamefont {K.}~\bibnamefont
  {Rajaswathi}}, \bibinfo {author} {\bibfnamefont {S.}~\bibnamefont
  {Bhuvaneswari}}, \bibinfo {author} {\bibfnamefont {R.}~\bibnamefont
  {Radha}},\ and\ \bibinfo {author} {\bibfnamefont {P.}~\bibnamefont
  {Muruganandam}},\ }\bibfield  {title} {\bibinfo {title} {{Dispersion
  engineering in spin-orbit-coupled spinor $F=1$ condensates driven by negative
  masses}},\ }\href {https://doi.org/10.1103/PhysRevA.108.033317} {\bibfield
  {journal} {\bibinfo  {journal} {Phys. Rev. A}\ }\textbf {\bibinfo {volume}
  {108}},\ \bibinfo {pages} {033317} (\bibinfo {year} {2023})}\BibitemShut
  {NoStop}%
\bibitem [{\citenamefont {Ma}\ and\ \citenamefont {Jia}(2019)}]{Ma2019}%
  \BibitemOpen
  \bibfield  {author} {\bibinfo {author} {\bibfnamefont {D.}~\bibnamefont
  {Ma}}\ and\ \bibinfo {author} {\bibfnamefont {C.}~\bibnamefont {Jia}},\
  }\bibfield  {title} {\bibinfo {title} {{Soliton oscillation driven by
  spin-orbit coupling in spinor condensates}},\ }\href
  {https://doi.org/10.1103/PhysRevA.100.023629} {\bibfield  {journal} {\bibinfo
   {journal} {Phys. Rev. A}\ }\textbf {\bibinfo {volume} {100}},\ \bibinfo
  {pages} {023629} (\bibinfo {year} {2019})}\BibitemShut {NoStop}%
\bibitem [{\citenamefont {Mithun}\ and\ \citenamefont
  {Kasamatsu}(2019)}]{Mithun_2019}%
  \BibitemOpen
  \bibfield  {author} {\bibinfo {author} {\bibfnamefont {T.}~\bibnamefont
  {Mithun}}\ and\ \bibinfo {author} {\bibfnamefont {K.}~\bibnamefont
  {Kasamatsu}},\ }\bibfield  {title} {\bibinfo {title} {{Modulation instability
  associated nonlinear dynamics of spin–orbit coupled Bose–Einstein
  condensates}},\ }\href {https://doi.org/10.1088/1361-6455/aafbdd} {\bibfield
  {journal} {\bibinfo  {journal} {J. Phys. B: At. Mol. Opt. Phys.}\ }\textbf
  {\bibinfo {volume} {52}},\ \bibinfo {pages} {045301} (\bibinfo {year}
  {2019})}\BibitemShut {NoStop}%
\end{thebibliography}%

\end{document}